\documentclass{aa}

\usepackage{lscape}
\usepackage[usenames, dvipsnames]{color}
\usepackage{booktabs, caption, makecell}
\usepackage{epstopdf}
\usepackage{gensymb}
\usepackage{threeparttable}
\usepackage{geometry}
\usepackage{hyperref}
\hypersetup{    colorlinks=true, linkcolor=blue, filecolor=blue,   urlcolor=blue, citecolor=blue, }

\usepackage{amssymb}
\usepackage{pifont}
\newcommand{\cmark}{\ding{51}}%
\newcommand{\xmark}{\ding{55}}%

\graphicspath{ {plots/}  }
\usepackage{epstopdf}

\definecolor{mypink3}{cmyk}{0.26, 1.0, 0.0, 0.56}
\definecolor{mygreen1}{rgb}{0.314, 0.588, 0.078}
\definecolor{gray}{rgb}{0.5, 0.5, 0.5}

\usepackage{color}

\usepackage{graphicx}
\usepackage{txfonts}
\begin{document}

   \title{Which molecule traces what: \\chemical diagnostics of protostellar sources}

   \author{Łukasz Tychoniec, \inst{1,2} Ewine F. van Dishoeck, \inst{2,3} Merel L.R. van 't Hoff, \inst{ 4}  Martijn L. van Gelder,\inst{2} Beno\^{i}t Tabone, \inst{2} \\ Yuan Chen, \inst{2}
  Daniel Harsono, \inst{5} Charles L. H. Hull,  \inst{6,7} Michiel R. Hogerheijde, \inst{2, 8} Nadia M. Murillo, \inst{9}  John J. Tobin \inst{10} 
          }
   \institute{European Southern Observatory, Karl-Schwarzschild-Strasse 2, 85748 Garching bei M\"unchen, Germany\\
                 \email{lukasz.tychoniec@eso.org}
                 \and
   Leiden Observatory, Leiden University, PO Box 9513, 2300RA, Leiden, The Netherlands
              \and
Max-Planck-Institut f{\"u}r Extraterrestrische Physik, Giessenbachstrasse 1, D-85748 Garching, Germany
\and
Department of Astronomy, University of Michigan, 500 Church Street, Ann Arbor, MI 48109, USA
\and
Institute of Astronomy and Astrophysics, Academia Sinica, No. 1, Sec. 4, Roosevelt Road, Taipei 10617, Taiwan, R. O. C.
\and
National Astronomical Observatory of Japan, NAOJ Chile, Alonso de 
C\'ordova 3788, Office 61B, 7630422, Vitacura, Santiago, Chile
\and
Joint ALMA Observatory, Alonso de C\'ordova 3107, Vitacura, Santiago, Chile
\and
Anton Pannekoek Institute for Astronomy, University of Amsterdam,
Science Park 904, 1098 XH Amsterdam, The Netherlands
\and
Star and Planet Formation Laboratory, RIKEN Cluster for Pioneering Research, Wako, Saitama 351-0198, Japan
\and
National Radio Astronomy Observatory, 520 Edgemont Road, Charlottesville, VA 22903, USA
}

  \abstract
  {
The physical and chemical conditions in Class 0/I protostars are fundamental in unlocking the protostellar accretion process and its impact on planet formation.}
   {The aim is to determine which physical components are traced by different molecules at sub-arcsecond scales (\textless 100 -- 400 au).}
   {We use a suite of Atacama Large Millimeter/submillimeter Array (ALMA) datasets in Band 6 (1 mm), Band 5 (1.8 mm) and Band 3 (3 mm) at spatial resolutions 0\farcs5 -- 3\arcsec\   for 16 protostellar sources. For a subset of sources, Atacama Compact Array (ACA) data at Band 6 with a spatial resolution of 6$''$  are added. The availability of both low- and high-excitation lines, as well as data on small and larger scales, are important to unravel the full picture.}
   {The protostellar envelope is well traced by  C$^{18}$O, DCO$^+$ and N$_2$D$^+$, with the freeze-out of CO governing the chemistry at envelope scales. Molecular outflows are seen in classical shock tracers like SiO and SO, but ice-mantle products such as CH$_3$OH and HNCO released with the shock are also observed. The molecular jet is a key component of the system, only present at the very early stages, and prominent not only in SiO and SO but also occasionally in H$_2$CO. The cavity walls show tracers of UV-irradiation such as hydrocarbons C$_2$H and c-C$_3$H$_2$ as well as CN. The hot inner envelope, apart from showing emission from complex organic molecules (COMs), also presents compact emission from small molecules like H$_2$S, SO, OCS and H$^{13}$CN, most likely related to ice sublimation and high-temperature chemistry.}  
   {Sub-arcsecond millimeter-wave observations allow to identify those (simple) molecules that best trace each of the physical components of a protostellar system. COMs are found both in the hot inner envelope (high excitation lines) and in the outflows (lower-excitation lines) with comparable abundances.  COMs can coexist with hydrocarbons in the same protostellar sources, but they trace different components.  In the near future, mid-IR observations with JWST--MIRI will provide complementary information about the hottest gas and the ice mantle content, at unprecedented sensitivity and at resolutions comparable to ALMA for the same sources.}

   \titlerunning{Which molecule traces what}
   \authorrunning{Tychoniec et al.}
   \maketitle

\section{Introduction}

The formation of Sun-like stars is set in motion by the collapse of a cold,
dense cloud. Many different physical processes take place in the protostellar stage --  first few $10^5$ yrs that are critical to the subsequent evolution of the star and its planetary system \citep{Lada1987}. The mass of the star and that of its
circumstellar disk are determined during this embedded phase
\citep{Hueso2005} and the first steps of planet formation must take
place then \citep{Greaves2010, Williams2012, ALMA2015,Manara2018,Harsono2018,Tobin2020,Tychoniec2018a,Tychoniec2020,SeguraCox2020}. At the
same time, on larger scales, the collapsing envelope is dispersed by
the energetic action of bipolar jets and winds emanating from the star-disk
system which create outflows of entrained gas and dust \citep{Arce2006,Offner2014}

Rotational transitions of molecules are a powerful tool to probe other
components of the system and can be used to infer densities,
temperatures, UV fields, chemical abundances and kinematics
\citep{vanDishoeck1998,Evans1999}. Until recently, studies of low-mass protostars have suffered from insufficient spatial resolution to disentangle these different physical
components. The advent of submillimeter interferometry opened the possibility to study protostellar systems at much smaller scales than with single-dish observations \citep[e.g.,][]{Chandler1996,Hogerheijde1999,Schilke1992, Wilner2000, Jorgensen2005b, Tobin2011}.

With the Atacama Large Millimeter/submillimeter Array (ALMA), it is possible to
image many molecular lines on the relevant physical scales with achievable observing times at sub-arcsecond resolution. Impressive ALMA studies of individual low-mass protostars
have been presented, focusing both on simple species (< 6 atoms) and complex molecules (> 6 atoms)
\citep[e.g.,][]{Sakai2014b, Jorgensen2016, LopezSepulcre2017,Lee2019,  Codella2018, LeeJE2019, Manigand2020, vanGelder2020, Bianchi2020}. The first ALMA surveys of complex molecules on a larger sample are taking place \citep{Yang2021} and of Class I disks \citep{Podio2020, Garufi2020}. The CALYPSO survey with NOEMA studied a larger sample of protostars but at more limited resolution and sensitivity \citep{Belloche2020, Maret2020, Podio2021}.

Here we present ALMA data of 16 protostellar sources covering rotational transitions of various molecules; we use these data to build a complete picture of what types of molecules trace which physical structures in protostars. This sample constitutes one of the largest combinations of high-resolution ALMA observations of Class 0/I protostars to date in ALMA Band 3, 5, and 6. Covering a broad range of protostellar properties
within the low-mass regime, the aim is to identify and describe key molecular tracers of future Sun-like stars and what physical components of star-forming sources they correspond to. Parts of the ALMA datasets presented here have been already published, focusing on different aspects: complex organic molecules \citep{vanGelder2020,Nazari2021}; Class I disks temperature structure \citep{vantHoff2020b}; outflows and high-velocity jets in Serpens \citep{Hull2016, Tychoniec2019}, molecular emission associated with magnetic fields \citep{Hull2017, LeGouellec2019}. In this work we make a comprehensive overview of these different datasets, making full use of those observations, with uniform analysis methods. This allows to reveal and systematise the molecular tracers of Class 0/I protostars. Out of 16 presented sources 11 are included in upcoming {\it James Webb Space Telescope} (JWST) observations with the Mid-Infrared Instrument (MIRI;  \citealt{Wright2015})  (\citeyear{JWST1290,JWST1257}) and Near-Infrared Spectrograph (NIRSpec) ( \citeyear{JWST1960,JWST1186,JWST1798,JWST2104})

\subsection{Physical components of a protostellar system}

Class 0 sources are defined by their strong excess of submillimeter
luminosity and very low bolometric temperatures \textless 70~K
\citep{Andre1993, Chen1995}. These sources are associated with
powerful outflows, and the envelope mass dominates the mass of the
entire system.  Class I sources are defined by having an infrared
spectral index indicating strong reddening \citep{Lada1987} with
bolometric temperatures of 70--650 K \citep{Chen1995}. Those systems
have already converted most of their envelope mass into disk and
protostar \citep{Crapsi2008, vanKempen2009a, Maury2011}. For the
typical envelope masses of sources presented here and average disk
masses found by \citet{Tychoniec2020}, the $M_{\rm disk}/M_{\rm
  env}\simeq$ 1\% for Class 0 and $\simeq$ 20\% for Class I, with
values up to 75--98\% in cases of rotationally supported disks \citep{Jorgensen2009}.

\begin{table*}
\caption{Targeted protostellar systems}             
\label{table:targets}      
\centering                          
\begin{tabular}{l l c c c c c  c c}        
\hline\hline                 
Source name & R.A. & Decl. & $d$ & Class & $L_{\rm bol}$ &   $T_{\rm bol}$ & $M_{\rm env}$ &  Ref. \\
 & (J2000) & (J2000) & (pc) & & (L$_{\odot}$) &   (K) & (M$_{\odot}$) &   \\

\hline                                   
Serpens SMM1 &  18:29:49.8 & +01:15:20.5 & 439 & 0 & 109 &   39  & 58 &  (1)\\

Serpens S68N & 18:29:48.1 & +01:16:43.3 & 439 & 0 & 6 &   58  & 10 &  (2) \\

Ser-emb 8 (N) & 18:29:48.7 & +01:16:55.5 & 439 & 0 & --- &  ---  & --- &  ---\\
Serpens SMM3 & 18:29:59.2   & +01:14:00.3  & 439 & 0 & 28 & 38 &  13  & (1) \\

BHR 71 & 12:01:36.3 &  -65:08:53.0 & 200 & 0 & 15 &  44  & 2.7  &  (1)\\

IRAS 4B & 03:29:12.0  &  +31 13 08.1 & 293 & 0 & 7 &  28  & 4.7 &  (1)\\

Per-emb-25 & 03:26:37.5   &  +30:15:27.8 & 293 & 0/I & 1.9 &  61  & 2.0 &  (2) \\

B1-c &  03:33:17.9   & +31:09:31.8 & 293 & 0 & 5 &  48  & 15 &  (2) \\

HH211-mm  & 03:43:56.8   & +32:00:50.2 & 293  & 0 & 2.8 &  27  & 19 &  (2) \\
L1448-mm & 03:25:38.9   & +30:44:05.3 & 293 & 0 & 13 & 47 &  15  & (2) \\

L1527 IRS & 04:39:53.9   &  +26:03:09.5   & 140 & 0/I & 1.6 & 79 & 0.12 & (3) \\

B5-IRS1 & 03:47:41.6 & +32:51:43.7 & 293 & I & 7 &  181  & 3.5 &  (2) \\
TMC1 & 04:41:12.7   & +25:46:34.8  &140 & I  & 0.9 &  101  & 0.14 &  (1) \\ 
IRAS 04302 & 04:33:16.5   & +22:53:20.4 &140 & I & 0.7 &  300  & 0.05 & (1) \\
L1489 IRS & 04:04:43.0   & +26:18:57.0 &140 & I  & 3.8 &  200  & 0.2 &  (1)\\
TMC1A & 04:39:34.9  &  +25:41:45.0 &  140 & I & 2.7  &  118 & 0.2 & (1)\\

\hline                                   

\end{tabular}
\begin{tablenotes}\footnotesize
\item{(1) \citealt{Kristensen2012}, (2) \citealt{Enoch2009}, (3) \citealt{Green2013} } 

\end{tablenotes}
\end{table*}

The different components of protostellar systems vary significantly in
their physical conditions, such as density and temperature, molecular
enrichment, and dynamics.  Our current knowledge about them is described briefly below to set the scene for the interpretation of our data.

{\it Envelope.} The envelope surrounding a protostar is the material
that fuels the accretion process onto the star and disk. The
physical conditions in the outer envelope on scales of a few 1000 au
are reminiscent of those of starless cores with heavy freeze-out, and their chemical
composition is directly inherited from the cloud out of which the star
is being born \citep{Caselli2012}. Systematic motions such as infall
or expansion can occur but otherwise they are characterized by low turbulence and
narrow (FWHM < 0.5 - 1 km s$^{-1}$) line profiles indicative of quiescent gas \citep{Jorgensen2002}. 

{\it Warm inner envelope}. In the innermost part of the envelope on scales of the
disk, temperatures rise above 100 K, so any water and complex organic
molecules (COMs) contained in ices are released from the grains back
into the gas where they are readily observed at submillimeter
wavelengths. This region with its unique chemical richness is called
the hot core, or to distinguish it from its high-mass counterpart, hot
corino \citep{Herbst2009}.

{\it Jets and outflows}. As the material is accreting from the envelope onto the disk, excess angular momentum has to be transported via a still elusive process to allow material to accrete on the growing protostar. Jets and outflows constitute compelling  candidates to extract angular momentum via magnetic fields. In the earliest stages when the mass loss is at its peak, the densities are high enough to form molecules in the internal shocks in the jet \citep{Bachiller1992, Tafalla2010}. Much slower (\textless 20 km s$^{-1}$) and less collimated gas moving away from the protostar is called an outflow. Their origin remains debated. Large scale outflows reveal bow-shock shells and cavities possibly driven by the fast intermittent jet \citep{Gueth1996, Gueth1999, Tychoniec2019}. Temperatures in shocked regions are much higher than in the surrounding envelope, up to a few thousand K, and sputtering of grain cores and ice mantles can further result in unique chemical signatures \citep{Arce2008, Flower2013}. 

{\it Outflow cavity walls.} These are the narrow zones in between the
cold dense quiescent envelope material and the lower-density warm cone
where outflows are propagating at large velocities. Cavity walls are
exposed to UV radiation from the accreting star-disk boundary layer,
which can escape through the outflow cavity without being extincted
\citep{Spaans1995}. This creates conditions similar to those found in
Photon Dominated Regions (PDRs), which occur throughout the
interstellar medium near sources of intense UV radiation
\citep{Hollenbach1997}.  In units of the interstellar radiation field
(ISRF, \citealt{Draine1978}), typical values of 10$^2$--10$^3$ are found
on scales of $\sim 1000$ au \citep{vanKempen2009c, Yildiz2012,Benz2016,
  Karska2018}.

{\it Young disk.} In the inner envelope, a protoplanetary disk starts
to form as the natural outcome of a rotating collapsing core
\citep{Ulrich1976, Cassen1981,Terebey1984}. 
A young disk should be rotating in Keplerian motion. At early stages it is difficult to identify whether the so-called embedded disk is
rotationally supported, since any molecular emission from the disk is entangled with that from the envelope. In recent years several
embedded disks have been identified to have Keplerian rotational
structure on scales of $\sim$100 au \citep{Tobin2012,Murillo2013,Ohashi2014, Yen2017}. Molecular tracers in young disks, apart from
providing the kinematic information, can probe their temperature
structure as well \citep{vantHoff2018a}.

This work is organized as follows. In Section \ref{section:observations}, the observations used in this work are presented, Section \ref{section:results} presents the results of this work including detections and morphology of the targeted molecules. In Section \ref{section:discussion} the results are discussed. with special focus on which molecular tracers are corresponding to each of the physical components. The focus is on a qualitative description, rather than quantitative analyses for which source specific models and more rotational transitions of a
given molecule would be needed. We summarize our
work in Section \ref{section:conclusions}.

\section{Observations}
\label{section:observations}

\subsection{Datasets}

Six different ALMA 12m datasets at Band 3, 5, and 6 are used in this work to cover 14 out of 16 sources. The spatial resolution of all ALMA 12m datasets is comparable (0\farcs3--0\farcs6), except for 2017.1.01174.S, where Band 3 observations are obtained at 3\arcsec. Additionally for 6 out of 16 sources in Band 6, ACA observations with 7m antennas were obtained at 6\arcsec\  resolution.

\begin{figure*}[h]
\centering
  \includegraphics[width=1\linewidth,trim={0cm 0cm 0cm 0cm}]
   {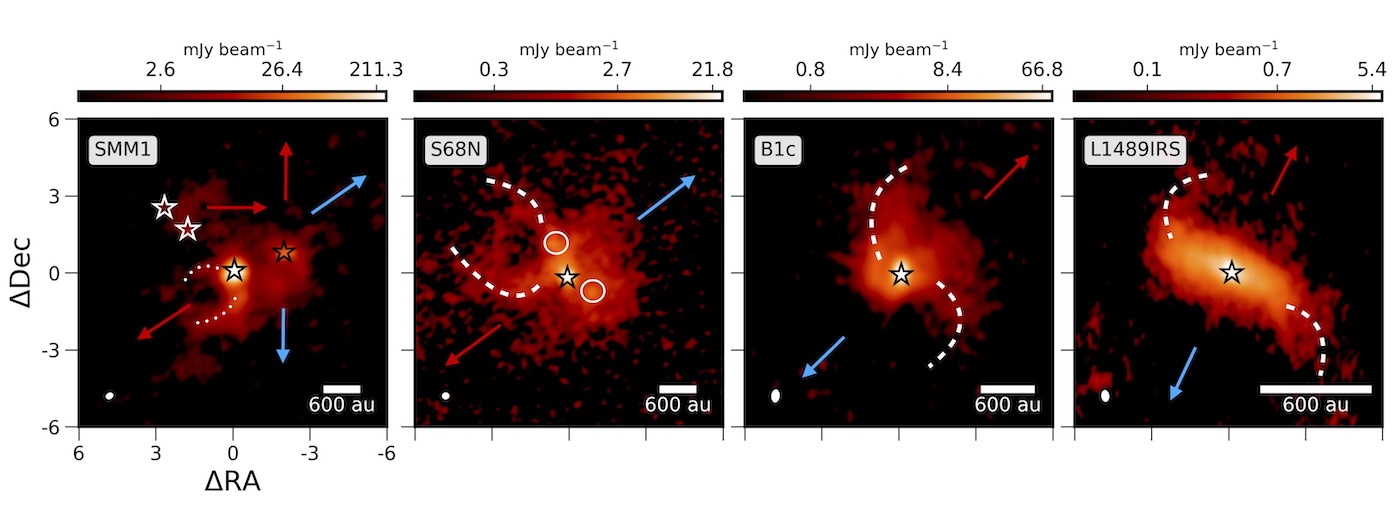}
  \caption{ Continuum emission at 1.3 mm of four example protostellar systems obtained with ALMA at 0\farcs5 resolution. Symbols of stars point to confirmed protostellar sources, circles show condensations of continuum emission, without confirmed protostellar nature, dotted lines show outflow cavity walls, and dashed lines show streams of envelope material. Arrows indicate outflow directions.}
 \label{fig:continuum_showcase_v1}
\end{figure*}

The targeted protostars are well-known objects located in different
star-forming regions. They span a range of properties within the
low-mass regime; the probed range of $L_{\rm bol}$, $T_{\rm bol}$ and
$M_{\rm env}$ is shown in Table  \ref{table:targets}, and are provided by a suite of observations across the infrared and submillimeter spectrum  \citep{Enoch2009,Kristensen2012,Green2013}.

The details of the observations  are summarized in Table \ref{table:observations}.  The spatial resolution allows to observe protostellar systems at solar-system scales; Band 5 and 6 observations provide a resolution of $\sim$0\farcs5, which corresponds to a 70--220 au diameter for sources in our sample. Thus, regions down to 35-110 au radius in the inner envelope are probed. The Band 3 data achieve moderate resolution of $\sim$3\arcsec\   which provides information on intermediate envelope scales of 500--1500 au. The ACA observations of six sources at 6\arcsec\  resolution probe envelope scales of 800--2000 au. 

\subsection{Spectral setup of the observations}

A collection of different datasets using different ALMA bands implies
varying spectral and spatial resolution as well as spectral coverage
across the analysis. This is the reason that throughout this paper the
sources shown in the figures differ when presenting detections and maps of different
molecules. In all cases, when the molecule is discussed, only those
sources where the given transition has been targeted are
discussed. All non-detections are explicitly
stated. Table \ref{table:table_mols} provides a list of targeted
molecular transitions, with sources that have a particular line covered and
detected or not detected.

\begin{figure*}[h]
\centering
  \includegraphics[width=0.95\linewidth,trim={2cm 1cm 2cm 0cm}]
  {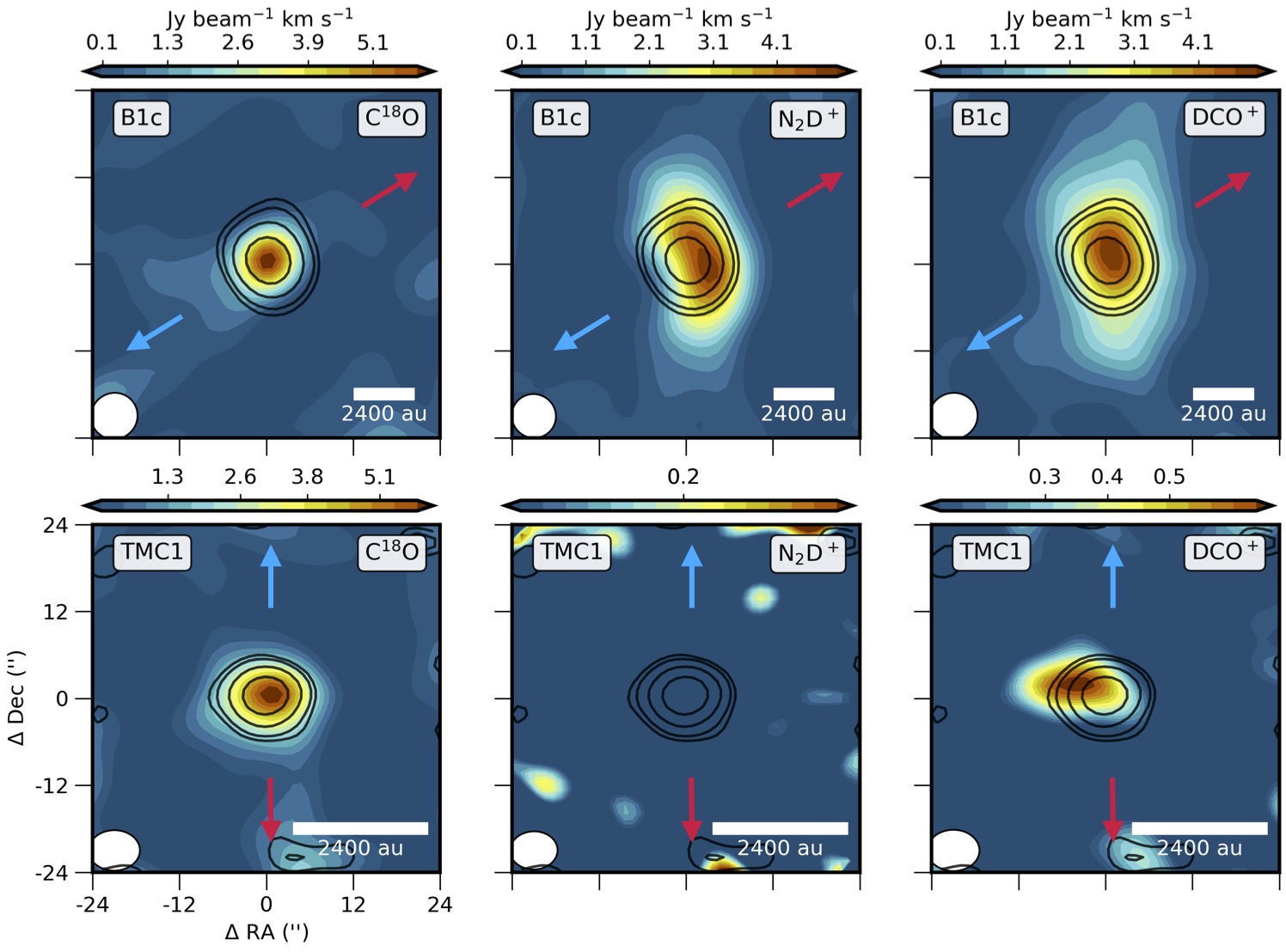}
  
  \caption{Maps of key envelope tracers toward B1-c (Class 0, top) and TMC1 (Class I, bottom) obtained with ACA. Contours represent continuum emission at 1.3 mm observed with ACA. Note different distances to B1-c and TMC1 resulting in different spatial resolutions of the maps. {\it Left:} C$^{18}$O 2 -- 1 {\it Middle:} N$_2$D$^+$ 3 -- 2. {\it Right:} DCO$^+$ 3 -- 2. All moment 0 maps are integrated from -2.5 to 2.5 km s$^{-1}$ w.r.t $\varv_{\rm sys} $.}
 \label{fig:envelope_plot_no1_v1}
\end{figure*}

The ALMA observations presented here target different spectral setups across Band 3, 5, and 6. This allows to probe the strongly varying physical scales and conditions. In particular, our Band 3 data grants access to lines at very low excitation levels that enable tracing more extended material. Dust is less optically thick in Band 3 compared with Band 6 which potentially allows to peek inside the densest inner regions. Band 6 offers a usual set of tracers of outflows: CO, SiO and SO. 

Cold outer envelopes with temperatures \textless~20~K are probed
with low $E_{\rm up}$ transitions. Additionally, non-thermal processes
such as sputtering of material from the grains in the outflow, will
also be seen in low $E_{\rm up}$ due to their lower critical
densities. On the other hand, thermal desorption from grains in
the innermost regions are best probed with lines with high $E_{\rm up}$. With the large span of frequencies of the observations, different transitions of the same
molecule can be detected and used to trace different components of the system (e.g., a HNCO line at $E_{\rm up}=$  15 K is available in Band 3 and lines at 70 and 125 K are covered in Band 6). 

\section{Results}
\label{section:results}

\subsection{Continuum emission from protostars }

Figure \ref{fig:continuum_showcase_v1} presents continuum emission maps toward four example protostars obtained with ALMA at $\sim$ 0\farcs5. The examples illustrate most characteristic features observed in the continuum maps. The continuum images for all sources are presented in  Fig. \ref{fig:continuum_no1_v1}.

The continuum emission observed at millimeter wavelengths (1.3 - 3 mm in our observations) traces thermal dust emission from the inner envelope and the embedded disk. In the Class 0 protostars, the central continuum source generally appears compact $R$ \textless 100 au (e.g., B1-c, Fig. \ref{fig:continuum_showcase_v1}). The example of SMM3 (Fig. \ref{fig:continuum_no1_v1}) shows a large resolved dust structure perpendicular to the outflow, but its classification as a disk is not certain. The fact that we observe primarily compact continuum emission towards Class 0 sources is consistent with observations of confirmed rotationally-supported disks \citep[e.g.,][]{Tobin2018, Maury2019, Tobin2020} and predictions of models \citep{Visser2010,Harsono2015b, Machida2016}. While we assume that this compact emission belongs to the young, embedded disk, we do not have spatial and kinematic resolution to confirm the presence of Class 0 Keplerian disks in our sample.

The extended continuum emission in Class 0 sources is consistent with a significant amount of envelope material surrounding the protostar.
In the case of the SMM1 system, presented in Fig. \ref{fig:continuum_showcase_v1} (left), the continuum clumps outside the central emission are  components of multiple protostellar systems, marked with stars, confirmed by the presence of individual molecular outflows  \citep{Hull2016, Hull2017}. Binary components are also seen in TMC1, BHR71, and IRAS 4B (Fig. \ref{fig:continuum_no1_v1}). In the case of S68N presented in Fig. \ref{fig:continuum_showcase_v1}, two emission peaks that stand out from the diffuse envelope emission are marked with circles, but their protostellar nature is not confirmed.

\subsection{Protostellar envelope}

We present molecules that trace the bulk of the protostellar envelope in Fig. \ref{fig:envelope_plot_no1_v1}. The protostellar envelope has a typical radius on the order of a few 1000 au  \citep{Jorgensen2002,Kristensen2012}.  Thus, the sub-arcsecond ALMA 12m array observations tend to resolve-out the envelope emission. For instance, the maximum recoverable scale (MRS) of ALMA Band 6 observations at 0\farcs4  presented here is 5\arcsec, which is between 600--2000 au diameter depending on distance to the source. For that reason, we discuss in this section mainly the ALMA-ACA observations obtained at lower spatial resolution (6\arcsec; 750--2500 au) for six sources in our datasets; Class 0 sources: B1-c, BHR71, Per-emb-25, SMM3, and IRAS 4B, and Class I source TMC1. The ACA can zoom-in on what was previously contained in a single-dish beam of 15--20\arcsec, while the MRS of ACA (30\arcsec) enables us to preserve sensitivity to large-scale emission.  The MRS of all observations presented here are reported in Table \ref{table:observations}.

In Fig. \ref{fig:envelope_plot_no1_v1} typical envelope tracers C$^{18}$O 2--1 ($E_{\rm up} = 16$ K),  DCO$^+$ 3--2 ($E_{\rm up} = 21$ K), and N$_2$D$^+$ 3--2 ($E_{\rm up} = 22$ K), observed at 6\arcsec\    resolution with the ACA, are presented toward example Class 0 and Class I sources -- B1-c and TMC1, respectively. The emission from the presented molecules exhibits similar behaviour for all Class 0 sources, therefore B1-c serves as a representative case; TMC1 is the only Class I source in the sample with 7m observations available. The maps for all sources for which these molecules have been targeted can be found in Appendix \ref{AppendixD}. All envelope tracers presented here are characterized by narrow line profiles with FWHM $\sim 1$ km s$^{-1}$.

The C$^{18}$O emission peak coincides with the continuum peak for our six sources and appears to be compact, less than 1000 au diameter for B1-c and TMC1. For B1-c and all Class 0 sources (Fig. {\ref{fig:envelope_co18_7m}), low-level extended C$^{18}$O emission is seen along the outflow direction. For the only Class I source targeted with the ACA, emission is marginally resolved in the direction perpendicular to the outflow.

In our observations N$_2$D$^+$ is seen extended in the direction perpendicular to the core major axis toward B1-c (Fig. \ref{fig:envelope_plot_no1_v1}) and other Class 0 sources except IRAS 4B (Fig.{\ref{fig:envelope_n2d+_7m}). In the case of IRAS 4B the emission from this molecule appears dominated by large-scale emission from the filament detected toward this source, connecting it with IRAS 4A \citep{Sakai2012}. The peak of the N$_2$D$^+$ emission is significantly shifted from the continuum peak in all cases, with a significant decrease in the inner regions in some cases (see BHR 71 in Fig. {\ref{fig:envelope_n2d+_7m}). Similar extended N$_2$D$^+$ emission in other Class 0 sources was seen by \cite{Tobin2013} based on lower resolution SMA and IRAM-30m data. For TMC1 the N$_2$D$^+$  molecule is not detected.  

The DCO$^+$ emission is seen extended  in a similar fashion to what is observed for N$_2$D$^+$. However, contrary to N$_2$D$^+$, DCO$^+$ is brightest on the continuum peak for all sources except TMC1 and Per-emb-25. For these two sources, the emission peak is offset by 1000--2000 au from the continuum source in the direction perpendicular to the outflow. In the Class I source TMC1, DCO$^{+}$ is present on much smaller scales ($< 2000$ au radius) than in Class 0 sources.

\begin{figure*}[h]
\centering
  \includegraphics[width=0.95\linewidth,trim={0cm 0cm 0cm 0cm}]
  {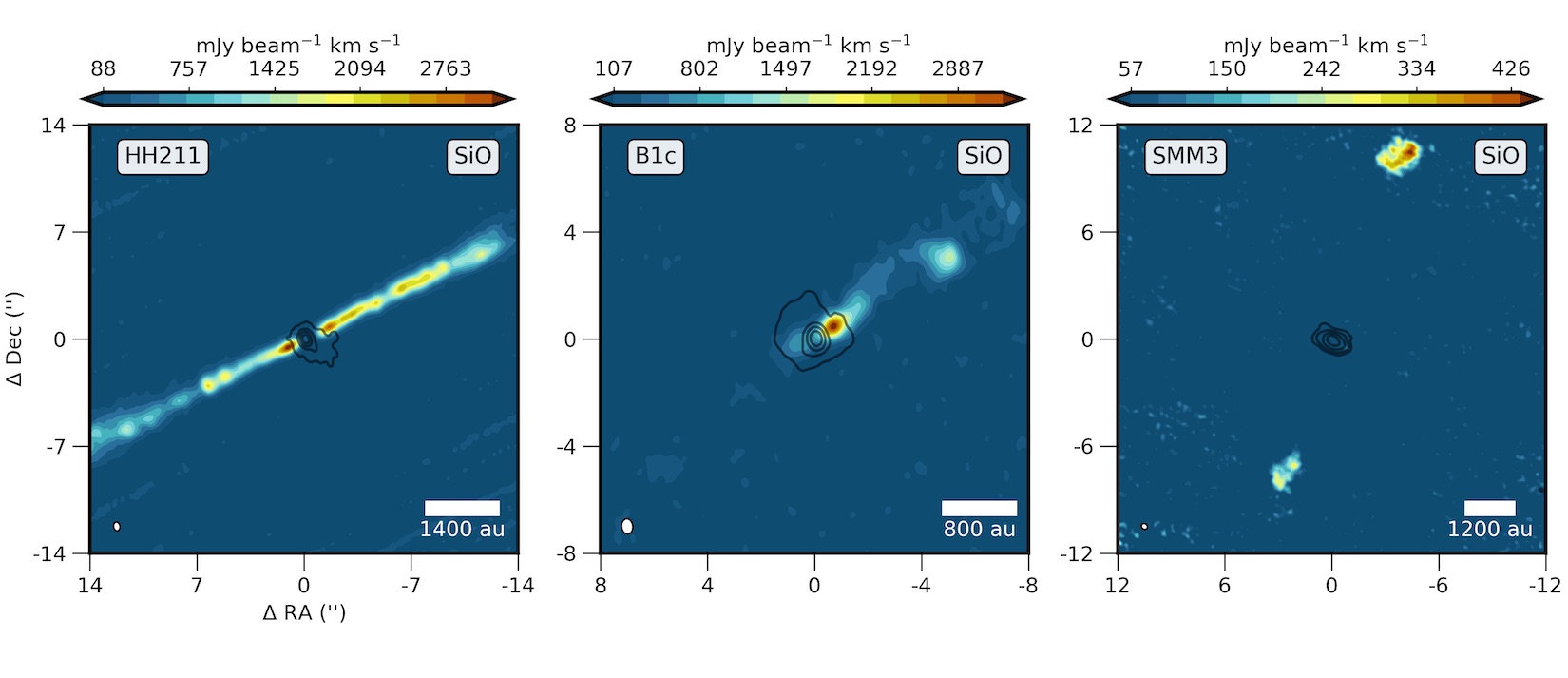}
\caption{Maps of the EHV jets observed in SiO. Moment 0 maps are presented in color scale with continuum emission at 1.3 mm presented in black contours, both obtained with 12m observations. SiO 4--3 map of HH211 integrated from -20 to -10 and from 10 to 20 km s$^{-1}$ w.r.t $\varv_{\rm sys}$; B1-c ntegrated from -70 to -40 and from 40 to 70 km s$^{-1}$ w.r.t $\varv_{\rm sys} $;  SMM3 integrated from -60 to -40 and from 20 to 35 km s$^{-1}$ w.r.t $\varv_{\rm sys}$. }
 \label{fig:ehv_no1}
\end{figure*}

\begin{figure*}[h]
\centering
  \includegraphics[width=0.95\linewidth,trim={2cm 0cm 2cm 2cm}]
  {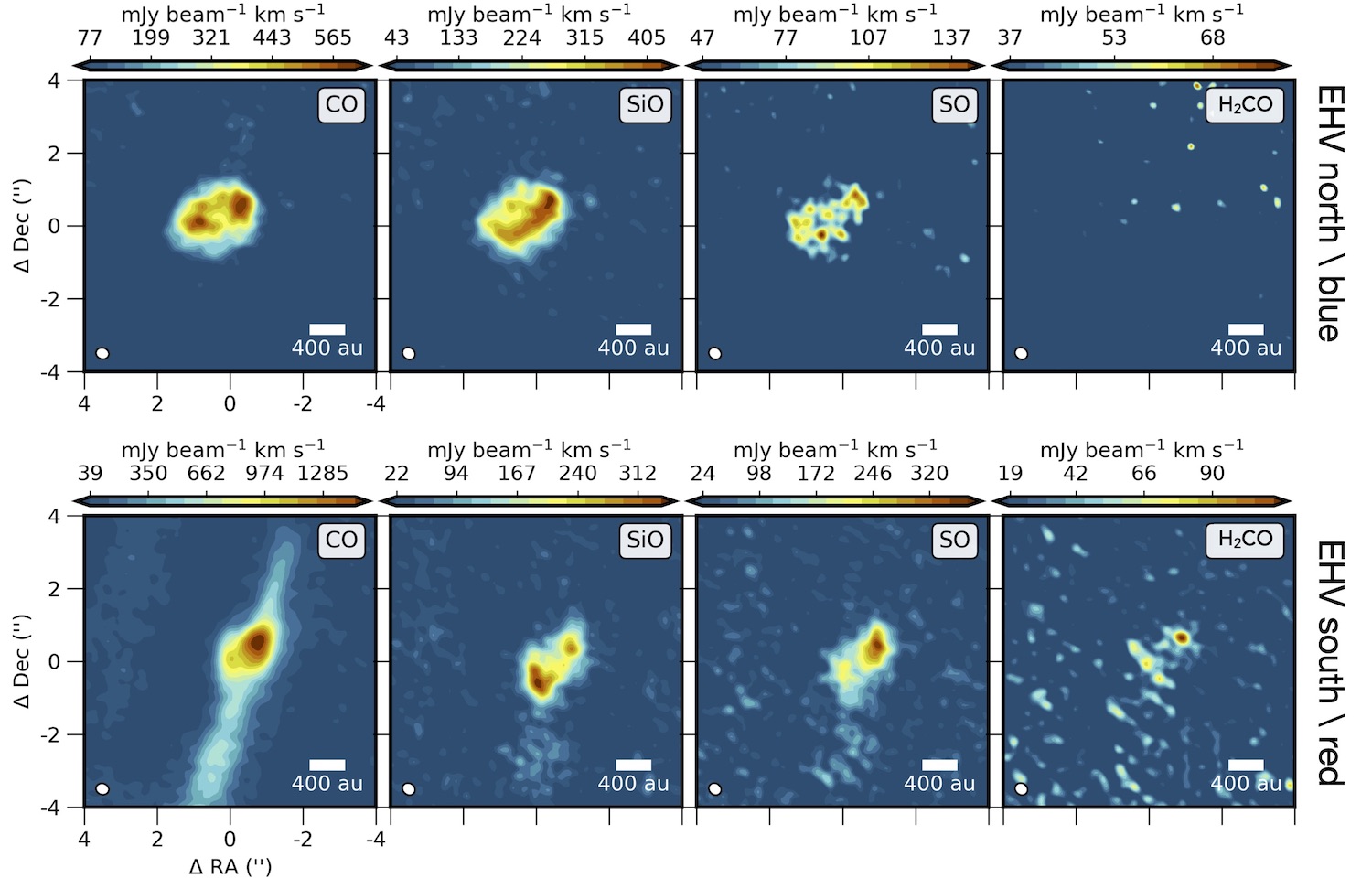}

  \caption{Zoom-in on molecular bullets from the SMM3 jet (see  Fig. \ref{fig:ehv_no1}). CO, SiO, SO, and H$_2$CO molecular transitions are presented. {\it Top:}  Norther/blueshifted bullet. Moment 0 maps are integrated from -60 to -40 km s$^{-1}$ w.r.t $\varv_{\rm sys}$. The map center is offset from the SMM3 continuum center by (--3\farcs7, +10\farcs3). {\it Bottom:}  Southern/redshifted bullet. Moment 0 maps are integrated  from 20 to 40 km s$^{-1}$.  The map center is offset from the SMM3 continuum center by (+2\farcs7, --7\farcs).}
 \label{fig:ehv_no2}
\end{figure*}

\begin{figure*}[h]
\centering
  \includegraphics[width=0.95\linewidth,trim={0cm 0cm 0cm 0cm}]
  {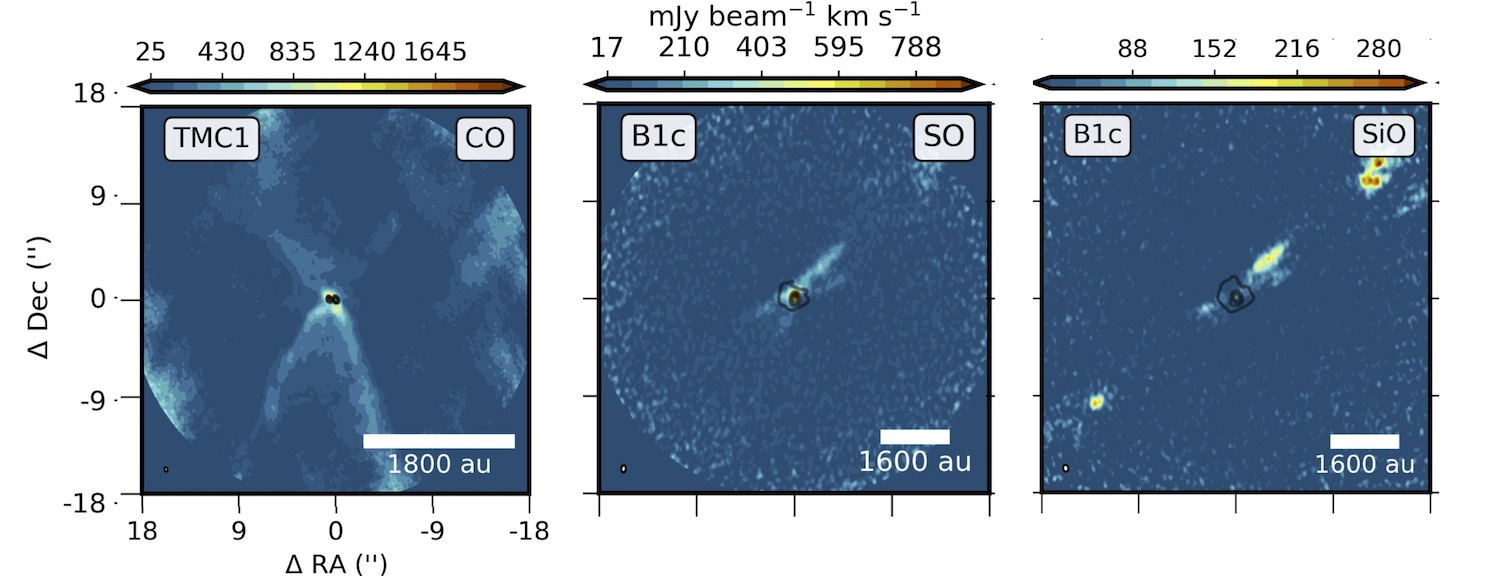}

  \caption{Low-velocity outflow in CO, SO, and SiO. Moment 0 maps of CO toward TMC1, SO and SiO map toward B1c are integrated from -10 to 10 km s$^{-1}$ w.r.t $\varv_{\rm sys}$.}
 \label{fig:lowvel}
\end{figure*}

\subsection{Outflows and jets}
\label{section:results_outflows}

Figure \ref{fig:ehv_no1} presents the extremely high-velocity (EHV)
molecular jet component for HH211, B1c, and SMM3 observed in
SiO with the ALMA 12m array; the L1448-mm EHV is shown in Fig.   \ref{fig:outflow_showcase_no1}. The latter and HH211 are well-known EHV sources
\citep{Guilloteau1992, Lee2007} while SMM3 and B1-c are new detections
of the jet component. HH211 shows SiO emission at low-velocities because the outflow is almost in the plane of the sky, but the high velocities are
evident from the large proper motion movements of the bullets
($\sim$115 km s$^{-1}$; \citealt{LeeCF2015}). CO, H$_2$CO, and SiO in
the EHV jets of Emb8N and SMM1 are presented in detail in
\cite{Hull2016} and \cite{Tychoniec2019}.

These data are particularly interesting as still only few molecules
tracing EHV jets have been identified to date (see \citealt{Lee2020} for
review). Apart from those presented here -- CO, SiO, SO, and H$_2$CO-- 
molecules such as HCO$^+$ and H$_2$O have been seen in this
high-velocity component \citep{Kristensen2012,LeeCF2014}. What is
especially important to highlight is the third detection of a H$_2$CO
bullet in SMM3 after IRAS 04166 \citep{Tafalla2010} and Emb8N
\citep{Tychoniec2019}. These detections mean that either a significant fraction of ice-coated dust is released with the jet, or that the H$_2$CO is efficiently produced in the jet through gas-phase chemistry.

In the case of L1448-mm and HH211, there are
several molecular bullets along the jet axis with velocities up to 100 km~s$^{-1}$, while both B1-c and SMM3
show a much simpler structure with two bullets detected on one side in
the former and a single pair of symmetrically placed bullets observed
in the latter case.  B1-c actually has a pair of bullets $\sim$~200 au from the continuum peak, with the other bullet at 2500 au only seen in the redshifted part of the jet. The emission in SiO and SO appears very similar (Fig. \ref{fig:sio_outflows_no2}, Fig. \ref{fig:so_outflow}). There are no other molecules tracing the high-velocity component toward this source. H$_2$CO and $^{12}$CO are not targeted with our ALMA 12m datasets toward B1-c. The SMM3 jet has two distinct high-velocity bullets at $\sim$~3200 au from the source which appear similar in CO, SiO and SO (see zoom-in  on Fig \ref{fig:ehv_no2}). Additionally, the redshifted bullet shows faint, but significant emission from H$_2$CO. No traces of H$_2$CO are found in the blueshifted outflow.

Figure \ref{fig:lowvel} presents low-velocity outflow tracers CO 2--1  ($E_{\rm up}=16$ K) for TMC1, SO 5$_6$--4$_5$  ($E_{\rm up}=35$ K), and  SiO 4--3 ($E_{\rm up}=21$ K) for B1-c. In Fig. \ref{fig:co_outflows} we present an overview of CO 2--1 emission for five sources obtained with the ALMA 12 m array at  0\farcs4 resolution. The sources show a variety of emission structures in the low velocity gas   (\textless 20 km s$^{-1}$). In all cases, we do not capture the entirety of the outflows as they extend beyond the primary beam of observations ($\sim$ 30\arcsec). SMM3 and Emb8N have very narrow outflow opening angles ($<$ 20 degrees), while SMM1, S68N and TMC1 present larger opening angles.

CO emission is especially prominent in the cavity walls, which can be related to both the limb brightening effect as well as higher (column) density of the material in the outflow cavity walls. This is especially highlighted in the Class I source, TMC1, where CO emission is almost exclusively seen in the outflow cavity walls. Even though CO is piling up in this region, it is observed at velocities up to 15 km s$^{-1}$ so it is clearly tracing the entrained material and not the envelope. The lower envelope density in Class I results in less material to be entrained in the outflow. In SMM1, three CO outflows from SMM1-a, SMM1-b, and SMM1-d  are overlapping \citep{Hull2016,Tychoniec2019}.

In Fig. \ref{fig:sio_outflows_no2}, SiO maps are presented: Band 5 SiO 4--3 ($E_{\rm up}=21$ K) and Band 6 SiO 5--4 ($E_{\rm up}=31$ K) data are shown in velocity ranges corresponding to the low-velocity outflow. In contrast to CO emission, the low-velocity SiO is mostly observed in clumps of emission instead of tracing the entirety of the outflowing gas. Several such clumps can be seen in the S68N source. In some cases, the clumps are relatively symmetric (Emb8N, B1-c), while monopolar emission is seen in other examples (L1448-mm, SMM1-d). In the case of SMM1-a and SMM1-b, very weak SiO emission at low velocities is observed. SiO emission in outflows is exclusively present in the Class 0 sources, while absent in the Class I sources, TMC1 and B5-IRS1, covered in these data sets.

Fig. \ref{fig:so_outflow} presents SO 5$_6$--4$_5$  ($E_{\rm up}=35$ K)  and SO 6$_7$--5$_6$ ($E_{\rm up}=47$ K)  observations in Band 6. The emitting regions of SO are comparable with those of SiO for the Class 0 sources. The cases of S68N and B1-c show that SO emission also peaks at the source position while SiO is absent there. Thus, SO and SiO do not always follow each other and some SO might be associated with hot core emission \citep{Drozdovskaya2018}. Important differences are observed for TMC1, where SO seems to be associated with the remainder of the envelope or the disk, while the SiO is not detected toward this source, as mentioned above.

HCN 1--0 ($E_{\rm up}=4$ K) and H$^{13}$CN 2--1 ($E_{\rm up}=12$ K) maps are presented in Fig. \ref{fig:hcn_outflow} and \ref{fig:h13cn}, respectively. HCN is clearly seen in outflowing material enhanced in similar regions as low-velocity SiO. For Emb8N the HCN emission has been associated with intermediate velocity shock \citep{Tychoniec2019}. In B1-c and L1448-mm weak extended emission along the outflow direction is detected but H$^{13}$CN strongly peaks on source. In the case of HH211, H$^{13}$CN is seen only in the outflow, with a geometry consistent with the outflow cavity walls, but with velocity profiles that are consistent with the outflowing material.

\begin{figure*}[h]
\centering
  \includegraphics[width=0.95\linewidth,trim={2cm 1cm 2cm 0cm}]
  {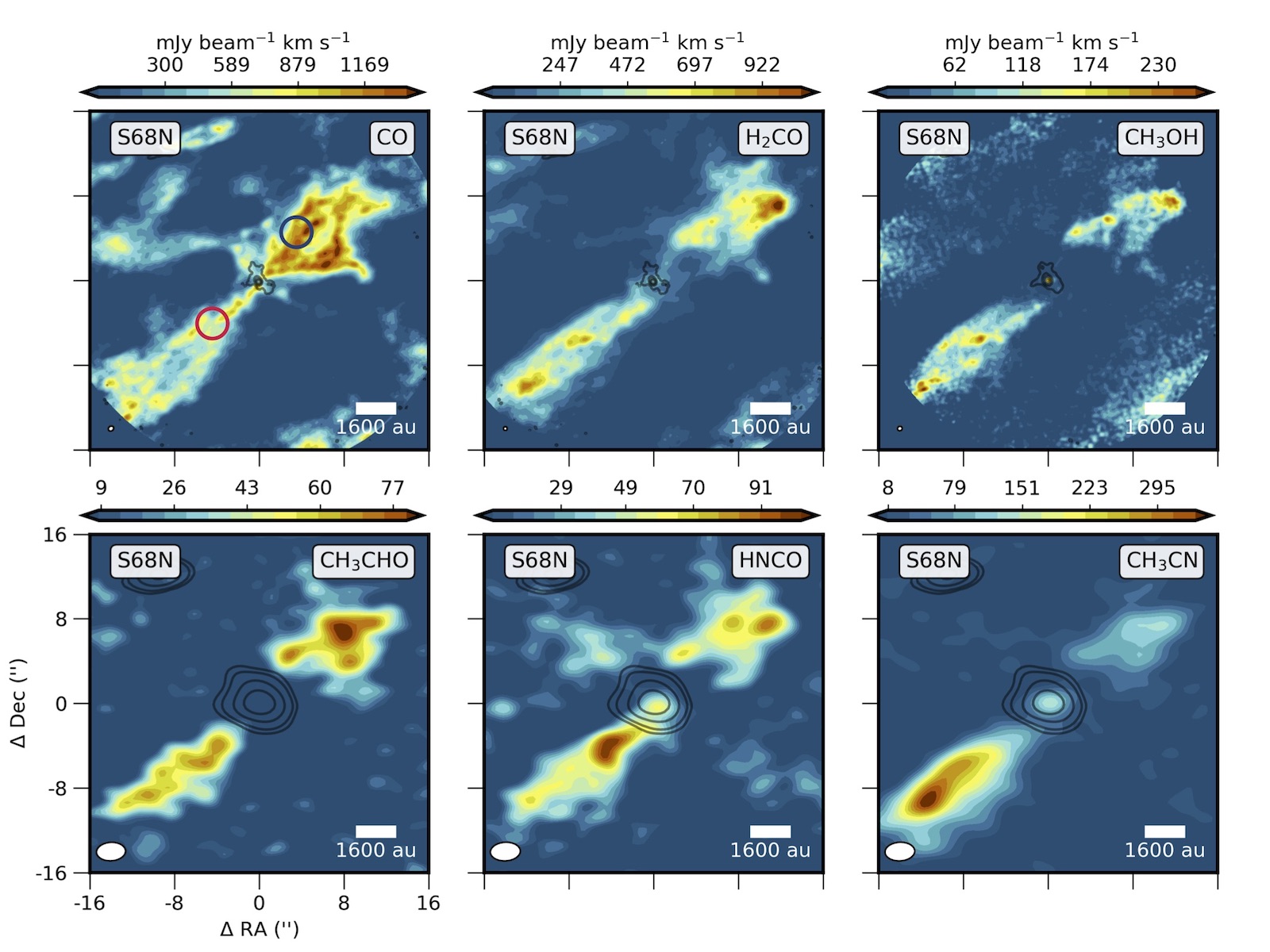}
  \caption{Maps of the ice mantle tracers toward the S68N outflow, with the CO low-velocity outflow map for reference. {\it Top:} CO, H$_2$CO and CH$_3$OH moment 0 maps obtained in Band 6 at 0\farcs5 resolution. Circles show regions from which spectra were obtained for analysis in Section 7.1.  {\it Bottom:} CH$_3$CHO, HNCO, and CH$_3$CN moment 0 maps obtained in Band 3 at 2\farcs5 resolution. The emission is integrated from -10 to -1 km s$^{-1}$ and from 1 to 10 km s$^{-1}$ w.r.t  $\varv_{\rm sys} $.} 
 \label{fig:icemantle_no1}
\end{figure*}

\begin{figure*}[h]
\centering
  \includegraphics[width=0.95\linewidth,trim={2cm 0cm 2cm 0cm}]
  {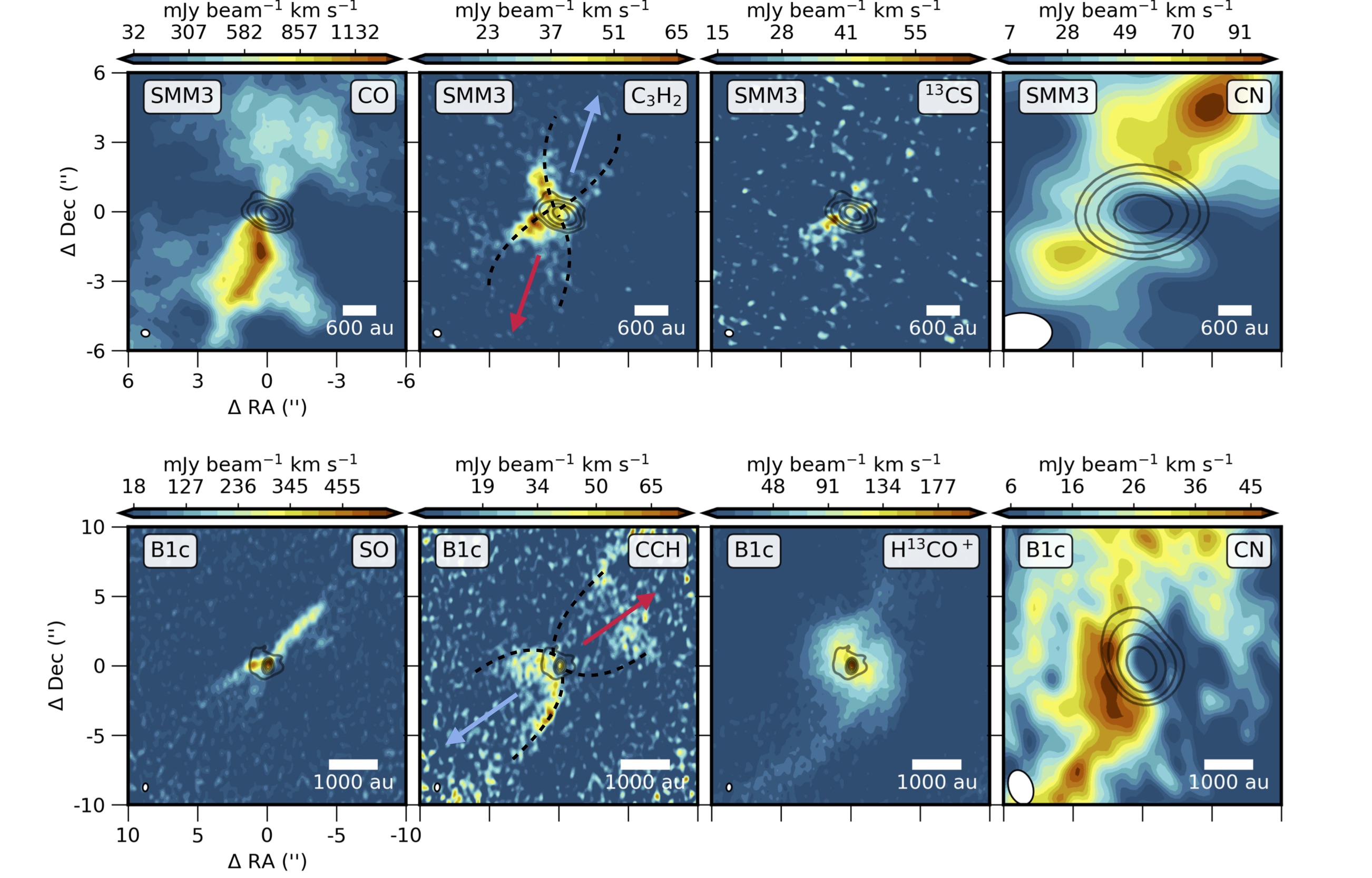}
\caption{Maps of the outflow cavity wall tracers toward SMM3 and B1c, with low-velocity outflow map for reference. {\it Top:} Moment 0 maps toward SMM3 of CO 2--1, c-C$_3$H$_2$ 6$_{1,6}$--5$_{0,5}$ , and $^{13}$CS 5--4  obtained in Band 6 at 0\farcs5 resolution and CN 1--0 in Band 3 at 3\arcsec  with continuum emission at the same band and resolution in black contours. {\it Bottom:} Moment 0 maps toward B1c of SO 6$_{7}$--5$_{6}$, C$_2$H 3$_{2.5,3}$--2$_{1.5,1}$  and H$^{13}$CO$^+$ 3--2  obtained at 0\farcs5 and CN 1--0  at 3\arcsec with continuum emission at the same band and resolution in black contours. The emission is integrated from -5 to -1 km s$^{-1}$ and from 1 to 5 km s$^{-1}$ w.r.t  $\varv_{lsr} $. Outflow directions and delineated cavity walls are shown in C$_2$H and C$_3$H$_2$ maps.} 
 \label{fig:hydrocarbons_no1}
\end{figure*}

Ice-mantle tracers are a different class of molecules detected in low-velocity protostellar outflows. They are produced and entrained through interactions between the jet and the envelope. Here we present ALMA 12m array observations in Band 6 at 0\farcs5 resolution for CO 2 -- 1  ($E_{\rm up}=17$ K), CH$_3$OH 2$_{1,0}$ -- 1$_{0,0}$ ($E_{\rm up}=28$ K), and  H$_2$CO 3$_{0,3}$ -- 2$_{0,2}$ ($E_{\rm up}=21$ K) , and in Band 3 at 3\arcsec\ for CH$_3$CN  6$_1$ -- 5$_1$  ($E_{\rm up}=26$ K), CH$_3$CHO 6$_{1,6,0}$ -- 5$_{1,5,0}$ ($E_{\rm up}=21$ K), and HNCO 5$_{0,5}$ -- 5$_{0,4}$ ($E_{\rm up}=16$ K). Fig. \ref{fig:icemantle_no1} compares maps of integrated emission from those molecules with those of CO for S68N. All ice-mantle tracers detected in the outflow are observed in their low-energy transitions. 
Additional maps for S68N are presented in the Appendix (Fig. \ref{fig:fig9_extra}): Band 6  0\farcs5  resolution image of CH$_3$CHO 14$_{0,14}$--13$_{0,13}$ ($E_{\rm up}=96$ K) and H$_2$CCO 13$_{1,13}$--12$_{1,12}$  ($E_{\rm up}=101$ K), which is overlapping with NH$_2$CHO 12$_{2,10}$--12$_{2,10}$  ($E_{\rm up}=92$ K), and the Band 3 image at 3\arcsec\   resolution of CH$_3$OCHO 10$_{0,10}$--9$_{0,9}$ ($E_{\rm up}=30$ K).  

Around the frequency of H$_2$CCO 3$_{1,13}$--12$_{1,12}$  ($E_{\rm up}=101$ K) line an extended emission can be seen in the outflow, this is however coincident with the NH$_2$CHO which is only 4 km s$^{-1}$ apart (Fig. \ref{fig:fig9_extra}).  NH$_2$CHO is more commonly observed in the shocked regions than H$_2$CCO \citep{Ceccarelli2017, Codella2017}; it is possible that those lines are blended. More low-energy transitions would be required to confirm the identification of these lines.

The velocities observed for ice-mantle tracers in the outflow are \textless 15 km s$^{-1}$ with respect to the systemic velocity. This is slower than the CO and SiO outflow line wings which have velocities up to 20--30 km s$^{-1}$. On the other hand, the lines are clearly broader than those of molecular tracers of UV-irradiated regions that trace passively heated gas (see Section \ref{section:outflow_cavity_walls}). 

Ice-mantle tracers are also detected in SMM3 and B1-c, two Class 0 protostars (Fig. \ref{fig:161} and \ref{fig:151}). For SMM3 only a lower $A_{ij}$ transition of CH$_3$OH with $E_{\rm up}=$ 61 K was targeted and not detected towards this source. B1-c has its methanol emission confused with the high-velocity SO emission, whereas emission from other COMs in the outflow is weak and appears only on the redshifted part of the outflow. Thus, S68N is the best case to study the composition of shock-released ice mantles.

\subsection{Outflow cavity walls}

In this section, we highlight key molecules detected in the outflow cavity walls. It is challenging to precisely distinguish cavity walls from the outflowing material. The velocity of the gas in the cavity walls should be lower than in the outflow, as the cavity wall contains envelope material at rest but which is  passively heated by UV radiation. We first discuss maps of species associated with cavity walls and then their line profiles.

\begin{figure*}[h]
\centering
  \includegraphics[width=0.95\linewidth,trim={0cm 0cm 0cm 0cm},draft=False]
  {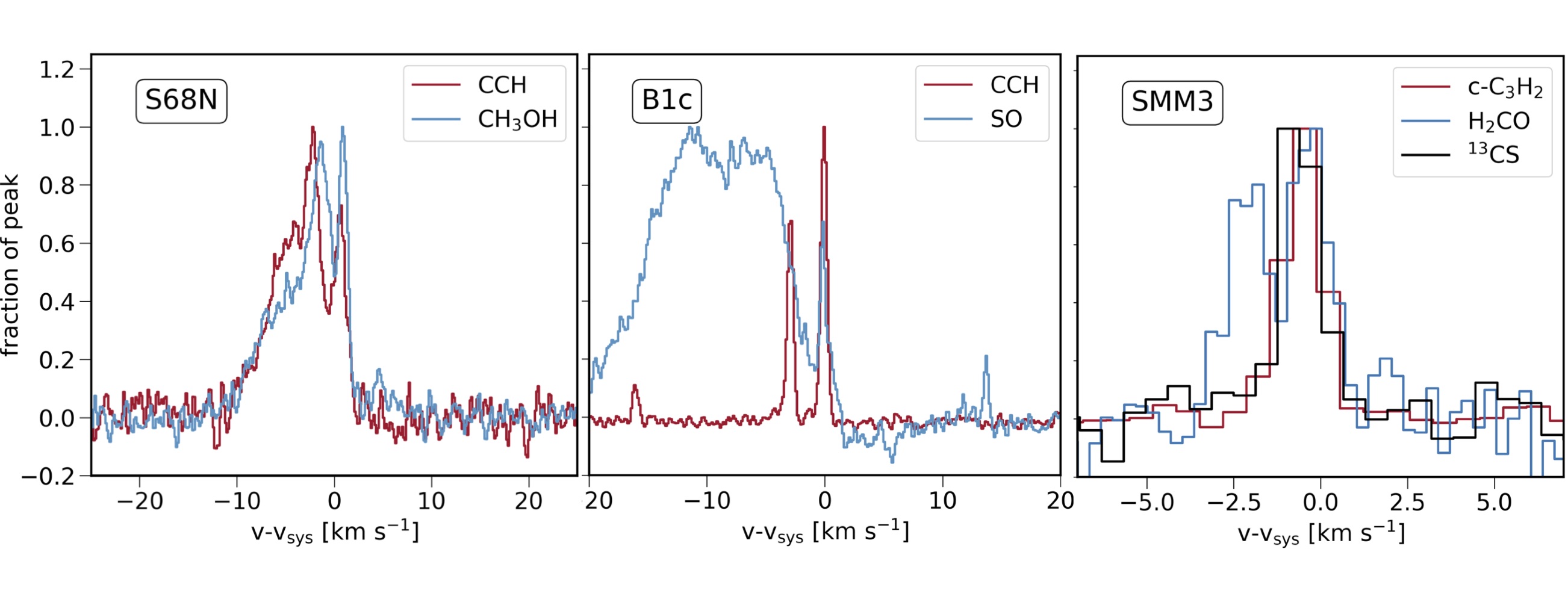}

  \caption{Spectra obtained at the cavity wall positions for hydrocarbons (red) and ice-mantle tracers (blue). {\it Left:} C$_2$H 3$_{2.5, 2}$--2$_{1.5, 1}$ ($E_{\rm up}=$ 25 K) and CH$_3$OH 2$_{1,0}$--1$_{0,1}$ ($E_{\rm up}=$ 28 K) spectra for S68N, {\it Middle:}   C$_2$H 3$_{2.5, 2}$--2$_{1.5, 1}$ ($E_{\rm up}=$ 25 K ) and SO 6$_7$--5$_6$ ($E_{\rm
  up}=48$ K)  spectra for B1-c, {\it Right:}  c-C$_3$H$_2$ $4_{4,1}$--$3_{3,0}$ ($E_{\rm
  up}=32$ K), H$_2$CO 3$_{2,1}$--2$_{2,0}$ ($E_{\rm
  up}=68$ K), and $^{13}$CS 5--4 ($E_{\rm
  up}=33$ K) spectra for SMM3.}
 \label{fig:spectra_coms_outflow_no1}
\end{figure*}

\begin{figure}[h]
\centering
    \includegraphics[width=0.95\linewidth,trim={0cm 0cm 0cm 0cm}]
  {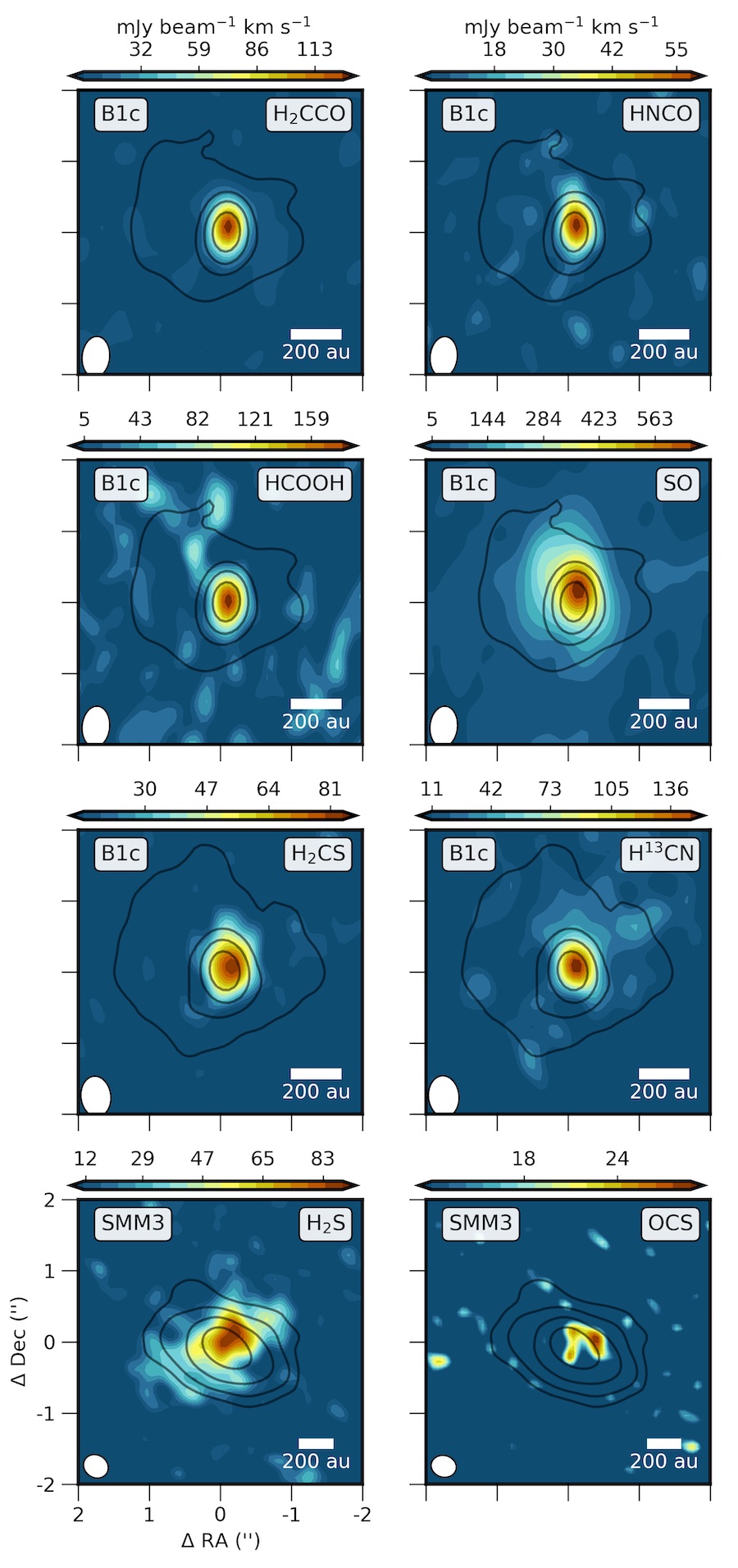}
  
  \caption{Compact emission for B1-c and SMM3 for various molecules tracing the warm inner envelope (hot core). Moment 0 maps shown in colorscale integrated from -3 to 3  km s$^{-1}$ w.r.t $\varv_{\rm sys}$. 1.3 mm continuum presented in contours.} 
 \label{fig:hotcore_v1}
\end{figure}

Figure \ref{fig:hydrocarbons_no1} presents integrated emission maps of key tracers discussed in this section for two examples: SMM3 and B1-c. Plots for the remaining sources (S68N, Emb8N, and TMC1) are presented in the Appendix in Fig. \ref{fig:hydrocarbons_no2}. Key tracers of the outflow cavity walls are simple unsaturated
hydrocarbon molecules: c-C$_3$H$_2$ $4_{4,1}$--$3_{3,0}$ ($E_{\rm
  up}=32$ K) for SMM3 and TMC1, and C$_2$H $3_{2.5,3}$--$2_{1.5,2}$
($E_{\rm up}=25$ K) for S68N, B1-c, and Emb8N, as seen in maps
obtained with ALMA Band 6 at 0\farcs5 resolution.

In Emb8N and SMM3 the emission from C$_2$H and c-C$_3$H$_2$, respectively, is symmetric; it appears similar in extent and shape on both sides of the continuum source. Comparison with CO emission, which traces the bulk of the outflowing gas, indicates that the hydrocarbons are located in the outflow cavity walls close to the source. A higher energy transition of c-C$_3$H$_2$ 7$_{2,6}$--7$_{1,7}$ $E_{\rm up}$=61 K is seen towards SMM3 closer to the protostar compared with the lower-energy transition. This reflects the increase of temperature of the cavity walls closer to the source.
For B1-c, the emission from C$_2$H is U-shaped suggestive of a cavity wall, stronger on the blueshifted side of the outflow, which could be either a projection effect or an asymmetry in the envelope structure. For B1-c no ALMA CO observations exist to compare with the bulk of the outflow at comparable resolution; however, the other outflow tracer, SO, confirms the outflow direction and rough extent of the outflow cavity walls. Moreover, the shape of the cavity walls is consistent with the appearance of the CO 3--2 outflow observed with SMA toward this source at 4\arcsec\   resolution \citep{Stephens2018}.

S68N presents a chaotic structure (Fig. \ref{fig:hydrocarbons_no2}, top), but C$_2$H is found elongated in the outflow direction. While it is difficult to identify the cavity wall, the C$_2$H emission surrounds the CO outflow emission. The C$_2$H  emission toward this source is asymmetric, with stronger emission in the blueshifted part of the outflow. Emission of c-C$_3$H$_2$ toward the Class I source TMC1  (Fig. \ref{fig:hydrocarbons_no2}, bottom) is not directly related to the cavity walls, but is extended perpendicular to the outflow, which suggests that c-C$_3$H$_2$ traces the envelope or extended disk material. 

Figure \ref{fig:hydrocarbons_no1} also shows CN 1--0 ($E_{\rm up}$ = 5 K) observed at 3\arcsec\  resolution in Band 3 for SMM3 and B1-c. Compared with C$_2$H, CN is tracing similar regions. In S68N CN has a similar extent as C$_2$H but not over the full extent of the outflow traced by CO. In B1-c, the CN emission has a similar shape of the cavity wall cone as seen in C$_2$H, but also a significant contribution from larger scales is detected. In all cases the CN emission avoids the central region, which likely results from on-source absorption by the foreground CN molecules. 

In some cases, like SMM3, the extent of the CN is broader than hydrocarbons, more comparable with CO, but the narrow linewidths suggest that this emission is still associated with passively irradiated envelope rather than with the entrained outflow.

TMC1 presents a high-resolution example of CN emission (Fig. \ref{fig:hydrocarbons_no2}). The offset between CO and CN reveals a physical
structure of the inner regions of the protostellar system: the
entrained outflow traced with CO appears closer to the jet axis, while
CN highlights the border between the outflow cavity wall and quiescent
envelope. CN is sensitive to UV radiation, as it can be produced with atomic C and N, whose abundances are enhanced in PDRs, with UV photodissociation of HCN contributing as well
\citep{Fuente1993,Jansen1995a,Walsh2010,Visser2018}.

H$^{13}$CO$^{+}$ emission is presented for B1c (Fig. \ref{fig:hydrocarbons_no1}), and S68N and Emb8N (Fig. \ref{fig:hydrocarbons_no2}) observed in Band 6 at 0\farcs4 resolution. The bulk of the emission from this molecule appears to be related to the cold envelope, however streams of material can be seen in B1-c and S68N. The streams of gas observed in  H$^{13}$CO$^+$ are coincident with the cavity wall observed in C$_2$H. As H$^{13}$CO$^+$ is expected to probe the dense envelope, the similarity of the morphology of the traced material between H$^{13}$CO$^+$ and C$_2$H and CN shows that the envelope material is UV-irradiated. $^{13}$CS observed in Band 6 with the 12m array is detected for SMM3 (Fig. \ref{fig:hydrocarbons_no1}). The morphology of  $^{13}$CS emission is very similar to that of c-C$_3$H$_2$.

In Fig. \ref{fig:spectra_coms_outflow_no1} spectra of ice mantle
tracers, CH$_3$OH and H$_2$CO, and the hydrocarbon molecules, C$_2$H and
C$_3$H$_2$ are presented. All spectra are shifted by their source
velocity to zero km s$^{-1}$. In case of S68N, it is seen that C$_2$H
and CH$_3$OH have very similar line profiles indicating that they
trace similar material. The width of $\sim 10$ km$^{-1}$ suggests
that this material is entrained with the outflow. A narrow component
appears to be superposed at systemic velocities. Note that fine splitting of C$_2$H blends
the spectra although the other transition at +2 km s$^{-1}$ does
not affect the blueshifted velocity component. In contrast, B1c shows
only remarkably narrow C$_2$H line profiles with a FWHM of $\sim$ 2 km
s$^{-1}$, and SMM3 has similarly narrow c-C$_3$H$_2$ and $^{13}$CS lines compared with broader H$_2$CO emission (Fig. \ref{fig:spectra_coms_outflow_no1}, right).

\begin{figure*}[h]
\centering
  \includegraphics[width=0.95\linewidth,trim={3cm 1cm 3cm 0cm}]
  {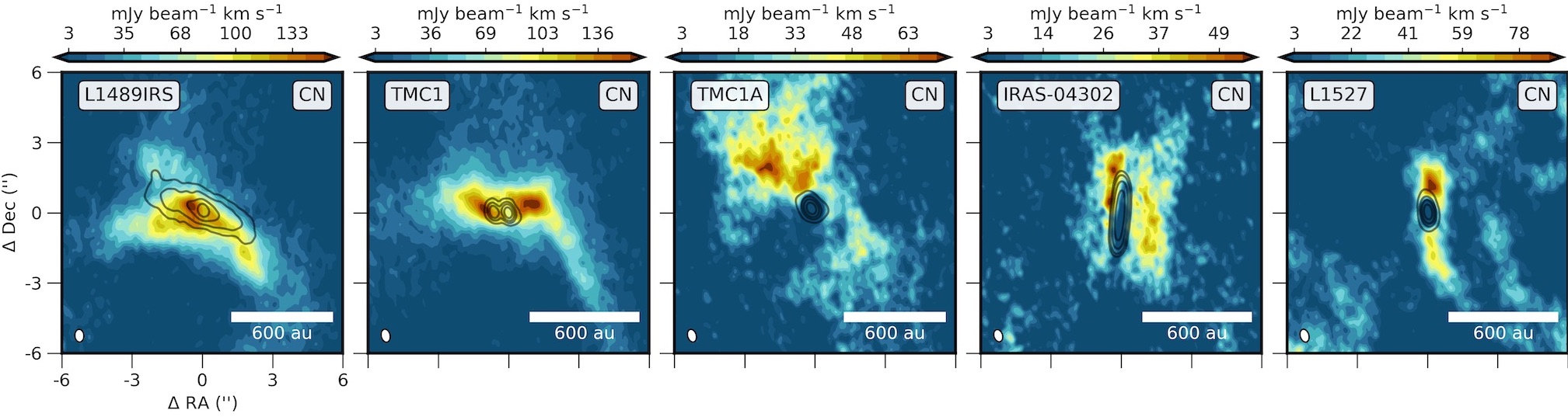}
     \includegraphics[width=0.95\linewidth,trim={3cm 1cm 3cm 0cm}]
    {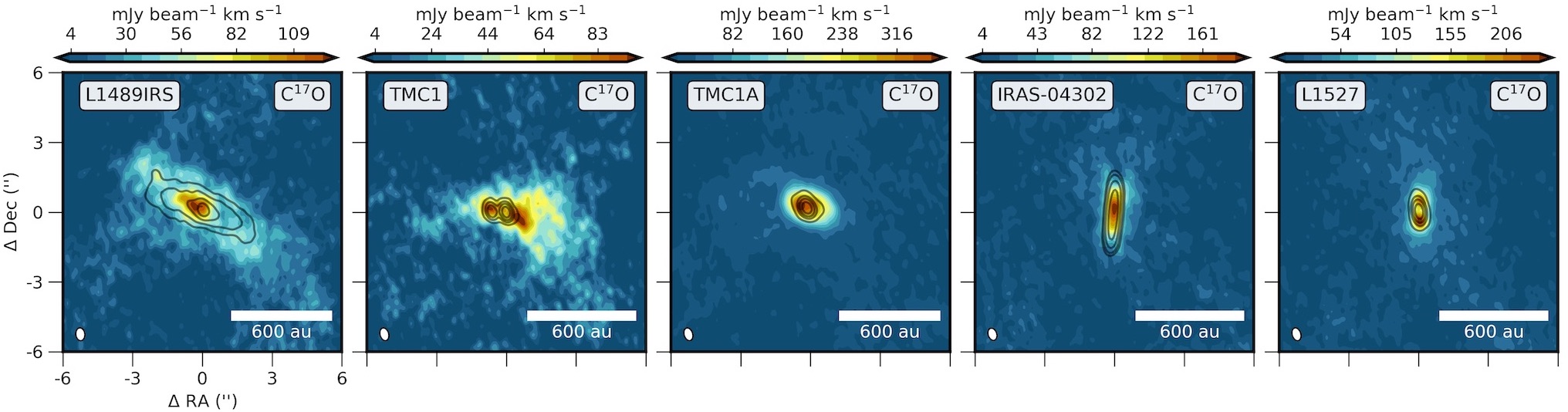}
  \caption{Images of Class I disks observed with ALMA 12m in Band 6 \citep{vantHoff2020b}. {\it Top:} Moment 0 maps of C$^{17}$O at 0\farcs4 resolution integrated from  -10 to 10 km s$^{-1}$ w.r.t  $\varv_{\rm sys} $.  {\it Bottom:} Moment 0 maps of CN integrated from -2 to 2 km s$^{-1}$ w.r.t  $\varv_{\rm sys} $. Continuum contours in black.}  
 \label{fig:disks_no1_v1}
\end{figure*}

\subsection{Inner envelope}
\label{section:results_inner_env}

Fig. \ref{fig:hotcore_v1} shows emission from H$_2$CCO, HNCO, t-HCOOH,
SO, H$_2$CS, and H$^{13}$CN for the Class 0 protostar B1-c, and H$_2$S
and OCS lines for SMM3; all observed with the ALMA
12m array at 0\farcs5. Several molecules tracing the inner hot envelope are detected in 7m data and are presented in Fig. \ref{fig:102}. CCS is detected for SMM3, BHR71, and IRAS 4B, while non-detected for B1-c, TMC1, and Per-emb-25. OCS and H$_2$S (Fig. \ref{fig:102}) for all targeted in 7m Class 0 sources while non-detected for Class I sources.

The SO 6$_7$--5$_6$ $E_{\rm up}=48$ K line is seen to
peak on the central source for B1-c and S68N (Fig. \ref{fig:so_outflow}).  SO has already been discussed in the outflow (Section \ref{section:low-velocity_outflow}), but it is also
prominent in the inner envelope. While the spatial resolution does not
allow to disentangle the hot core emission from the small-scale
outflow on a few hundred au scale, there is a difference between these
two sources and SMM3 and Emb8N. The latter two sources show a
substantial decrease in SO intensity towards the continuum peak, i.e.,
they have prominent SO emission in the outflow, but not from the hot
core. This suggests that sources like B1-c and S68N, which are
bright in SO toward the continuum emission peak, have an additional
component responsible for SO emission. This is highlighted by the
narrower lines of SO toward the continuum peak compared with the
outflow in S68N (Fig. \ref{fig:so_spectra_v1}). The narrow component, visible
in spectra taken on-source, has a width of $\sim$ 5 km s$^{-1}$. The main component of the spectrum taken in the blueshifted outflow has a similar width but has a more prominent line wing up to 20 km s$^{-1}$.

TMC1 clearly shows SO emission toward both components of the
binary system, slightly offset from their peak positions
(Fig. \ref{fig:so_outflow}). There is also a molecular ridge present in SO close to the disk-envelope interface.

HNCO and HN$^{13}$CO are detected toward  B1-c and S68N peaking on source in higher $E_{\rm up}$ transitions \citep{Nazari2021}.  For lines with $E_{\rm up} < $90 K an extended component is also detected in the outflow. SMM3 and Emb8N have no detections of HNCO on source, but for SMM3 this molecule appears in the outflow. For all Class I sources where the relatively strong HNCO 11$_{0,11}$--10$_{0,10}$ line ($A_{ij}=2\times10^{-4}$ s$^{-1}$, $E_{\rm up}$=70 K)  was targeted, it was not detected.

OCS is detected toward SMM3 peaking in the center; S68N and SMM1 show centrally peaked O$^{13}$CS detection, a minor isotopologue signalling a high abundance of OCS (Fig. \ref{fig:o13cs_appendix}). In all cases the emission is moderately resolved; of size $\sim$ 200 au in case of SMM3 and detected up to 500 au away from source for S68N and SMM1.

H$_2$S shows strong emission toward SMM3 and is also weakly present in TMC1. For SMM3 the emission is resolved along the outflow direction and perpendicular to the expected disk axis. Those are the only two
sources for which H$_2$S 12m array data were taken. Additionally, the 7m data
presented in the Appendix (Fig. \ref{fig:102}) show prominent, centrally peaked H$_2$S
emission for four more Class 0 sources. 

H$_2$CS is detected for B1-c, S68N and L1448-mm through a line with
$E_{\rm up}=$ 38 K. Another transition with $E_{\rm up}=46 $ K is found
in Class I disks: IRAS-04302, L1489, and TMC1A. In B1-c, the emission is marginally
resolved, while in IRAS-04302 the molecule is clearly seen across the
midplane, indicating sublimation from icy grains at temperatures of at least 20 K
\citep{vantHoff2020b,Podio2020}.

H$_2$CCO is detected for B1-c and S68N. For the lower $E_{\rm up}=100$~K transition, the molecule is also detected in the outflow.  For Class I sources, the transition at comparable energy is not detected. HCOOH is detected for B1-c and S68N with low-energy transitions that are seen both on source and in the outflow, while the higher energy line ($E_{\rm up}=83$~K) is seen only on source. H$^{13}$CN is detected in B1-c and L1448-mm on source, additionally to the outflow component.

Disks are commonly observed in Class I sources \citep{Harsono2014,Yen2017}, as the envelope clears out.  Fig. \ref{fig:disks_no1_v1} presents maps of
the C$^{17}$O 2--1 ($E_{\rm up}=$ 16 K) and CN 2--1 ($E_{\rm up}=$16
K) lines toward Class I disks and L1527-IRS, which is identified as a Class 0/I object, observed with ALMA 12m at 0\farcs3
resolution. The C$^{17}$O and H$_2$CO emission for disks in Taurus
using these data are analyzed in detail by \cite{vantHoff2020b}. Here
  we discuss CN in comparison with C$^{17}$O. C$^{17}$O is observed concentrated towards the continuum emission for all disks, and is a much cleaner tracer of the disk than any other more
abundant CO isotopologues, although even C$^{17}$O still shows some trace
emission from the surrounding envelope. 

\begin{figure*}[h]
\centering
  \includegraphics[width=0.95\linewidth,trim={0cm 0cm 0cm 0cm},draft=False]
{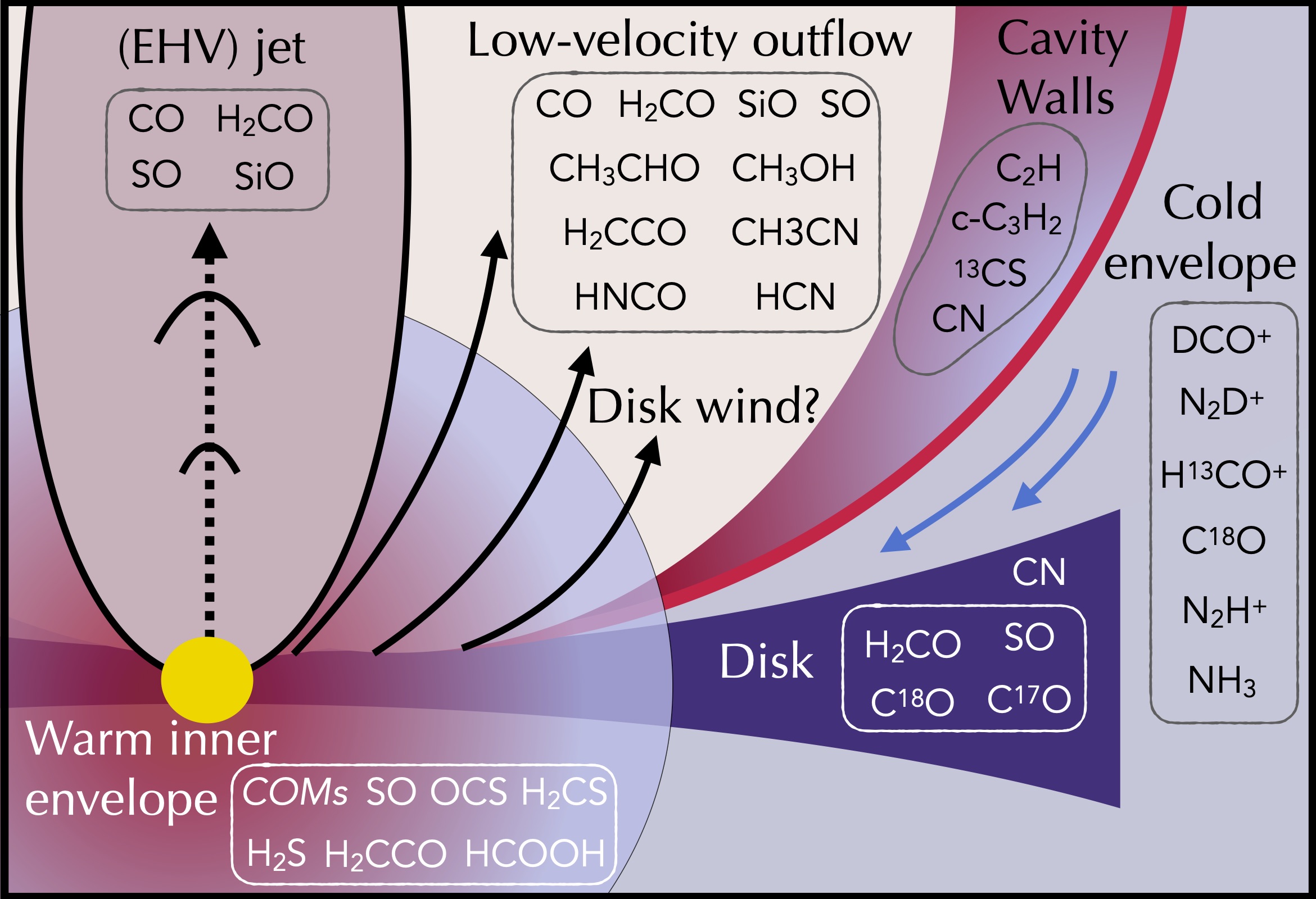}
  \caption{Summary cartoon presenting key molecular tracers of different components, limited to molecules presented in this work except for NH$_3$ and N$_2$H$^{+}$ which are important large scale envelope tracers. CN molecule is shown on top of the disk as it traces the disk atmosphere and not the midplane. Protostar indicated in the center in yellow.} 
 \label{fig:cartoon_summary}
\end{figure*}

\section{Discussion}
\label{section:discussion}

Figure \ref{fig:cartoon_summary} summarizes current understanding about which molecule traces which component in Class 0/I protostars. Table \ref{table:table_components_mols}
 lists all molecules discussed in this work with indication of the physical component that they trace. Below, we will address several implications of the results. We note that the discussion is confined to the molecules targeted and detected in our datasets, but is by no means complete. Other relevant tracers not discussed here are NH$_3$ and N$_2$H$^{+}$, which are tracing the envelope well \citep[e.g.,][]{Tobin2011, Chen2007}. PO and PN have been recently suggested as a cavity walls or outflow tracers, but have very low abundances \citep{Bergner2019}. SO$_2$ is an important tracer of the warm gas in the disk-envelope interface \citep{ArturV2019}. Finally, this work focuses on low-energy transitions observed with ALMA $E_{\rm up} <$100 K with the exception for hot core tracers.
 
\begin{table*}
\centering
\caption{Molecular tracers of physical components exclusive to this work}
\label{table:table_components_mols}
\begin{tabular}{lrrrccccc}
\hline \hline
 Molecule  & Envelope & Disk & Hot & Outflow & Jet & Cavity  \\
   &  &  & core &  & &   \\
\hline
CO & \cmark  & --- & --- & \cmark & \cmark & \cmark \\
C$^{18}$O & \cmark  & \cmark & --- & \cmark & --- & \cmark \\
$^{13}$CO & \cmark  & \cmark & --- & \cmark & --- & \cmark \\
C$^{17}$O & \cmark  & \cmark &  \cmark & --- & --- &  \cmark \\

H$^{13}$CO$^{+}$ & \cmark  & --- & --- & \cmark & --- & \cmark \\
H$_2$CO & \cmark  & \cmark & \cmark & \cmark & \cmark & --- \\
H$_2$CS & ---  & \cmark & \cmark & --- & --- & --- \\

SO & ---  & \cmark & \cmark & \cmark & \cmark & --- \\
SiO & ---  & --- & --- & \cmark & \cmark & --- \\

DCO$^+$ & \cmark  & --- & --- & --- & --- & --- \\
N$_2$D$^+$ & \cmark  & --- & --- & --- & --- & --- \\

OCS & ---  & --- & \cmark & --- & --- & --- \\
O$^{13}$CS & ---  & --- & \cmark & --- & --- & --- \\

HNCO & ---  & --- & \cmark & \cmark & --- & --- \\
H$_2$CCO & ---  & --- & \cmark & \cmark & --- & --- \\
HCOOH & ---  & --- & \cmark & \cmark & --- & --- \\
$^{13}$CS & \cmark  & --- & --- & --- & --- & \cmark \\
CN & \cmark  & \cmark & --- & --- & --- & \cmark \\
HCN & ---  & --- & --- & \cmark & --- & --- \\
H$^{13}$CN  & ---  & --- & \cmark & \cmark & --- & --- \\

C$_2$H & \cmark  & --- & --- & \cmark & --- & \cmark \\
c-C$_3$H$_2$ & ---  & --- & --- & --- & --- & \cmark \\

CH$_3$CHO & ---  & --- & \cmark & \cmark & --- & --- \\
CH$_3$CN & ---  & --- & \cmark & \cmark & --- & --- \\

CH$_3$OCHO & ---  & --- & \cmark & \cmark & --- & --- \\
CH$_3$OH & ---  & --- & \cmark & \cmark & --- & --- \\
$^{13}$CH$_3$OH & ---  & --- & \cmark & --- & --- & --- \\

\hline
\end{tabular}
\end{table*}

\subsection{Cold envelope}

\subsubsection{C$^{18}$O}

C$^{18}$O is a good tracer of high column density material because of its low critical density. However, it becomes less abundant as soon as the dust temperature drops below the CO freeze-out temperature ($\sim$ 20--25 K). There is also a density threshold: CO freeze-out only occurs at densities above $\sim 10^4$--$10^5$ cm$^{-3}$, because at lower densities the timescales for freeze-out are longer than the lifetime of the core \citep{Caselli1999,Jorgensen2005a}.

For 4 out of 6 sources presented here at 6\arcsec\  resolution, \cite{Kristensen2012} performed modeling of the SED and sub-mm spatial extent using the DUSTY code \citep{Ivezic1997}. The results provide, among other properties, a temperature structure throughout the envelope and the radius at which the temperature drops below 10 K, which is considered as the border between envelope and the parent cloud. \cite{Kristensen2012} obtained radii of 3800, 5000, 6700, and 9900 au for IRAS4B, TMC1, SMM3, and BHR71, respectively. The compact C$^{18}$O emission observed on-source and its
non-detection over the full expected extent of the protostellar
envelope can be explained by CO freeze-out occurring already within
the inner 1800 -- 2500 au radius, which is the spatial resolution
of our observations for Class 0 sources. This upper limit on the CO
snowline is consistent with CO snowlines typically observed and
modeled toward other Class 0 protostars \citep{Jorgensen2004d,
  Anderl2016, Hsieh2019}. 

Equation 1 from \cite{Frimann2017} allows to calculate the expected CO snow
line for the current luminosity of the example sources B1-c and TMC1
presented in Fig. \ref{fig:envelope_plot_no1_v1} in the absence of an
outburst. For B1-c the snowline is expected to be at
200--400 au radius depending on the assumptions of the sublimation
temperature (larger radii for 21 K and smaller for 28 K), while the
TMC1 CO snowline is expected to be at 100--200 au. For the most luminous 
source with C$^{18}$O 7m observations available, the expected radius
is at 400--750 au. Therefore clearly in all cases the expected
CO snowline is well within the 7m beam.  In high-resolution studies, the CO emission is often
seen at greater distances than expected from the current luminosities
of those protostars. This is attributed to accretion bursts of
material which increase their luminosities resulting in a shift of the
observed CO emission radius up to a few times its expected
value (but usually still within a 1000 au radius)
\citep{Jorgensen2015, Frimann2017,Hsieh2019}. 

\subsubsection{N$_2$D$^+$ and DCO$^{+}$}

Both DCO$^{+}$ and N$_2$D$^+$ are considered cold gas
tracers \citep[e.g.,][]{Qi2015}. N$_2$D$^{+}$ is efficiently destroyed by CO in the gas-phase,
therefore freeze-out of CO results in N$_2$D$^+$ being retained in the
gas-phase at larger radii of the envelope, where temperatures are
lower. This behaviour has been demonstrated in several other protostellar sources by \cite{Tobin2011,Tobin2013,Tobin2019}.

Both DCO$^+$ and N$_2$D$^+$ are produced through reactions with H$_2$D$^ +$.  At cold
temperatures the H$_2$D$^+$ abundance is enhanced through the
H$_3^+$ + HD $\rightarrow$ H$_2$D$^+$ + H$_2$ reaction, which is exothermic by
230 K. As the reverse reaction is endothermic, low temperatures
increase H$_2$D$^+$. Additionally, both H$_3^+$ and
H$_2$D$^+$ are enhanced in gas where CO has been depleted. However, the CO
molecule is still needed for the production of DCO$^+$ through the
H$_2$D$^+$ + CO reaction. Therefore, DCO$^{+}$ is expected to be most
abundant around the CO snowline \citep{Jorgensen2004d,Mathews2013}.
Warmer production routes through CH$_2$D$^+$ + CO are also possible
\citep{Wootten1987, Favre2015, Carney2018}.

The difference between DCO$^{+}$ and N$_2$D$^{+}$ chemistry is reflected in morphology of both molecules in the dense regions close to the continuum peak. As DCO$^+$  requires gas-phase CO for its formation, it peaks close to the CO snowline, which is within the resolution of our observations ($\sim$1800--2500 au radius), while N$_2$D$^+$ is only located where CO is not present in the gas phase. Therefore we observe a significant decrease of N$_2$D$^+$ in the inner envelope. If the warm production of DCO$^+$ is triggered in the inner regions, this will additionally produce DCO$^+$ within the beam of our observations, hence DCO$^+$ does not decrease in the inner envelope. The extent of the DCO$^+$ and N$_2$D$^+$ emission in each source is comparable, ranging from $\sim$ 5000 au in B1-c and BHR71 to 1500 au in TMC1, suggesting that their outside radii trace the region where CO becomes present again in the gas phase due to the low density.
 
The morphology of the emission from cold gas tracers such as DCO$^{+}$ and N$_2$D$^+$  is sensitive to the density and temperature profile of the system, which can be affected by system geometry (i.e., outflow opening angle, disk flaring angle, flattening of the envelope). DCO$^{+}$ has been shown to increase its abundance in the cold shadows of a large embedded disk \citep{Murillo2015}.  Emission from  DCO$^{+}$ and N$_2$D$+$ is consistent with a picture of a dissipating envelope in Class I sources, resulting in less dense, warmer gas surrounding the protostar. TMC1 has an order of magnitude lower envelope mass compared to the Class 0 sources (Table \ref{table:targets}; \citealt{Kristensen2012}). This causes the  extent of the cold and dense region to shrink, preventing  N$_2$D$^+$ from being detected, and limiting the extent of  DCO$^+$ emission. The dense gas toward TMC1 is clearly present only in the flattened structure surrounding the binary system, likely forming a young, embedded disk. In fact this source is suggested to have a rotationally-supported circumbinary disk \citep{Harsono2014, vantHoff2020b}. The geometry of the disk can create favourable conditions for the  DCO$^+$ enhancement in the cold shadows of the disk.

Other relevant molecules that trace the quiescent envelope material but are not presented here are HCO$^+$ and  H$^{13}$CO$^+$ (\citealt{Hogerheijde1997, Jorgensen2007, Hsieh2019}, van 't Hoff et al. in prep.). These molecules have been shown to probe the material outside of the water snowline  \citep{Jorgensen2013,vantHoff2018}. As water sublimates at temperatures $\sim 100 $ K, much higher than CO, HCO$^+$ can be seen throughout the envelope, except for the warmest inner regions. N$_2$H$^+$ is tracing the envelope material and CO snowline and has been shown to peak closer to the central protostar than N$_2$D$^+$ \citep{Tobin2013}. Their ratio can potentially be used as an evolutionary tracer of protostars \citep{Emprechtinger2009}. NH$_3$ is also observed to map similar components as N$_2$H$^+$ \citep{Tobin2019}. 

In summary,  the quiescent envelope material is traced by dense and/or cold gas tracers. Chemical interactions result in N$_2$D$^+$ tracing the outer envelope where CO is frozen-out, whereas DCO$^{+}$ is seen both in the outer envelope as well as in the inner regions, tracing the unresolved CO snowline. C$^{18}$O is a good tracer of dense ($n$ $> 10^5$ cm$^{-3}$) and warm  ( $T$ $>$ 30 K) regions in the inner 2000 au radius of the protostellar systems. The evolution of the physical conditions from Class 0 to Class I is evident as the envelope becomes less dense and the protostellar luminosity can heat up dust and gas more easily.

\subsection{Outflows and jets}

Outflowing material from protostellar systems is best analyzed with kinematic information. In the following section, we will discuss three different componfents of protostellar outflows observed with the ALMA 12m array:  1) the high-velocity jet  (\textgreater 30 km s$^{-1}$), 2) the low-velocity outflow (\textless 30 km s$^{-1}$), 3) the gas that results from the interaction with the outflow -- ice sputtering products at velocities close to that of ambient material, but with linewidths significantly broader (up to 15 km s$^{-1}$) than the quiescent envelope tracers. The three components are presented in Fig. \ref{fig:outflow_showcase_no1} with examples of S68N, a representative case of a prominent outflow and ice sputtering, and L1448-mm, a source with a prototypical high-velocity jet.
 
\subsubsection{Extremely high-velocity (EHV) jet}
\label{section:extremely_high}

There are 6 out of 7 Class 0 sources targeted at high-resolution strengthen the conclusion that EHV jets are more common
than previously thought in Class 0 sources \citep[][]{Podio2021}. It is argued that the molecular jet tracers have a very different physical origin than the protostellar outflows; contrary to the low-velocity outflow, which consists mostly of entrained envelope material, the EHV jet is expected to be directly launched from the innermost region of the system \citep{Tafalla2010, Lee2020}.  The high-velocity jet is comprised of atomic material which readily forms molecules in the high-density clumps \citep[internal working surfaces;][]{Raga1990, Santiago-Garcia2009} that are resulting from shocks in the jet, which are produced by the velocity variations of the ejection. This in turn means that by observing the high-velocity bullets, one gains insight on the variability of the accretion process \citep{Raga1990,Stone1993}. The new EHV sources B1c and SMM3 show bullet spacings of 1200 au in B1c and 3200 au in SMM3, which can be converted using the terminal velocity of the jets (not corrected for inclination) to the dynamical ages of 80 and 250 years, respectively. If the central mass of the protostar can be estimated this can be used to provide the orbital period of the component causing the variability \citep{Lee2020}.

The fact that the EHV jet tracers are dominated by O-bearing molecules has been associated with a low C/O ratio in the jet material \citep{Tafalla2010}. For high mass-loss rates, molecules are produced efficiently in the jet from the launched atomic material \citep{Glassgold1991, Raga2005, Tabone2020a}. Additionally the ratio of SiO-to-CO can indicate the presence of dust in the launched material, which can in turn inform about the jet launching radius, i.e., whether it is inside or outside the dust sublimation radius \citep{Tabone2020a}. The new detections of high-velocity jets suggest that this process may be occurring in every young Class 0 object. Studying large samples of objects with ALMA and combining with multi-transition observations can unveil the atomic abundances of the inner regions, which are difficult to measure otherwise directly \citep{McClure2019}.  

The presence of H$_2$CO in the jet could result from gas-phase production through the reaction of CH$_3$ with O. Alternatively, if the icy grains were launched with the jet, they could be sputtered in the internal working surfaces at high velocities \citep{Tychoniec2019}.

In summary: O-bearing species such as CO, SiO, SO, and H$_2$CO observed at high velocities are  excellent tracers of the chemistry within the protostellar jet. Those molecules most likely formed in the internal working surfaces from the material carried away from the launching region of the jet.

\begin{figure*}[h]
\centering
  \includegraphics[width=0.95\linewidth,trim={0cm 0cm 0cm 0cm}]
  {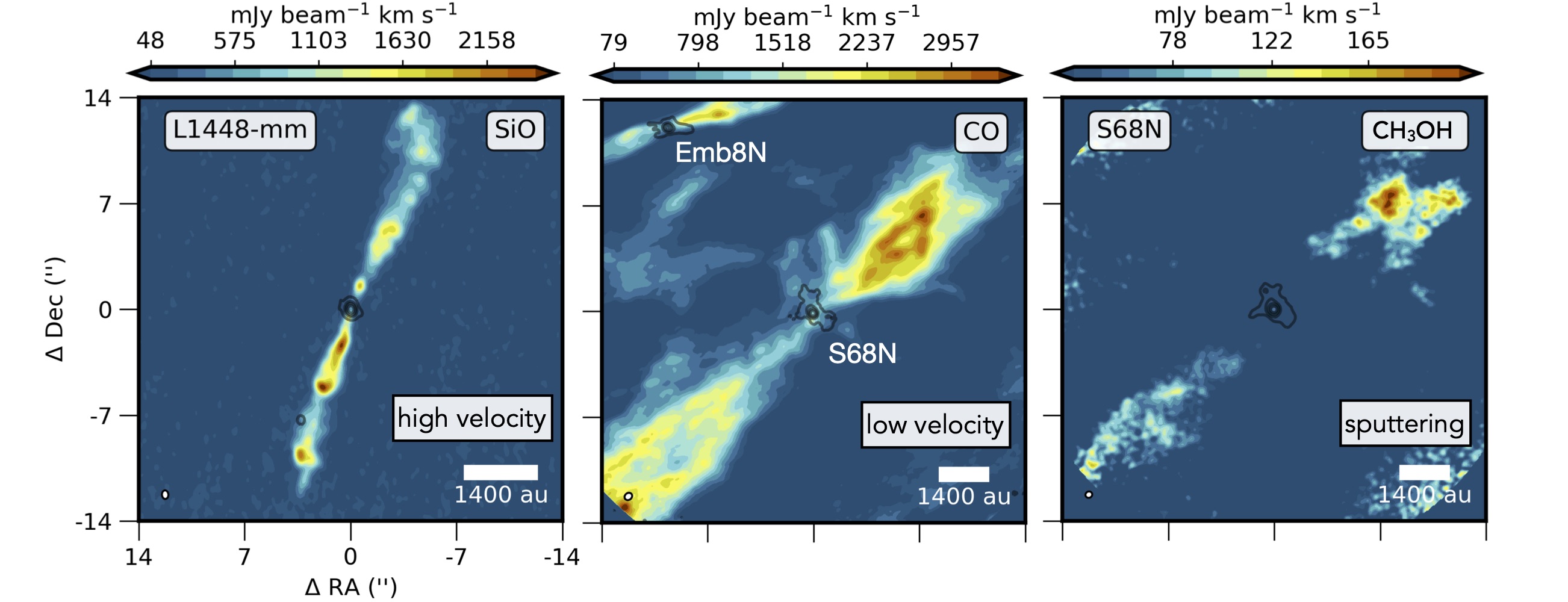}
  \caption{Maps of three different components of the outflow. Moment 0 maps are presented in color scale with continuum emission at 1.3 mm in black contours, both obtained with ALMA 12m observations. 
   {\it Left:} An extremely high-velocity (EHV) molecular jet illustrated with the SiO (4--3) map for L1448-mm integrated from -70 to -50 and from 50 to 70 km s$^{-1}$ w.r.t $\varv_{\rm sys} $.
  {\it Middle:} Low-velocity outflow illustrated with the CO 2--1 map for S68N integrated from -15 to -3 and from 3 to 15 km s$^{-1}$ w.r.t $\varv_{\rm sys} $. {\it Right:} Ice mantle content released with shock sputtering presented with the CH$_3$OH (2$_{1,0}$ -- 1$_{0,0}$) map for S68N integrated from -8 to -1 and from 1 to 8 km s$^{-1}$ w.r.t $\varv_{\rm sys} $. The CO outflow from Ser-emb-8N is present at the edge of the map.}
 \label{fig:outflow_showcase_no1}
\end{figure*}

\subsubsection{Low-velocity outflow}
\label{section:low-velocity_outflow}

The CO molecule traces the bulk of the gas as it is the most abundant molecule detectable in the sub-mm regime. It  serves as an indicator of the outflow extent and gas morphology, as it is not affected by chemical processing in shocks. It can also be used to quantify the total mass-loss rates, using the dense cloud abundance ratio of CO/H$_2$ $\sim$ 10$^{-4}$. A well-known correlation between molecular outflows and protostellar luminosities indicates a strong link between the accretion and ejection processes \citep{Cabrit1992, Bontemps1996, Mottram2017}. The decrease in accretion and total envelope mass with evolution of the system also results in fainter, less powerful outflows. The weak emission from the Class I source TMC1, which is an order of magnitude lower in intensity with respect to outflows from Class 0 protostars, is consistent with this trend.

SiO is a molecule that is enhanced by several orders of magnitude in shocks compared with gas in cold and dense clouds where most of the Si is locked in the grains \citep[e.g.,][]{Guilloteau1992, Dutrey1997}. Shocks release Si atoms from the grains by means of sputtering and grain destruction, leading to subsequent reactions with OH, another product of shocks, forming SiO  \citep{Caselli1997,Schilke1997,Gusdorf2008,Gusdorf2008a}. Thus, SiO is much more prominent in high-velocity gas, where grains are more efficiently destroyed (see Section \ref{section:extremely_high}), than in low-velocity shocks.

SO is enhanced in shocks through reactions of atomic S released from the grains with OH, as well as through H$_2$S converted to SO with atomic oxygen and OH \citep{Hartquist1980,Millar1993}. Shocks could also explain the emission toward TMC1, where weak accretion shocks onto the disk could enhance the SO abundances \citep{Sakai2014a,Yen2014, Podio2015}.
Overall, there is a clear decrease of both SiO and SO low-velocity
emission from Class 0 to Class I. Either the less powerful jet cannot
destroy the grains and create conditions for the production of SiO and
SO and/or the much less dense envelope and outflow cavity walls do not
provide enough dust grains for creating large column densities of those
molecules; additionally, the excitation conditions might change significantly
 with evolution of the protostellar system, hampering the
detection of even low-$J$ SO and SiO transitions with critical
densities in the $10^5-10^6$ cm$^{-3}$ range.

HCN is associated with the most energetic outflows \citep{Jorgensen2004, Walker-Smith2014} and is enhanced at high temperatures in shocks. H$^{13}$CN is likely associated with both the hot inner regions (not detected in HCN due to optical thickness of the line) and the low-velocity outflow \citep{Tychoniec2019, Yang2020}. The geometry of H$^{13}$CN emission seen in HH211 resembles a cavity wall: it could be a result of CN produced by UV-photodissociation and subsequent production of HCN via the H$_2$ + CN reaction, which requires high temperatures \citep{Bruderer2009,Visser2018}. The fact that H$^{13}$CN is seen at outflow velocities shows that shocks are required to produce HCN, which is likely released at the cavity walls and then dragged with the outflow.

\subsubsection{Shock sputtering products}
\label{section:shock_sputtering}

Ice-mantle tracers are detected in outflows of SMM3, B1-c and S68N protostars. As indicated in Section \ref{section:results_outflows} S68N is the best test-case to study the composition of shock-released ice mantles. The morphology of the emission of ice-mantle tracers in S68N is somewhat uniform. CH$_3$CN and HNCO are clearly brighter on the redshifted part of the outflow (south-east) and CH$_3$CHO are brighter in the blueshifted (north-west) side. CH$_3$OH and H$_2$CO show an even distribution between the two lobes. The peak intensity for all species occurs at significant distances from the source ($\sim$ 5000 au) and in some cases the emission drops below the detection limit closer to the source. This is contrary to the CO emission, which can be traced all the way back to the central source. In all tracers the emission is also detected at the continuum position, however, this emission has a narrow profile and results from thermal sublimation of ices in the hot core of S68N \citep{vanGelder2020}.

The velocities observed for CH$_3$OH and other ice-mantle tracers are high enough for these molecules to be material near the outflow cavity walls, where ice mantles could be sputtered. These molecules therefore most likely trace low-velocity entrained material with a considerable population of ice-coated grains that are sputtered in the shock \citep{Tielens1994,Buckle2002, Arce2008, JimenezSerra2008, Burkhardt2016}. Thermal desorption of molecules from grains is not likely as dust temperatures  at distances of few times 10$^3$ au from the protostar are well below the desorption temperatures for most COMs. The fact that there is no enhancement of these tracers closer to the source also argues against emission being related to high temperature.

\subsection{Outflow cavity walls}
\label{section:outflow_cavity_walls}

Emission from hydrocarbons and CN in the outflow cavity walls is directly related to the exposure of those regions to the UV radiation from the accreting protostar. Both c-C$_3$H$_2$ and C$_2$H have been prominently observed in PDRs such as the Horsehead Nebula and the Orion Bar tracing the layers of the cloud where UV-radiation photodissociates molecules, which helps to maintain high atomic carbon abundance in the gas-phase that is needed to build these molecules. C$_2$H is enhanced in the presence of UV radiation at cloud densities \citep{Fuente1993, Hogerheijde1995} and c-C$_3$H$_2$ usually shows a good correlation with C$_2$H \citep{Teyssier2004}. Both molecules have efficient formation routes involving C and C$^+$, although  models with only  PDR chemistry tend to underpredict their abundances, especially for c-C$_3$H$_2$. A  proposed additional mechanism is the top-down destruction of PAHs \citep{Teyssier2004, Pety2005, Pety2012, Guzman2015}; the spatial coincidence of PAH emission bands with hydrocarbons in PDRs is consistent with that interpretation \citep{vanderWiel2009}. 

It is instructive to compare the conditions between classic PDRs and outflow cavity walls around low-mass protostars.The $G_0$ value for Orion Bar is estimated at 2.6 $\times$ 10$^4$ \citep {Marconi1998}, while the Horsehead Nebula has a much more moderate radiation field of $~ 10^2$ \citep{Abergel2003}. The UV radiation field around low-mass protostars measured by various tracers is 10$^2$--10$^3$ at $\sim$ 1000 au from the protostar \citep{Benz2016, Yildiz2015, Karska2018}, therefore the PDR origin of small hydrocarbons is plausible. The top-down production of hydrocarbons due to PAHs destruction does not appear to be an efficient route here, as PAHs are not commonly observed in low-mass protostellar systems \citep{Geers2009}, and the UV-fields required for this process are above 10$^3$ \citep{Abergel2003}.

The difference in morphology of hydrocarbons between Class 0 and Class I systems -- outflow cavity walls in Class 0 versus rotating disk-like structure in Class I -- is most likely related to the evolution of the protostellar systems. Class 0 sources have a dense envelope and the UV radiation can only penetrate the exposed outflow cavity walls, while for Class I it is likely much easier for both the UV radiation from the accreting protostar and the interstellar radiation to reach deeper into the envelope or disk. This is consistent with emission from well-defined cavity walls seen toward those sources.

The case of moderate $\sim 10$ km s$^{-1}$ velocity material observed
toward S68N indicates that the C$_2$H line does not in all sources
trace exclusively the quiescent cavity walls. The profile is consistent with the
observed morphology of the line (see Fig. \ref{fig:hydrocarbons_no2}) -- emission is seen up to a few thousand au from the source and its shape does not resemble a
cavity wall as clearly as in other sources. The narrow component centered at systemic velocity seen
in Fig. \ref{fig:spectra_coms_outflow_no1} indicates that while the broad
component might be dominating the emission, the UV-irradiated cavity
wall also contributes to the emission observed for S68N. 

S68N could be a very young source, as the chaotic structure of its
outflow and envelope indicates \citep{LeGouellec2019}. The high
abundances of freshly released ice-mantle components described in
Section \ref{section:shock_sputtering} are consistent with this interpretation. It is also
possible that UV radiation produced locally in shocks is causing the
enhancement of C$_2$H emission at higher velocities.

The $^{13}$CS molecule, as a high-density tracer likely traces the material piling up on the cavity walls pushed by the outflow. This emission is usually slow ($\pm$ 2  km s$^{-1}$ ) indicating that this is not outflowing gas but rather envelope material on the outflowing cavity walls. The non-detection of $^{13}$CS in 12m data (Fig. \ref{fig:hydrocarbons_no2}) towards the Class I source TMC1 is consistent with the dissipating envelope as the source evolves, hence no high-density material is seen in the remnant cavity walls, even though they are still highlighted by the CO emission. In the remaining Class 0 sources with the $^{13}$CS detections -- Emb8N, SMM1 and S68N (Fig. \ref{fig:fig_E2}), -- emission also follows other cavity wall tracers. SMM1 is the only source that shows higher velocity structure (> 7 km s$^{-1}$) of the  $^{13}$CS emission on one side of the cavity wall. As this is spatially coincident with the high-velocity jet observed in CO \citep{Hull2016}, this emission might be related to the material released with the jet from the envelope.

To summarize, we observe the hydrocarbons C$_2$H and c-C$_3$H$_2$, as
well as CN, H$^{13}$CO${^+}$ and $^{13}$CS in the outflow cavity walls
of Class 0 protostars. C$_2$H, as the most abundant of the hydrocarbons presented here, is also seen prominently across the envelope at
velocities comparable to the low-velocity outflow whereas c-C$_3$H$_2$ appears as a clean tracer of the quiescent, UV-irradiated gas in the cavity walls in the Class 0 sources.

\subsection{Warm inner envelope}

\subsubsection{Compact emission}
\label{section:compact_emission}

The inner regions of young protostellar systems are characterized by
high temperatures which result in a rich chemistry as molecules that
form efficiently in ices on grains in cold clouds sublimate into the
gas phase.  Complex organic molecules (COMs) detected in these
datasets are discussed quantitatively in detail elsewhere, for both
O-bearing by \cite{vanGelder2020} and for N-bearing species by \citep{Nazari2021} In this section we focus on smaller molecules that also
trace the innermost hot core regions and therefore are likely abundant in ices. This includes several small S-bearing molecules.

Both B1-c and S68N are sources characterized as hot cores
\citep{Bergner2017, vanGelder2020}, which means that the conditions in
their inner regions are favourable for release of molecules from the
ice mantles. For the well-studied case of IRAS16293-2422,
\cite{Drozdovskaya2018} have also identified a hot core component of
SO based on isotopologue data, in addition to an SO component in the large scale outflow. Overall, SO observed in protostars appears to be related mostly to evaporation of grain mantle.

The SO appearance close to the central protostar could be related to accretion shocks onto the disk, which are weaker than shocks that cause the SO emission seen in the outflows (Section 4); in accretion shocks,  SO can be released from the icy mantles with the infalling material \citep{Sakai2014a}.  However, narrow linewidths of SO toward TMC1 seem to rule out the accretion shock scenario, and points to emission along the cavity walls \citep{Harsono2020}.

HNCO emission has been modeled by \cite{HernandezGomez2018} and suggested to be a superposition of both warm inner regions of the envelope as well as the colder, outer envelope. Its similar behaviour to sulphur-bearing species, also observed in our work, is proposed to be related to the fact that O$_2$ and OH are involved in formation of  species like SO and HNCO \citep{RodriguezFernandez2010}.

H$_2$S is expected to be the dominant sulphur carrier in
ices \citep{Taquet2020}. However, it has not yet been detected in ice absorption spectra
to date \citep{Boogert2015}. The weak emission from H$_2$S in dark
clouds has been modeled as a result of the photodesorption of ices at
the outside of the cloud (like in the case of H$_2$O,
\citealt{Caselli2012b}), while chemical desorption is important for
grains deeper inside the cloud but outside the water snowline
\citep{NavarroAlmaida2020}. These models are consistent with 
H$_2$S ice containing most of the  sulphur. 
Multiple lines of H$_2$CS are a
powerful tool to probe the warm > 100 K, innermost regions of the
protostellar systems \citep{vantHoff2020a}. 

While B1-c and S68N are characterized as hot cores with many COM lines detected \citep[][]{vanGelder2020}, and L1448-mm has warm water in the inner regions \citep{Codella2010}, SMM3 does not appear to have significant emission from COMs. Moreover, simple molecules associated with the hot cores for B1-c, such as SO and H$_2$CO, are only seen outside of the SMM3 central source. While it is  possible that the optically thick continuum prevents a detection of COMs in its inner envelope \citep{DeSimone2020}, differences in chemistry or physical structure (e.g., a large cold disk, see Section \ref{section:results_inner_env}) between the SMM3 and the hot core sources are also possible.
The fact that emission from OCS and H$_2$S is centrally peaked (Fig. \ref{fig:hotcore_v1}, bottom row) suggests that continuum optical depth is not an issue, although both those species could be a result of grain destruction or ices sputtering, therefore not necessarily coming from the midplane but rather from the surface of a disk-like structure. The additional detection of CCS in 7 data towards SMM3, which is on the other hand not detected in B1-c could hint at different chemical composition of the two protostellar systems. 



\subsubsection{Embedded disks}

In the case of large edge-on
disks like IRAS-04302 and L1527, the vertical structure of the emission can be
probed, as well as the radius where CO freeze-out occurs. Overall,
Class I disks are warmer than their Class II counterparts with CO freeze-out taking place only in the
outermost regions \citep{Harsono2015,vantHoff2018,vantHoff2020b,Zhang2020}.

We detect CN in all observed Class I disks, although it does not trace the midplane of the disk. In the near edge-on example of  IRAS-04302, the CN emission originates from the upper layers of the disks, in the same direction as the outflow, which is perpendicular to the disk in this source. This opens a possibility that the emission is also related to the irradiated residual cavity walls in those sources. TMC1A is a clear example where CN is tracing the same material as probed by \cite{Bjerkeli2016} in CO, which is attributed to a disk wind. TMC1 and L1527 show CN oriented in the same direction as the disk; in TMC1 there is also a clear filament structure on larger scales irradiated by the UV from the protostar, seen also in other tracers.

In comparison with the C$^{17}$O emission, which traces the midplane disk, CN thus appears in the upper layers and in the outflow, therefore in most cases the two molecules are mutually exclusive. This picture is consistent with the bulk density traced by CO and the irradiated layers of the disk and envelope exposed to UV traced by CN. Recent observations of a sample of Class I sources including IRAS-04302 by \cite{Garufi2020} is consistent with CN not tracing the disk midplane.

In the younger Class 0 or borderline Class 0/I sources characterization of the disk is much more difficult because of the strong envelope emission. Nevertheless, several Keplerian disks have been identified with observations of CO
isotopologues like C$^{18}$O and $^{13}$CO
\citep{Tobin2012,Murillo2013}.
Our data allow us to investigate this for the case of SMM3. Fig. \ref{fig:smm3_disk} shows the red and blue-shifted emission from C$^{18}$O toward SMM3. There is a clear rotational signature in the direction perpendicular to the outflow on scales of a few hundred au. However, to unambiguously identify the disk and its radius, higher spatial and spectral resolution data are necessary.

\begin{figure*}[h]
\centering
  \includegraphics[width=0.99\linewidth,trim={1cm 0cm 0cm 0cm}]
 {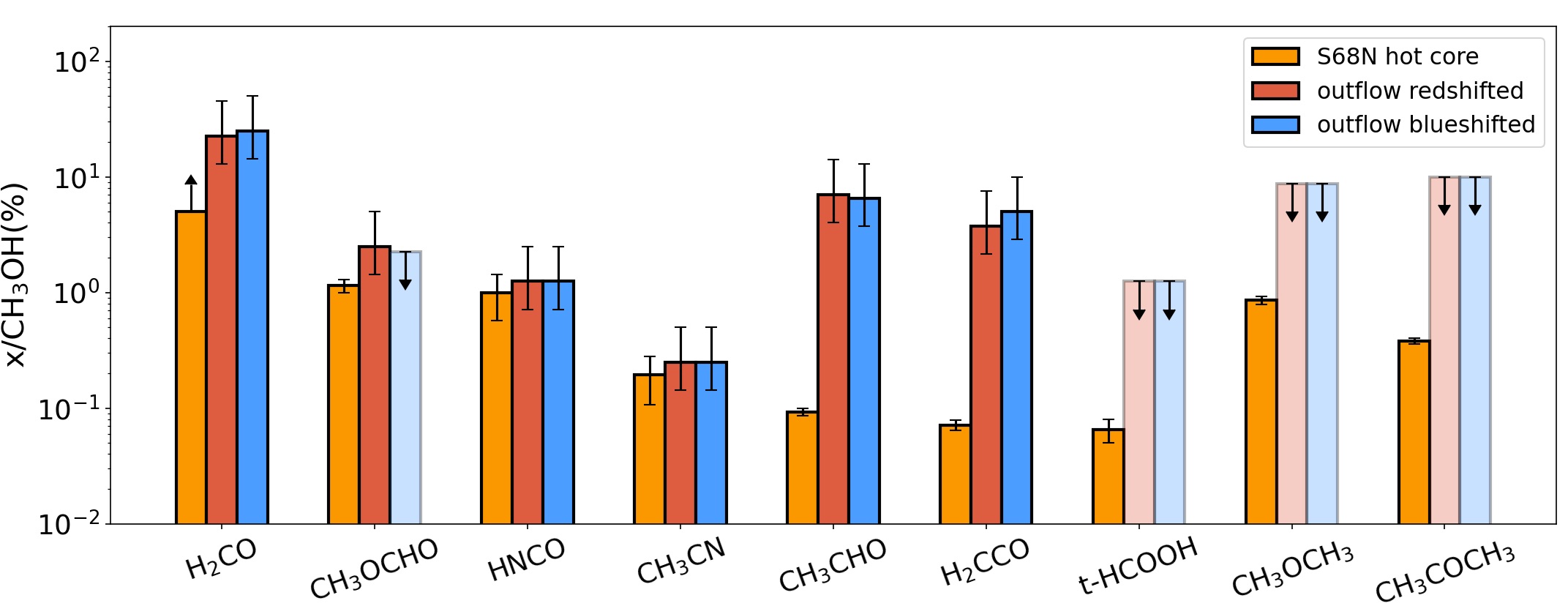}
  \caption{Ratio of complex organic molecules with respect to CH$_3$OH in S68N. The ratios are shown for a 3\arcsec\   region in both redshifted and blueshifted part of the outflow. The abundance ratios in the hot cores are taken from \cite{vanGelder2020} and \cite{Nazari2021}. The errorbars mark the lower and upper limit on the ratios, as calculated from the optically thick CH$_3$OH emission (lower limit) and $^{13}$CH$_3$OH (upper limit). All outflow abundances obtained from the Band 6 data at 0\farcs45 resolution except for  CH$_3$OCHO and CH$_3$CN which were derived from Band 3 data at 3\arcsec. } 
 \label{fig:barplot}
\end{figure*}

\subsection{COMs in outflow versus hot core}

The complex organic molecules are already detected in the prestellar stage of star formation \citep{Bacmann2012, Scibelli2020}, where they are efficiently produced on the surfaces of icy grains \citep[e.g.,][]{Watanabe2004,Oberg2009}. COMs can be subsequently released back into the gas through various non-thermal desorption processes and/or be re-formed by gas-phase reactions. In the inner envelope, temperatures are high enough that thermal desorption of ices is enabled. In the outflow cavity walls, sputtering by shocks can release ices from the grains. In the hot core, COMs are detected through high excitation lines, while in the outflow, they are observed primarily through transitions with low $E_{\rm up}$. Therefore, comparing the observations of molecular complexity in outflows with hot cores can unveil if there is any warm temperature processing of the ices and whether some molecules also have a gas-phase production route. 



Therefore, we compare abundance ratios between different ice
mantle species and methanol at three positions for S68N: one on
the source, obtained from \cite{vanGelder2020} and \cite{Nazari2021}, and
one each in the blueshifted and redshifted part of the outflow, which are calculated from the fluxes obtained in this work.  We
measure the abundance of the species in a 3\arcsec\   region centered on
CH$_3$OH peak on the blueshifted and redshifted regions. The size of the
region is based on the spatial resolution of the Band 3 data. The regions are
indicated in Fig. \ref{fig:icemantle_no1}. From these regions, we extracted spectra and calculated column densities using the spectral analysis tool CASSIS  \footnote{http://cassis.irap.omp.eu/}.

For the CASSIS model, we assume $T_{\rm ex}$ = 20 K, typical of subthermally excited molecules with large dipole moments in outflow gas. The FWHM of the lines was fixed at 3 km s$^{-1}$, which is the width of the CH$_3$OH line. With those parameters, an LTE calculation provides column densities for the observed line intensities under the assumption of optically thin emission (see \cite{Tychoniec2019} for a discussion on uncertainties in this method.)


Only a single CH$_3$OH line   2$_{1,0}$--1$_{0,1}$ ($E_{\rm up}$ = 28 K) is detected in the outflow of S68N. Escape probability calculations with RADEX \citep{Tak2007} show that even at the lower densities in outflows this CH$_3$OH line is likely optically thick. No optically thin CH$_3$OH isotopologue is detected in the outflow; therefore, we provide an upper limit for the $^{13}$CH$_3$OH 2$_{1,1}$-1$_{0,1}$ ($E_{\rm up}=$28 K). The upper limit on this $^{13}$CH$_3$OH line of $1\times 10^{14}$ cm$^{-2}$ translates to an upper limit on the CH$_3$OH column density of $7\times 10^{15}$ cm$^{-2}$ assuming $^{12}$C / $^{13}$C = 70 \citep{Milam2005}. 


The column density calculated from the flux measurement of the CH$_3$OH line gives  of $2\times 10^{15}$ cm$^{-2}$; thus, the methanol abundance can be underestimated by a factor of 4. With the information available, we assume that the CH$_3$OH column density is between $2\times 10^{15}$ cm$^{-2}$ and $7\times 10^{15}$ cm$^{-2}$ and we compare abundances of other molecules with respect to CH$_3$OH using this range of values. The hot core methanol column density of S68N in \cite{vanGelder2020} is corrected for optical thickness with CH$_3^{18}$OH.

Figure \ref{fig:barplot} compares the abundance ratios of various species with respect to methanol for S68N. First, it shows that the relative abundances are remarkably similar on both sides of the outflow, well within the uncertainties. Second, most abundances relative to CH$_3$OH are found to be comparable to the S68N hot core within our uncertainties and those of
\cite{vanGelder2020} and \cite{Nazari2021}, noting that due to lack of optically thin  CH$_3$OH line detected in the outflow, our uncertainties are larger. The greatest difference is seen for CH$_3$CHO and H$_2$CCO, which could be attributed to the additional gas-phase formation in shocks. While the difference is well beyond the conservative uncertainties assumed here, this result should be confirmed with a larger sample of sources with different properties. The jump in CH$_3$CHO abundance with respect to methanol in the outflow  has been also reported for two other sources - L1157 and IRAS4A \citep{DeSimone2020b}.

For the well-studied L1157 outflow, \cite{Codella2020} find
CH$_3$CHO/CH$_3$OH is 0.5\%. which they find in agreement with ratios
in hot cores toward different protostars, but not including L1157
itself. In the case of S68N we uniquely show a comparison between the
emission from molecules in the outflows and hot core for the same
source.  While modeling by \cite{Codella2020} shows that the similar
spatial origin of CH$_3$OH and CH$_3$CHO does not imply that they are
both solely grain sputtering products, an agreement between hot cores
and outflows could mean that similar processes are responsible for
emission in both regions. 


It is interesting that the SMM3 outflow shows ice mantle tracers in
the outflow but not on the source. Thus, the lack of hot core
emission is likely due to the physical conditions in the inner regions such as a large disk (Section \ref{section:results_inner_env}) or
continuum optical depth \citep{DeSimone2020} and not to the lack of
complex molecules on the grains. Since methanol is not detected, we cannot
provide abundance ratios for this outflow. B1-c - another source
with an outflow containing ice mantle products, has methanol lines
overlapping with the high-velocity SO outflow, therefore precise
abundance measurements are not possible.

{\it JWST} will be able to provide information on the ice content due
to rich absorption spectra in the mid-IR. The abundances of COMs in
the gas-phase provided by ALMA \citep{vanGelder2020,Nazari2021} can then be directly compared with the ice
content. So far, the ice content observed on cloud scales with {\it
  Spitzer} does not show a correlation with the gas content but those
data do not probe thermally desorbed ices \citep{Perotti2020}.

\subsection{Carbon-chains and other hydrocarbons vs COMs}

Early observations at low spatial resolution suggested that large
carbon-chains such as C$_5$H, HC$_7$N, and HC$_9$N and complex organic
molecules are mutually exclusive, and therefore the two were proposed
to be tracers of two different categories of protostellar sources
driven by different chemistry, namely warm carbon-chain chemistry
(WCCC) and hot core chemistry \citep{Sakai2013}. Warm carbon-chain
chemistry is thought to occur above the sublimation temperature of CH$_4$ $\sim$ at
30 K.  The proposed difference then lies in the WCCC sources
collapsing more rapidly than the hot core sources, which prevents CO
to accumulate on the ices, leaving a higher CH$_4$ ice content. Thus,
in WCCC sources, there would be less CO to form more complex organic
molecules such as CH$_3$OH, and an underabundance of COMs is indeed
observed. This scenario has been supported by the lower deuteration observed in
WCCC sources \citep{Sakai2009a}. With ALMA observations at
higher resolution and sensitivity, sources harboring both COMs and
small hydrocarbons, c-C$_3$H$_2$, and C$_2$H, have now been observed
\citep{Imai2016, Oya2017}.


\begin{figure}[h]
\centering
  \includegraphics[width=0.92\linewidth,trim={0cm 0cm 0cm 0cm}]
  {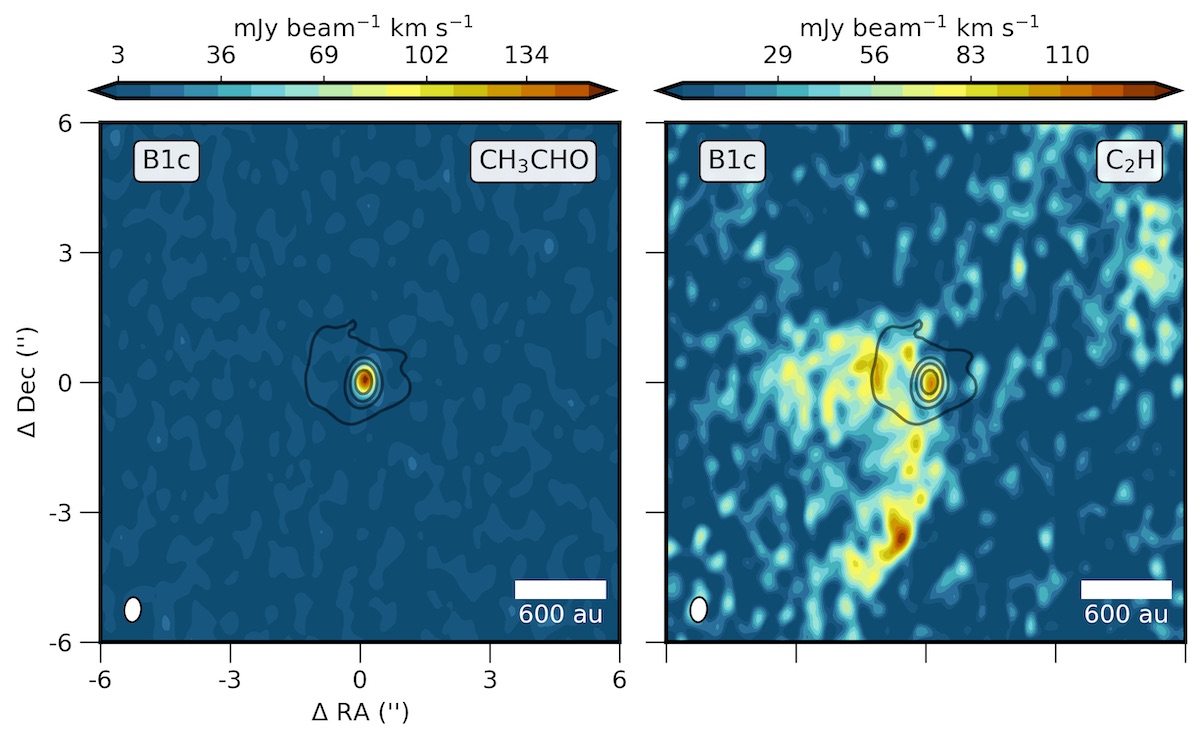}
  \caption{Emission maps of an example COM and hydrocarbon toward B1-c. {\it Left:} CH$_3$CHO moment 0 map, {\it Right:} C$_2$H moment 0 map; both integrated from -2 to 2 km s$^{-1}$ w.r.t  $\varv_{\rm sys} $.} 
 \label{fig:coms_vs_hydrocarbons}
\end{figure}

Our data also show detection of small hydrocarbons and COMs in the
same sources (Fig. \ref{fig:coms_vs_hydrocarbons}).  However, our high resolution maps now show those
hydrocarbons to originate in the UV irradiated outflow cavity walls
(Section 6). Both c-C$_3$H$_2$ and C$_2$H can be solely explained by
PDR chemistry \citep{Guzman2015, Cuadrado2015} even in moderate UV
fields of low-mass protostars and therefore WCCC chemistry does not
need to be invoked for the sources presented in this work. The PDR
origin is consistent with our observed geometry
at the edges of the cavities, where lower extinction allows the UV
radiation from the protostellar accretion to easily reach this
region. It also explains the confinement of this emission to the inner
2000 au of the envelope as the UV radiation decreases with distance
from the source. A deep search for larger hydrocarbons like HC$_5$N in
cavity walls could help to understand their origin.

\subsection{Class 0 vs Class I in different components}

Many processes take place within the protostellar
lifetime of a few $10^5$ yr that alter the chemical composition and
physical conditions. First, the envelope is dissipating, and the radius
where key molecules are in the gas-phase is becoming smaller
\citep{Jorgensen2005a}. With high-resolution studies, it is possible
to probe different snowlines in unprecedented detail now
\citep{vantHoff2018,Hsieh2019}. There is a clear evolutionary trend observed in temperature -- Class 0 disks are generally warmer than Class II disks, with Class I sources in between, which has major impact on the richness of their spectra \citep{vantHoff2020b,vantHoff2020a}. This is also observed in our sample where all Class 0 show either complex organic molecules or small molecules related to grain evaporation but no evidence of those molecules is found in Class I sources (Section \ref{section:compact_emission}). Table \ref{table:evolution_summary} summarizes the evolution of the molecular composition of protostellar systems.

Complex organic molecules are a key tracer of chemistry at the early stages. We see them in the outflows only in Class 0 sources, likely due to the decrease of the outflow force and mass and envelope density. In the inner regions, physical conditions such as temperature and density also favour the early stages for the abundances of COMs. In rare cases of rich chemistry in Class I disks, it is usually attributed to accretion outbursts heating up the disk \citep{vantHoff2018c,Lee2019}. 

Molecular EHV jets are only present in the Class 0 phase \citep{Nisini2015}. In our sample 6 out of 7 sources have high-velocity component and none is detected in Class I (Section \ref{section:extremely_high}), whereas the lower velocity outflows in these stages are rich in ice mantle tracers (Section \ref{section:shock_sputtering}). The Class I jets are invisible in molecular tracers, and their outflows show CO primarily, with much less contribution from shock tracers such as SO and SiO (Section \ref{section:low-velocity_outflow}).

Cavity walls are seen prominently in the Class 0 sources in molecules
whose abundances are sensitive to UV radiation. In Class I sources, a
flared disk and weak outflow remain. While there is less envelope to
attenuate the UV radiation, there is also less material to irradiate and accretion luminosity of older sources decrease,
which results in less prominent emission from UV tracers on large
scales in Class I (Section \ref{section:outflow_cavity_walls}). Another example of a transition of molecular appearance between Class 0 and I is SO: this molecule traces the outflow in Class 0 and the disk-envelope interface in Class I primarily.



\begin{table*}
\caption{Summary of evolution of chemical tracers}             
\label{table:evolution_summary}      
\centering                          
\begin{tabular}{p{2cm} p{7cm} p{7cm}}        
\hline\hline                 
Component & Class 0 & Class I \\

\hline                                   
Envelope & Cold, dense envelope results in cold tracers (DCO$^+$, N$_2$H$^+$, N$_2$D$^+$) & Envelope dissipates, extent of cold tracers is much smaller \\
\hline
Jet & O-bearing molecules (CO, SiO, SO, H$_2$CO) are present in high-velocity bullets & Molecular jet disappears, seen only in atomic and ionised gas \\
\hline
Outflow & Ice sputtering (CH$_3$OH, HNCO) and grain destruction (SO, SiO) tracers are present & Decreased outflow mass and less dense envelope result in no tracers of sputtering and grain destruction, only faint CO remains\\
\hline
Cavity walls & Prominent signs of UV-irradiated cavity walls (CN, C$_2$H, c-C$_3$H$_2$) & No prominent signs of hydrocarbons in cavity walls, CN still present \\
\hline
Hot core & COMs and simple tracers of ice sublimation and high-temperature chemistry  (H$_2$S, OCS) & Small extent plus disk shadowing results in less complexity, except for outbursting sources\\
\hline
Disk & Warm disk with COMs, dust obscuration & Colder disk molecules, Keplerian rotation seen in H$_2$CO, C$^{18}$O, CN in the disk surface.\\

\hline                                   

\end{tabular}
\end{table*}

\section{Conclusions}
\label{section:conclusions}

In this work, we have presented an overview of the molecular tracers of the physical components in protostellar Class 0 and Class I systems. Table \ref{table:table_components_mols} and Fig. \ref{fig:cartoon_summary} present an overview of the results, which are summarized below. 

\begin{itemize}
\item Protostellar envelopes are primarily traced with the dense gas tracers such as C$^{18}$O and CO, and snowline tracers such as DCO$^{+}$, N$_2$D$^+$, H$^{13}$CO$^+$. For this component, it is essential to use observations that are sensitive to large scales of $\sim$ 5000 au.

\item Protostellar outflows are separated into different components. High-velocity molecular jets are traced by O-bearing molecules CO, SiO, SO, and H$_2$CO. Entrained outflow material can be probed by low-velocity CO, SiO, SO, H$_2$CO, HCN as well as molecules released from ices by sputtering, such as CH$_3$OH,  CH$_3$CHO, HNCO, CH$_3$CN, and CH$_3$OCHO.

\item Outflow cavity walls are pronounced in UV irradiation tracers: C$_2$H, c-C$_3$H$_2$, and CN, and high-density tracers such as $^{13}$CS. 

\item Hot cores, the inner warm envelopes, are probed predominantly by COMs and simple molecules released from ice mantles and produced by high-temperature chemistry such as SO, H$_2$S, H$_2$CS, OCS, H$^{13}$CN, HCOOH, and HNCO.

\item Embedded disks, which are most clearly seen in Class I sources, can be seen in C$^{17}$O and H$_2$CO while CN traces their upper layers. CO isotopologues can be used to probe disks in Class 0 sources. However, the confusion with the envelope emission requires a detailed analysis of the kinematics to confirm that a Keplerian disk is present. 

\end{itemize}

Hydrocarbons and COMs are found to  co-exist routinely in protostellar
sources: the former are present in the UV-irradiated cavity walls and
the latter in the hot core and in the outflows. It is plausible that
PDR chemistry for the formation of hydrocarbons is sufficient to
explain the presence of molecules such as C$_2$H and c-C$_3$H$_2$, and
that no warm carbon-chain chemistry is required, which relies on
abundant CH$_4$ in ices. Observations of more complex carbon chains in
cavity walls and envelopes are necessary to understand whether the WCCC
plays a role there. {\it JWST} will be able to probe the CH$_4$ ice
content at  spatial scales comparable to ALMA, which will show if there are any differences in the ice
composition between sources.

Throughout this study, the availability of both low-lying molecular
lines to probe the cold extended envelope and outflow, as well as
high-excitation lines to trace compact hot cores has been a key for the
analysis. The occurrence of the same molecules in different physical
components provides an opportunity to study the processes
involved in their formation and excitation. For example, ice mantle
tracers sputtered from the grains in outflows give useful insight on
the ice composition, whereas comparison with abundances of thermally
released COMs in the inner region is a powerful tool to probe whether
gas-phase formation routes contribute to complex molecules formation.

{\it JWST}-MIRI will provide unprecedented resolution and sensitivity
in mid-IR that will allow probing the emission from hot gas in shocks
as well as ice absorption features. {\it JWST} observations can then be combined with our
understanding of the kinematics and spatial origin of molecules
revealed by ALMA.

{\it Acknowledgements. }The authors are grateful to the anonymous referee for comments that improved the quality of the paper.
This paper makes use of the following ALMA data: ADS/JAO.ALMA\# 2017.1.01174.S, ADS/JAO.ALMA\# 2017.1.01350.S, ADS/JAO.ALMA\# 2017.1.01371.S, ADS/JAO.ALMA\# 2017.1.01413.S,  ADS/JAO.ALMA\# 2013.1.00726.S, ADS/JAO.ALMA\# 2016.1.00710.S.  ALMA is a partnership of ESO (representing its member states), NSF (USA) and NINS (Japan), together with NRC (Canada), MOST and ASIAA (Taiwan), and KASI (Republic of Korea), in co-operation with the Republic of Chile. The Joint ALMA Observatory is operated by ESO, AUI/NRAO and NAOJ. Astrochemistry in Leiden is supported by the Netherlands Research School for Astronomy (NOVA). Ł.T.  acknowledges the ESO Fellowship Programme. M.L.vG. acknowledges support from the Dutch Research Council (NWO) with project number NWO TOP-1 614.001.751. C.L.H.H. acknowledges the support of the NAOJ Fellowship and JSPS KAKENHI grants 18K13586 and 20K14527. The National Radio Astronomy Observatory is a facility of the National Science Foundation operated under cooperative agreement by Associated Universities, Inc.
This research has made use of NASA's Astrophysics Data System Bibliographic Services. This research made use of NumPy \citep{harris2020array}; Astroquery \citep{2019AJ....157...98G}; Astropy, a community-developed core Python package for Astronomy \citep{2018AJ....156..123A, 2013A&A...558A..33A}; Pandas \citep{McKinney_2010, McKinney_2011}; and Matplotlib \citep{Hunter2007}.

\bibliography{mybib}

\begin{appendix}

\section{Tables}

\begin{sidewaystable}
\caption{Specifications of observations}             
\label{table:observations}      
\centering                          
\begin{tabular}{p{2.5cm}p{2.5cm}p{1.1cm}p{2cm}p{0.6cm}p{1.5cm}p{1.8cm}p{1.8cm}p{1.8cm}p{10cm}}    
\hline\hline                 
Project ID & Configuration & $\lambda$ & Res. & MRS$^{\rm a}$  & Calibration$^{\rm b}$  & Bandpass & Phase & Flux & Targets  \\
\hline                                   
2017.1.1350.S & Band 6 (C-4) &  1.3 mm & 0\farcs40 $\times$ 0\farcs29 & 5\arcsec &  5.4.0-68 & J0423-0120  & J0336+3218 &  J0510+1800 & SMM3, TMC1\\
2017.1.1350.S & Band 6 (7m) &  1.3 mm & 6\farcs4 $\times$ 6\farcs1 & 25\arcsec & 5.4.0-68 & J0423-0120  & J0336+3218 &  J0510+1800 & SMM3, TMC1, IRAS4B, BHR71, Emb25, B1c\\

2017.1.1174.S & Band 6 (C-4) &  1.3 mm & 0\farcs58  $\times$ 0\farcs39 & 6\arcsec  & 5.1.1 & J1751+0939  & J1830+0619 & J1751+0939 & B1c, S68N, SMM3 \\

2017.1.1174.S & Band 3 (C-2) &  3 mm & 2\farcs8  $\times$ 1\farcs8 & 16\arcsec & 5.1.1 & J0238+1636  & J0336+3218 & J0238+1636 & B1c, S68N \\

2017.1.1371.S & Band 5 (C-5) &  2 mm & 0\farcs6 $\times$ 0\farcs4 & 4\arcsec & 5.1.1 & J0237+2848  & J0336+3218 & J0237+2848 & L1448, B1c, B5IRS1, HH211\\

2017.1.1413.S & Band 6 (C-4) &  1.3 mm & 0\farcs42 $\times$ 0\farcs28 & 6\arcsec &  5.4.0 & J0510+1800  & J0438+3004 & J0510+1800 & TMC1, 04302, L1527, TMC1A, L1489\\

2013.1.00726.S & Band 6 (C-4, C-2) &  1.3 mm & 0\farcs45 $\times$ 0\farcs35 & 10\arcsec & 4.2.2 & J1733-1304  & J1751+0939 & Titan & SMM1, S68N, Emb8N \\


2016.1.00710.S & Band 3 (C-5) &  3 mm & 0\farcs5 $\times$ 0\farcs4 & 7\arcsec & 4.7.38335 &  J1751+0939  & J1838+0404& J1838+0404 & SMM1, S68N, Emb8N \\

\hline                                   

\end{tabular}
\tablefoot{
\tablefoottext{a}{MRS - maximum recoverable scale at a given configuration and wavelength}
\tablefoottext{b}{Version of CASA used for calibration}
}

\end{sidewaystable}

 \onecolumn

\begin{landscape}
\begin{longtable}{p{1.4cm}p{3.8cm}p{1.3cm}p{0.8cm}p{1.3cm}p{0.7cm}p{0.7cm}p{0.7cm}p{0.6cm}p{0.7cm}p{0.7cm}p{0.7cm}p{0.6cm}p{0.7cm}p{0.7cm}p{0.9cm}p{0.9cm}p{0.9cm}p{0.9cm}p{0.9cm}p{0.9cm}}
\caption{Molecular transitions targeted}
\label{table:table_mols}

\\
\hline
Molecule & Transition & Frequency & E$_{\rm up}$ & $A_{ij}$ & Serpens &  &  &  & Perseus &  &  &  & Taurus &  &  &  &  \\
  &  & GHz & K & s$^{-1}$ & S68N & SMM3 & SMM1 & 8N & B1c & L1448 & HH211 & B5 & 04302 & TMC1 & TMC1A & L1527 & L1489 \\
\hline 
\hline

\endfirsthead
\caption{Continued.}
\\
\hline
Molecule & Transition & Frequency & E$_{\rm up}$ & $A_{ij}$ & Serpens &  &  &  & Perseus &  &  &  & Taurus &  &  &  &  \\
  &  & GHz & K & s$^{-1}$ & S68N & SMM3 & SMM1 & 8N & B1c & L1448 & HH211 & B5 & 04302 & TMC1 & TMC1A & L1527 & L1489 \\
\hline
\hline

\endhead

CO & 2$_{}$--$1_{}$ & 230.5380 & 17 & 6.9e-7 & \cmark & \cmark & \cmark & \cmark & --- & --- & --- & --- & --- & \cmark & --- & --- & --- \\
C$^{18}$O & 2$_{}$--$1_{}$ & 219.5604 & 16 & 6.1e-7 & \cmark & \cmark & \cmark & \cmark & --- & --- & --- & --- & --- & \cmark & --- & --- & --- \\
$^{13}$CO & 2$_{}$--$1_{}$ & 220.3987 & 16 & 6.1e-7 & --- & \cmark &  --- & --- & --- & --- & --- & --- & \cmark & --- & --- & --- \\
C$^{17}$O & 2$_{1.5}$--1$_{1.5}$ & 224.7147 & 16 & 4.5e-7 & --- & --- & --- & --- & --- & --- & --- & --- & \cmark & \cmark & \cmark & \cmark & \cmark \\
H$^{13}$CO$^{+}$ & 1$_{}$--0$_{}$ & 86.7543 & 4 & 3.9e-5 & \xmark & --- & \cmark & \xmark & --- & --- & --- & --- & --- & --- & --- & --- & --- \\
 & 2$_{}$--1$_{}$ & 173.5067 & 13 & 3.7e-4 & --- & --- & --- & --- & \cmark & \cmark & \cmark & \cmark & --- & --- & --- & --- & --- \\
 & 3$_{}$--2$_{}$ & 260.2553 & 25 & 1.3e-3 & \cmark & --- & --- & \cmark & \cmark & --- & --- & --- & --- & --- & --- & --- & --- \\

H$_2$CO & 3$_{0,3}$--2$_{0,2}$ & 218.2222 & 21 & 2.8e-4 & \cmark & --- & \cmark & \cmark & --- & --- & --- & --- & --- & --- & --- & --- & --- \\
  & 3$_{1,2}$--2$_{1,1}$ & 225.6978 & 34 & 2.8e-4 & --- & --- & --- & --- & --- & --- & --- & --- & \cmark & \cmark & \cmark & \cmark & \cmark \\
 & 3$_{2,1}$--2$_{2,0}$ & 218.7601 & 68 & 1.6e-4 & --- & \cmark & --- & --- & --- & --- & --- & --- & --- & \cmark & --- & --- & --- \\
& 9$_{1,8}$--9$_{1,9}$ & 216.5687 & 174 & 7.2e-6 & --- & \xmark & --- & --- & --- & --- & --- & --- & --- & \xmark & --- & --- & --- \\
 
H$_2$CS & 5$_{1,4}$--4$_{1,3}$ & 174.3452 & 38 & 7.3e-5 & --- & --- & --- & --- & \cmark & \cmark & \xmark & \cmark & --- & --- & --- & --- & --- \\
 & 7$_{0,7}$--6$_{0,6}$ & 240.2669 & 46 & 2.1e-4 & --- & --- & --- & --- & --- & --- & --- & --- & \cmark & \xmark & \xmark & \xmark & \cmark \\

SO & 5$_{6}$--4$_{5}$ & 219.9494 & 35 & 1.4e-4 & --- & \cmark & --- & --- & --- & --- & --- & --- & --- & \cmark & --- & --- & --- \\
 & 6$_{7}$--5$_{6}$ & 261.8437 & 48 & 2.3e-4 & \cmark & --- & --- & \cmark & \cmark & --- & --- & --- & --- & --- & --- & --- & --- \\

SiO & 4$_{}$--3$_{}$ & 173.6883 & 21 & 2.6e-4 & --- & --- & --- & --- & \cmark & \cmark & \cmark & \xmark & --- & --- & --- & --- & --- \\
 & 5$_{}$--4$_{}$ & 217.1050 & 31 & 5.2e-4 & \cmark & \cmark & \cmark & \cmark & --- & --- & --- & --- & --- & \xmark & --- & --- & --- \\

DCO$^+$ & 3--2 & 216.1126 & 21 & 7.6e-4 & \cmark & \cmark & \cmark & \cmark & --- & --- & --- & --- & --- & \cmark & --- & --- & --- \\
N$_2$D$^+$ & 3$_{}$--2$_{}$ & 231.3218 & 22 & 7.1e-4 & --- & \cmark & --- & --- & --- & --- & --- & --- & --- & \xmark & --- & --- & --- \\

OCS & 19--18 & 231.0610 & 111 & 3.6e-5 & --- & \cmark & --- & --- & --- & --- & --- & --- & --- & \xmark & --- & --- & --- \\
O$^{13}$CS & 18--17 & 218.1990 & 99 & 3.0e-5 & \cmark& --- & \cmark & \xmark & --- & --- & --- & --- & --- & --- & --- & --- & --- \\

H$_2$S & 2$_{2}$--2$_{1}$ & 216.7104 & 83 & 4.8e-5 & --- &  \cmark & --- & --- & --- & --- & --- & --- & --- & \xmark & --- & --- & --- \\

HNCO & 5$_{0,5}$--4$_{0,4}$ & 109.9058 & 16 & 1.8e-5 & \cmark &  \cmark & --- & \xmark & \cmark & --- & --- & --- & --- & --- & --- & --- & --- \\
 & 11$_{0,11}$--10$_{0,10}$ & 241.7741 & 70 & 2.0e-4 & --- & --- & --- & --- & --- & --- & --- & --- & \xmark & \xmark & \xmark & \xmark & \xmark \\
 & 12$_{0,12}$--11$_{0,11}$ & 263.7487 & 82 & 2.6e-4 & \cmark & --- & --- & \xmark & \cmark & --- & --- & --- & --- & --- & --- & --- & --- \\
 & 12$_{1,12}$--11$_{1,11}$ & 262.7696 & 125 & 2.6e-4 & \cmark & --- & --- & \xmark & \cmark & --- & --- & --- & --- & --- & --- & --- & --- \\
H$_2$CCO & 13$_{1,13}$--12$_{1,12}$ & 260.1920 & 101 & 2.0e-4 & \cmark & --- & --- & \xmark & \cmark & --- & --- & --- & --- & --- & --- & --- & --- \\
 & 13$_{2,11}$--12$_{2,10}$ & 262.7609 & 141 & 2.0e-4 & \cmark & --- & --- & \xmark & \cmark & --- & --- & --- & --- & --- & --- & --- & --- \\
HCOOH & 5$_{0,5}$--4$_{0,4}$ & 111.7468 & 16 & 1.5e-5 & \cmark & --- & --- & \xmark & \cmark & --- & --- & --- & --- & --- & --- & --- & --- \\
 & 5$_{2,4}$--4$_{2,3}$ & 112.2871 & 29 & 1.3e-5 & \cmark & --- & --- & \xmark & \cmark & --- & --- & --- & --- & --- & --- & --- & --- \\
 & 5$_{3,3}$--4$_{3,2}$ & 112.4596 & 45 & 9.7e-6 & \cmark & --- & --- & \xmark & \cmark & --- & --- & --- & --- & --- & --- & --- & --- \\
 & 12$_{0,12}$--11$_{0,11}$ & 262.1036 & 83 & 2.0e-4 & \cmark & --- & --- & \xmark & \cmark & --- & --- & --- & --- & --- & --- & --- & --- \\

$^{13}$CS & 5--4 & 231.2207 & 33 & 2.5e-4 & \cmark & \cmark & \cmark & \cmark & --- & --- & --- & --- & --- & \xmark & --- & --- & --- \\

CN & $1_{0,0.5,0.5}$--$0_{0,0.5,1.5}$ & 113.1442 & 5 & 1.1e-5 & \cmark & \cmark & --- & \cmark & \cmark & --- & --- & --- & --- & --- & --- & --- & --- \\
 & $2_{0,1.5,2.5}$--$1_{0,1.5,2.5}$ & 226.3599 & 16 & 1.6e-5 & --- & --- & --- & --- & --- & --- & --- & --- & \xmark & \cmark & \cmark & \cmark & \cmark \\
 & $2_{0,1.5,0.5}$--$1_{0,0.5,1.5}$ & 226.6166 & 16 & 1.1e-5 & --- & --- & --- & --- & --- & --- & --- & --- & \xmark & \cmark & \xmark & \xmark & \cmark \\
  & $2_{0,1.5,2.5}$--$1_{0,0.5,1.5}$ & 226.6596 & 16 & 9.5e-5 & --- & --- & --- & --- & --- & --- & --- & --- & \cmark & \cmark & \cmark & \cmark & \cmark \\
 & $2_{0,1.5,0.5}$--$1_{0,0.5,0.5}$ & 226.6637 & 16 & 8.5e-5 & --- & --- & --- & --- & --- & --- & --- & --- & \cmark & \cmark & \cmark & \cmark & \cmark \\
 & $2_{0,1.5,1.5}$--$1_{0,0.5,0.5}$ & 226.6793 & 16 & 5.3e-5 & --- & --- & --- & --- & --- & --- & --- & --- & \cmark & \cmark & \cmark & \cmark & \cmark \\
 & $2_{0,2.5,2.5}$--$1_{0,1.5,1.5}$ & 226.8742 & 16 & 9.6e-5 & --- & --- & --- & --- & --- & --- & --- & --- & \cmark & \cmark & \cmark & \cmark & \cmark \\
 & $2_{0,2.5,3.5}$--$1_{0,1.5,2.5}$ & 226.8748 & 16 & 1.1e-4 & --- & --- & --- & --- & --- & --- & --- & --- & \cmark & \cmark & \cmark & \cmark & \cmark \\
 & $2_{0,2.5,1.5}$--$1_{0,1.5,0.5}$ & 226.8759 & 16 & 8.6e-5 & --- & --- & --- & --- & --- & --- & --- & --- & \cmark & \cmark & \cmark & \cmark & \cmark \\
 & $2_{0,2.5,1.5}$--$1_{0,1.5,1.5}$ & 226.8874 & 16 & 2.7e-5 & --- & --- & --- & --- & --- & --- & --- & --- & \cmark & \cmark & \cmark & \cmark & \cmark \\
 & $2_{0,2.5,2.5}$--$1_{0,1.5,2.5}$ & 226.8921 & 16 & 1.8e-5 & --- & --- & --- & --- & --- & --- & --- & --- & \cmark & \cmark & \cmark & \cmark & \cmark \\

HCN & 1--0 & 88.6318 & 4 & 2.4e-5 & \cmark & --- & \cmark & \cmark & --- & --- & --- & --- & --- & --- & --- & --- & --- \\

H$^{13}$CN  & 1--0 & 86.3423 & 4 & 2.2e-5 & \cmark & --- & \cmark & \cmark & --- & --- & --- & --- & --- & --- & --- & --- & --- \\
 & 2--1 & 172.6780 & 12 & 1.6e-4 & --- & --- & --- & --- & \cmark & \cmark & \cmark & \xmark & --- & --- & --- & --- & --- \\

CCH & $3_{2.5,2}$--$2_{1.5,2}$ & 262.0788 & 25 & 6.5e-6 & \cmark & --- & --- & \xmark & \cmark & --- & --- & --- & --- & --- & --- & --- & --- \\
 & $3_{2.5,3}$--$2_{1.5,2}$ & 262.0648 & 25 & 5.3e-5 & \cmark & --- & --- & \cmark & \cmark & --- & --- & --- & --- & --- & --- & --- & --- \\
 & $3_{2.5,2}$--$2_{1.5,1}$ & 262.0673 & 25 & 4.8e-5 & \cmark & --- & --- & \cmark & \cmark & --- & --- & --- & --- & --- & --- & --- & --- \\
 & $3_{3.5,3}$--$2_{2.5,2}$ & 262.0064 & 25 & 5.5e-5 & \cmark & --- & --- & \cmark & \cmark & --- & --- & --- & --- & --- & --- & --- & --- \\
 & $3_{3.5,4}$--$2_{2.5,3}$ & 262.0042 & 25 & 5.7e-5 & \cmark & --- & --- & \cmark & \cmark & --- & --- & --- & --- & --- & --- & --- & --- \\

c-C$_3$H$_2$  & 6$_{1,6}$--5$_{0,5}$ & 217.8221 & 39 & 5.9e-4 & --- & \cmark & --- & --- & --- & --- & --- & --- & --- & \cmark & --- & --- & --- \\
  & 7$_{2,6}$--7$_{1,7}$ & 218.7327 & 61 & 9.8e-5 & --- & \cmark & --- & --- & --- & --- & --- & --- & --- & \cmark & --- & --- & --- \\
& 7$_{4,3}$--7$_{3,4}$ & 112.4908 & 83 & 4.5e-5 & \cmark & \cmark & --- & \xmark & \cmark & --- & --- & --- & --- & --- & --- & --- & --- \\

CH$_3$CHO  & 6$_{1,6,1}$--5$_{1,5,1}$ & 112.2487 & 21 & 4.5e-5 & \cmark & \cmark & --- & \xmark & \cmark & --- & --- & --- & --- & --- & --- & --- & --- \\
& 6$_{1,6,0}$--5$_{1,5,0}$ & 112.2545 & 21 & 4.5e-5 & \cmark & \cmark & --- & \xmark & --- & --- & --- & --- & --- & --- & --- & --- & --- \\
 & 9$_{4,5}$--8$_{4,4}$ & 173.5191 & 78 & 1.4e-4 & --- & --- & --- & --- & \cmark & \cmark & \xmark & \xmark & --- & --- & --- & --- & --- \\
 & 14$_{0,14}$--13$_{0,13}$ & 262.9601 & 96 & 6.2e-4 & \cmark & --- & --- & \xmark & \cmark & --- & --- & --- & --- & --- & --- & --- & --- \\

CH$_3$CN  & 6$_{0}$--5$_{0}$ & 110.3835 & 19 & 1.1e-4 & \cmark & \cmark & --- & \xmark & \cmark & --- & --- & --- & --- & --- & --- & --- & --- \\
 & 6$_{1}$--5$_{1}$ & 110.3814 & 26 & 1.1e-4 & \cmark & \cmark & --- & \xmark & \cmark & --- & --- & --- & --- & --- & --- & --- & --- \\
 & 6$_{2}$--5$_{2}$ & 110.3750 & 47 & 9.9e-5 & \cmark & \cmark & --- & \xmark & \cmark & --- & --- & --- & --- & --- & --- & --- & --- \\
 & 6$_{3}$--5$_{3}$ & 110.3644 & 83 & 8.3e-5 & \cmark & \cmark & --- & \xmark & \cmark & --- & --- & --- & --- & --- & --- & --- & --- \\
 & 6$_{4}$--5$_{4}$ & 110.3495 & 133 & 6.2e-5 & \cmark & \xmark & --- & \xmark & \cmark & --- & --- & --- & --- & --- & --- & --- & --- \\

CH$_3$OH   & 2$_{1,0}$--1$_{0,1}$ & 261.8057 & 28 & 5.6e-5 & \cmark & --- & --- & \xmark & \cmark & --- & --- & --- & --- & --- & --- & --- & --- \\
 & 5$_{0}$--4$_{0}$ & 241.7914 & 35 & 6.1e-5 & --- & --- & --- & --- & --- & --- & --- & --- & \cmark & \xmark & \xmark & \xmark & \cmark \\
& 5$_{-1}$--4$_{-1}$ & 241.7672 & 40 & 5.8e-5 & --- & --- & --- & --- & --- & --- & --- & --- & \xmark & \xmark & \xmark &\xmark & \cmark \\
 & 5$_{2}$--4$_{2}$ & 241.9046 & 57 & 5.0e-5 & --- & --- & --- & --- & --- & --- & --- & ---  & \xmark & \xmark & \xmark & \xmark & \cmark \\
& 4$_{2}$--5$_{1}$ & 234.6834 & 61 & 1.9e-5 & --- & \xmark & --- & --- & --- & --- & --- & --- & --- & \xmark & --- & --- & --- \\

$^{13}$CH$_3$OH & 2$_{1,1}$--1$_{0,1}$ & 259.9865 & 28 & 5.5e-5 & \cmark & --- & --- & \xmark & \cmark & --- & --- & --- & --- & --- & --- & --- & --- \\
  & 5$_{2,3}$--4$_{1,3}$ & 261.8057 & 56 & 7.4e-5 & \cmark & --- & --- & \xmark & \cmark & --- & --- & --- & --- & --- & --- & --- & --- \\

CH$_3$OCHO & 9$_{5,5}$--8$_{5,4}$ & 110.8805 & 43 & 1.4e-5 & \cmark & \xmark & --- &\xmark & \cmark & --- & --- & --- & --- & --- & --- & --- & --- \\
 & 10$_{0,10}$--9$_{0,9}$ & 111.1699 & 30 & 2.0e-5 & \cmark & \xmark & --- & \xmark & \cmark & --- & --- & --- & --- & --- & --- & --- & --- \\
 & 9$_{1,8}$--8$_{1,7}$ & 111.6741 & 28 & 2.0e-5 & \cmark & \xmark & --- & \xmark & \cmark & --- & --- & --- & --- & --- & --- & --- & --- \\

\hline


\end{longtable}
\tablefoot{\cmark - transition detected, \xmark - transition not detected, -- transition not targeted
}
\end{landscape}

\section{Continuum images}

\begin{figure*}[h]
\centering
  \includegraphics[width=1\linewidth,trim={0cm 0cm 0cm 0cm}]
  {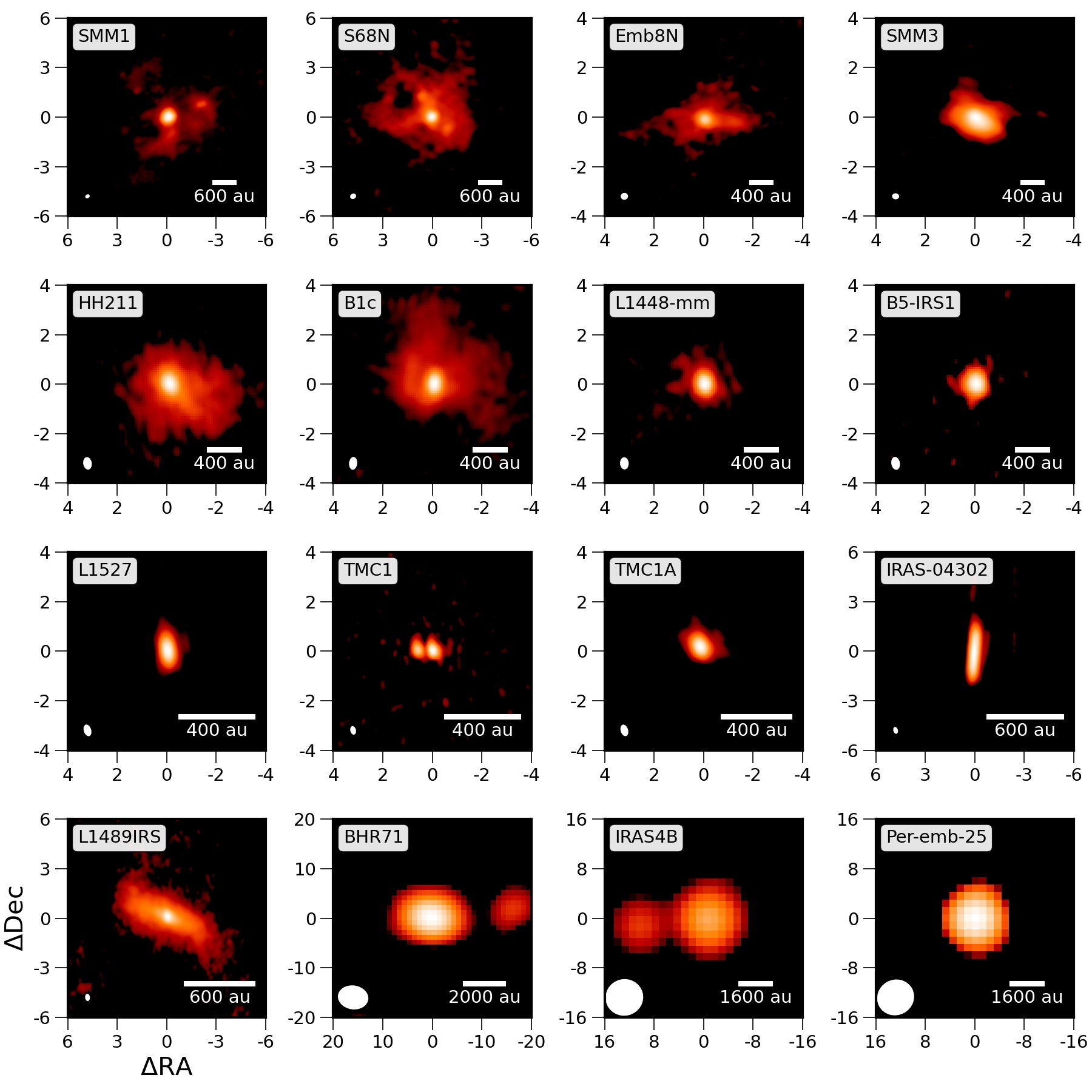}
  
  \caption{Continuum emission from the sources in the sample.  }
 \label{fig:continuum_no1_v1}
\end{figure*}

\newpage

\section{Envelope tracers} \label{AppendixD}


\begin{figure*}[h]
\centering

  \includegraphics[width=0.95\linewidth,trim={3cm 1cm 3cm 0cm}]
  {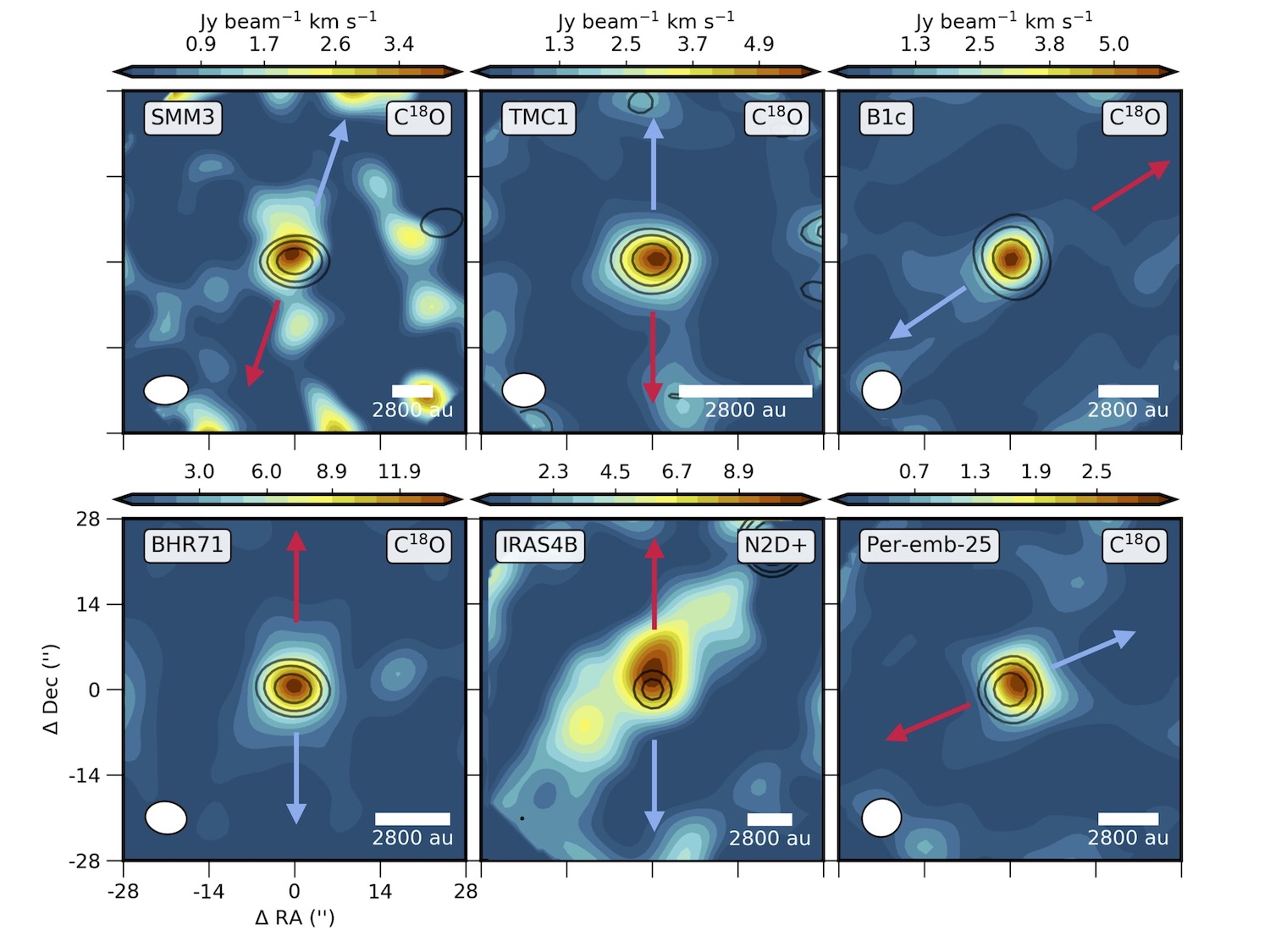}

  \caption{Moment maps of C$^{18}$O 2--1 ($E_{\rm up}$=16 K) at 6\arcsec\   resolution. Contours: continuum, color scale: C$^{18}$O integrated from -2 to 2 km s$^{-1}$ w.r.t  $\varv_{\rm sys}$} 
 \label{fig:envelope_co18_7m}
\end{figure*}

\begin{figure*}[h]

\centering

  \includegraphics[width=0.95\linewidth,trim={3cm 1cm 3cm 0cm}]
  {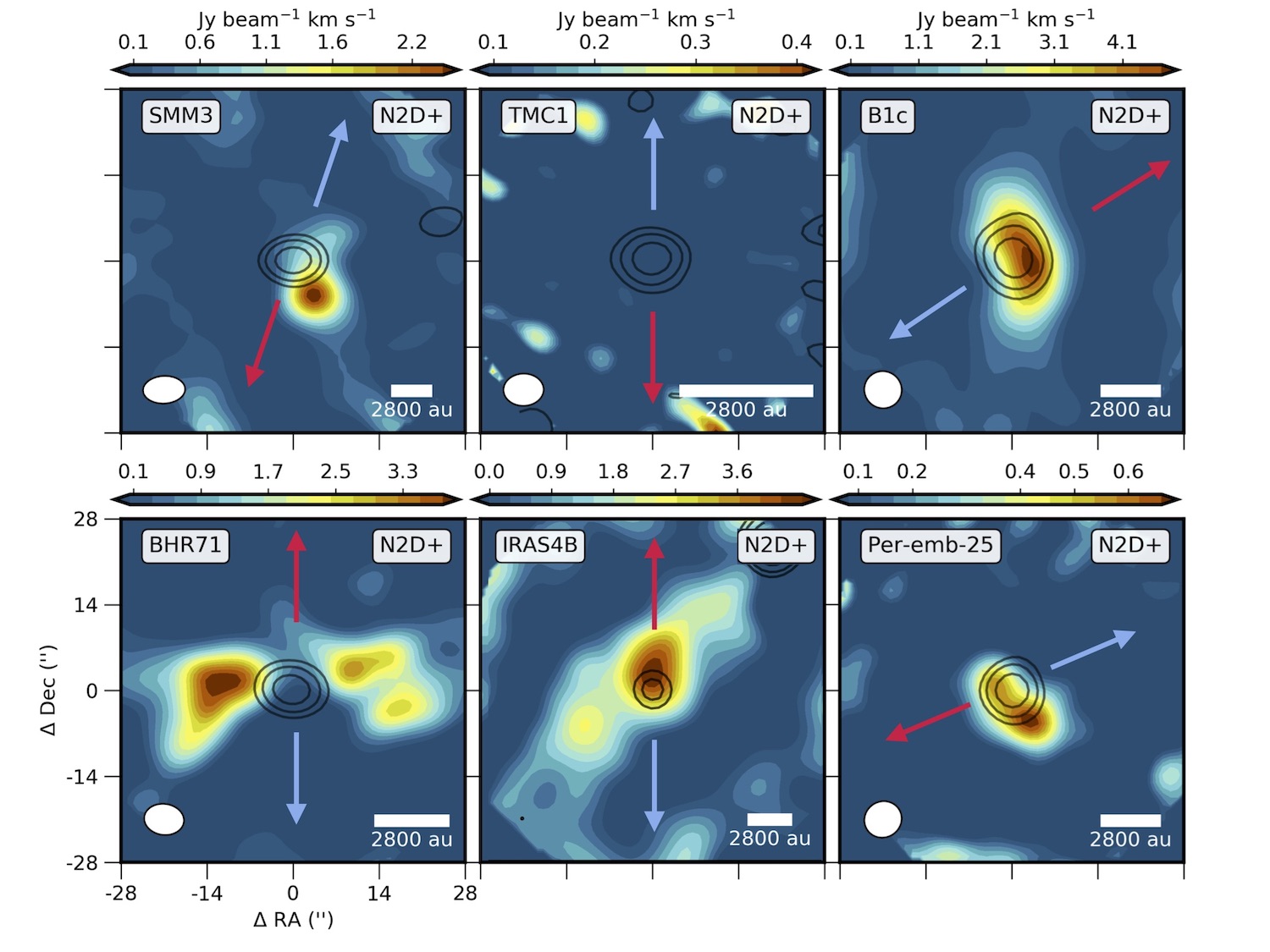}

  \caption{Moment maps of N$_2$D$^{+}$ 3--2 ($E_{\rm up}$=22 K)  at 6\arcsec\   resolution. Contours: continuum, color scale: moment 0 map integrated from -2 to 2 km s$^{-1}$ w.r.t  $\varv_{\rm sys}$ } 
 \label{fig:envelope_n2d+_7m}
\end{figure*}


\begin{figure*}[h]

\centering

  \includegraphics[width=0.95\linewidth,trim={3cm 1cm 3cm 0cm}]
  {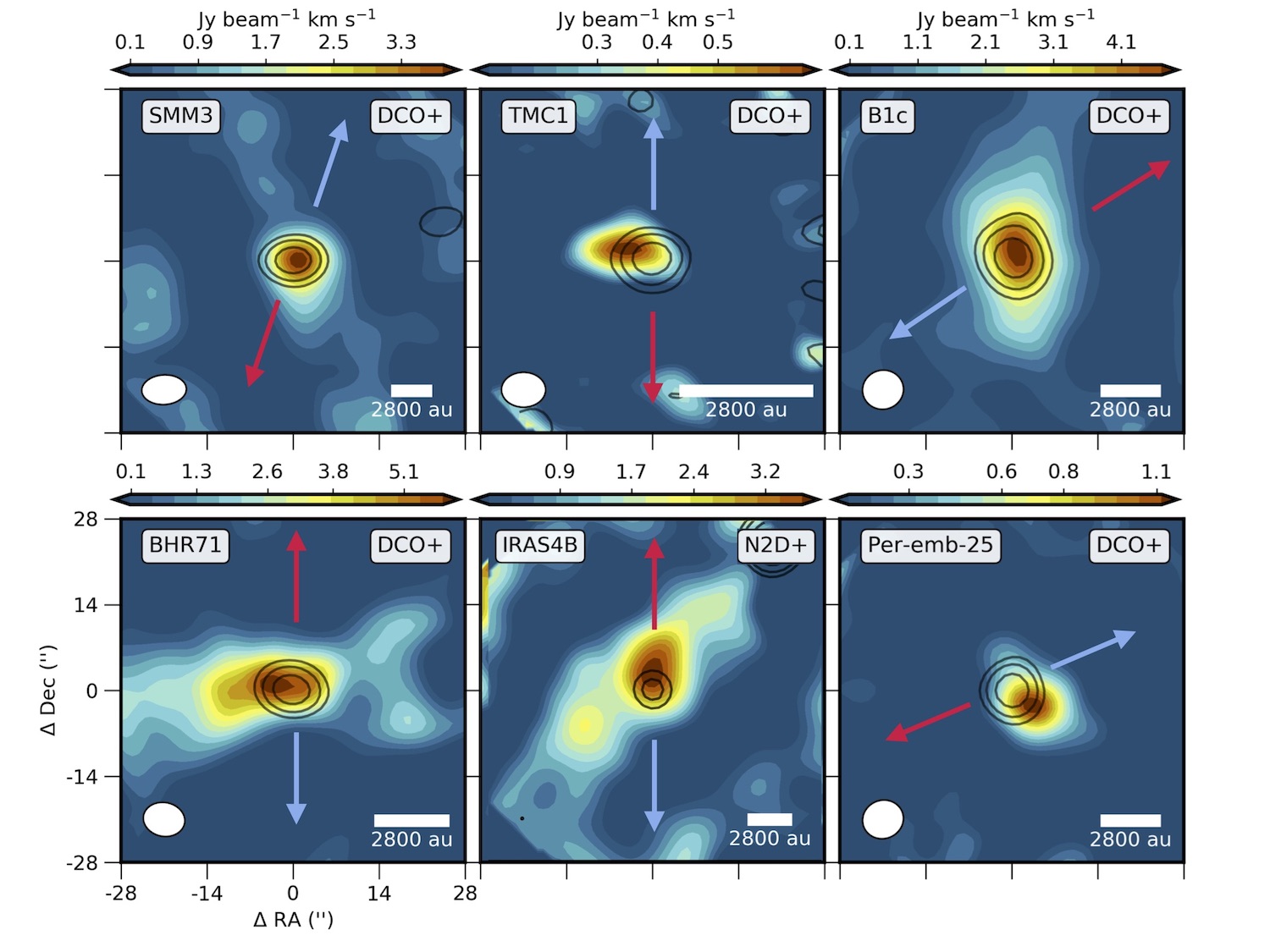}
  
  \caption{Moment maps of DCO$^{+}$ 3--2 ($E_{up}$=21 K) at 6\arcsec\   resolution. Contours: continuum, color scale: moment 0 map integrated from -2 to 2 km s$^{-1}$ w.r.t  $\varv_{\rm sys}$ } 
 \label{fig:envelope_dco+_7m}
\end{figure*}


\clearpage

\section{Outflow}

\begin{figure*}[h]
\centering
  \includegraphics[width=0.95\linewidth,trim={3cm 1cm 3cm 0cm}]
  {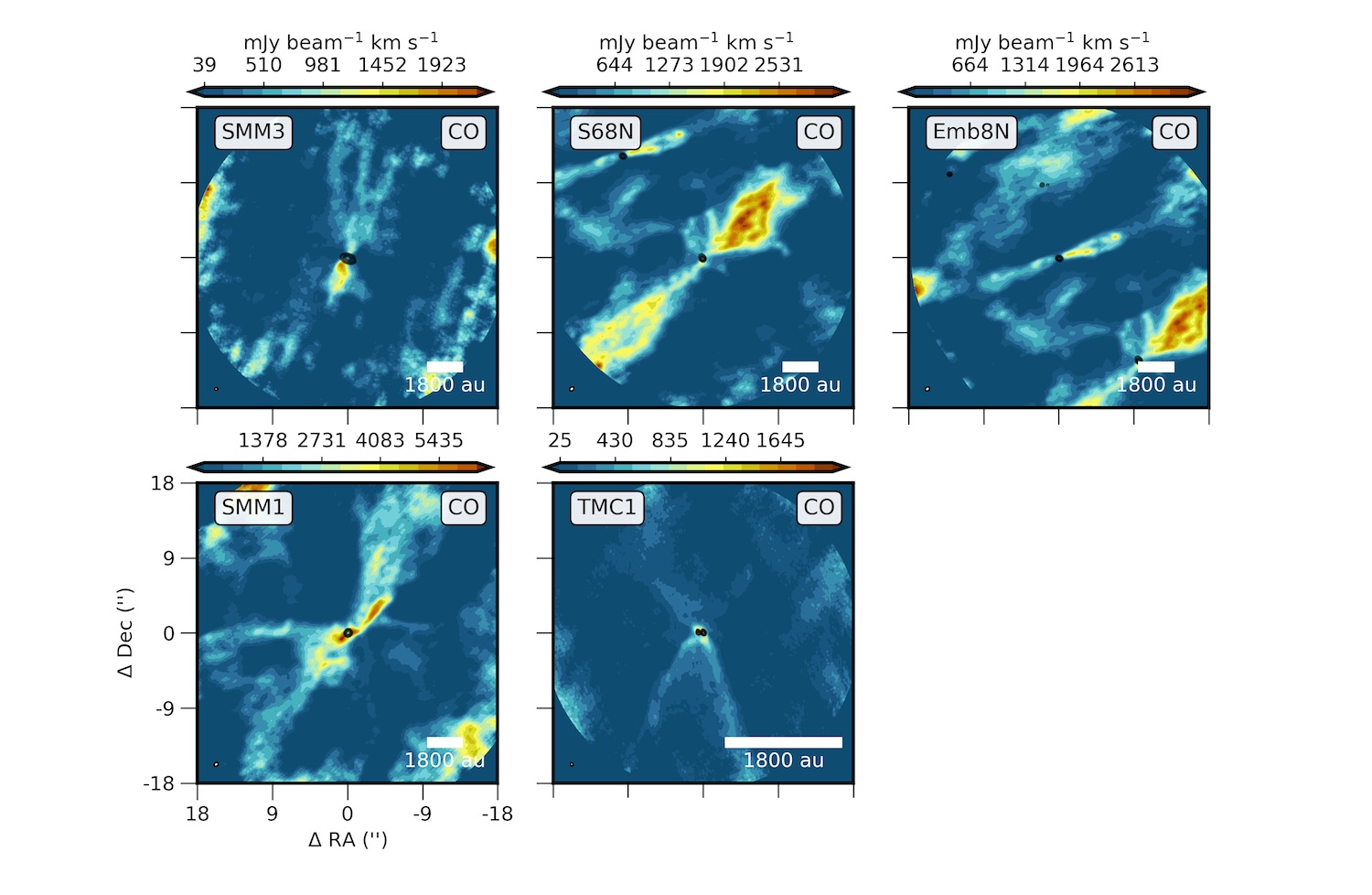}
  \caption{Moment maps of low-velocity CO 2--1 ($E_{\rm up}$=17 K) outflow at 0\farcs4 resolution. Contours: continuum, color scale: moment 0 map integrated from -10 to 10 km s$^{-1}$ w.r.t  $\varv_{\rm sys}$ }  \label{fig:co_outflows}
\end{figure*}

\begin{figure*}[h]
\centering
  \includegraphics[width=0.99\linewidth,trim={0cm 0cm 0cm 0cm}]
  {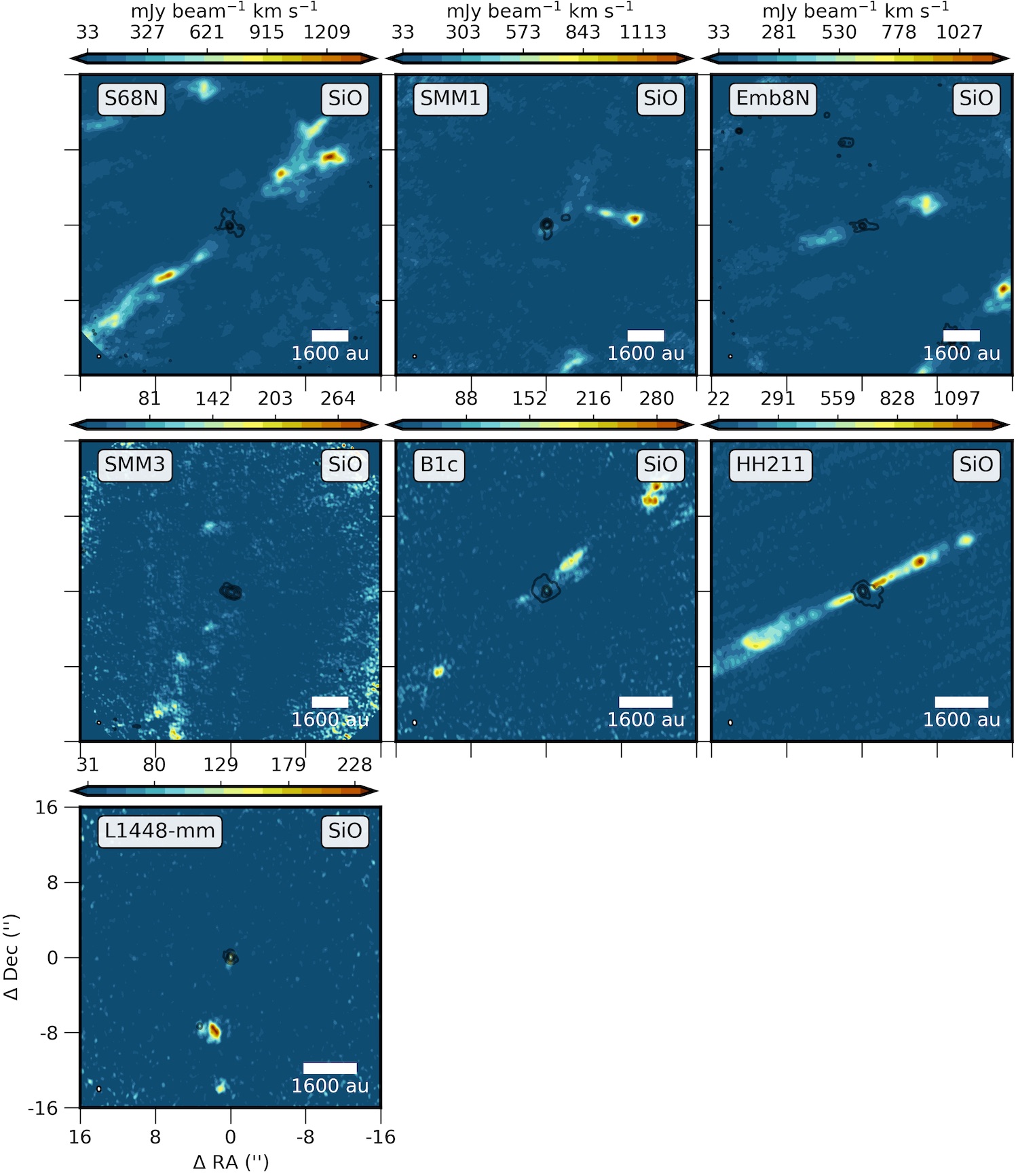}
  \caption{Moment maps of low-velocity SiO  5$_{}$--4$_{}$ ($E_{\rm up}$=34 K) and 4$_{}$--3$_{}$ ($E_{\rm up}$=21 K) outflow at 0\farcs4 resolution. Contours: continuum, color scale: moment 0 map integrated from -15 to 15 km s$^{-1}$ w.r.t  $\varv_{\rm sys}$ }  \label{fig:sio_outflows_no2}
\end{figure*}

\begin{figure*}[h]
\centering

  \includegraphics[width=1.1\linewidth,trim={0cm 0cm 0cm 0cm}]
  {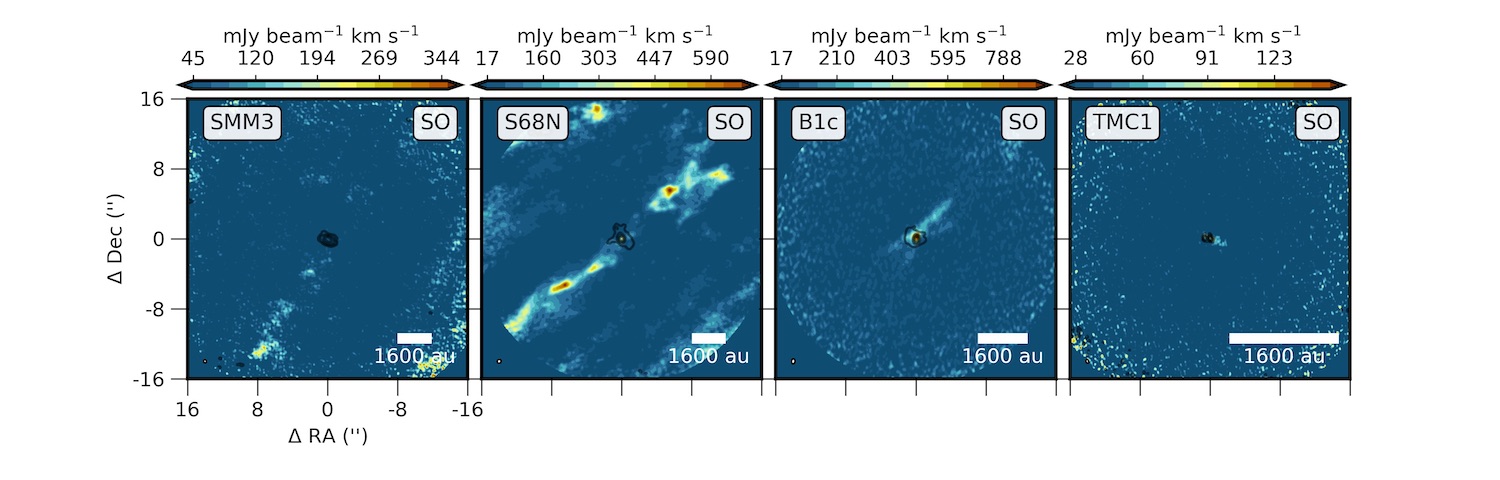}
  \caption{Moment maps of low-velocity SO 5$_{6}$--4$_{5}$  ($E_{\rm up}$=35 K) and 6$_{7}$--5$_{6}$  ($E_{\rm up}$=48 K)  outflow at 0\farcs4 resolution. Contours: continuum, color scale: moment 0 map integrated from -10 to 10 km s$^{-1}$ w.r.t  $\varv_{\rm sys}$ }  \label{fig:so_outflow}
  
\end{figure*}

\begin{figure*}[h]
\centering
     \includegraphics[width=1.1\linewidth,trim={0cm 0cm 0cm 0cm}]
  {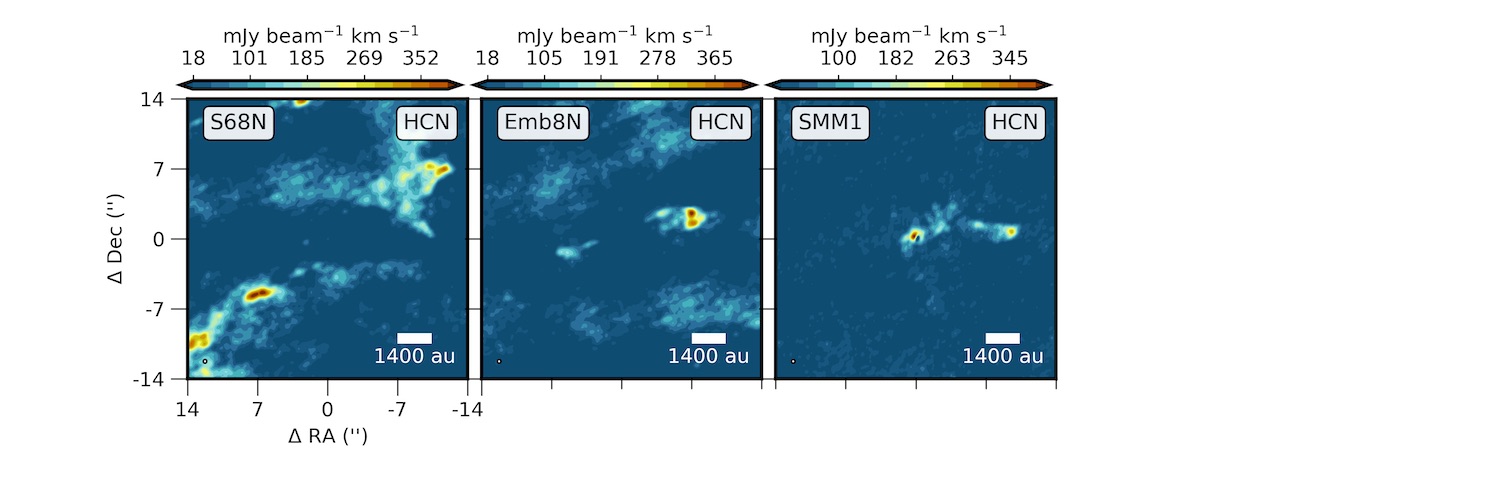}

    \caption{Moment maps of HCN 1--0  ($E_{\rm up}$=4 K) at 0\farcs5 resolution.  Color scale: moment 0 map integrated from -15 to 15 km s$^{-1}$ w.r.t  $\varv_{\rm sys}$ } 
  
 \label{fig:hcn_outflow}
  
\end{figure*}

\begin{figure*}[h]
\centering
    \includegraphics[width=1.1\linewidth,trim={0cm 0cm 0cm 0cm}]
  {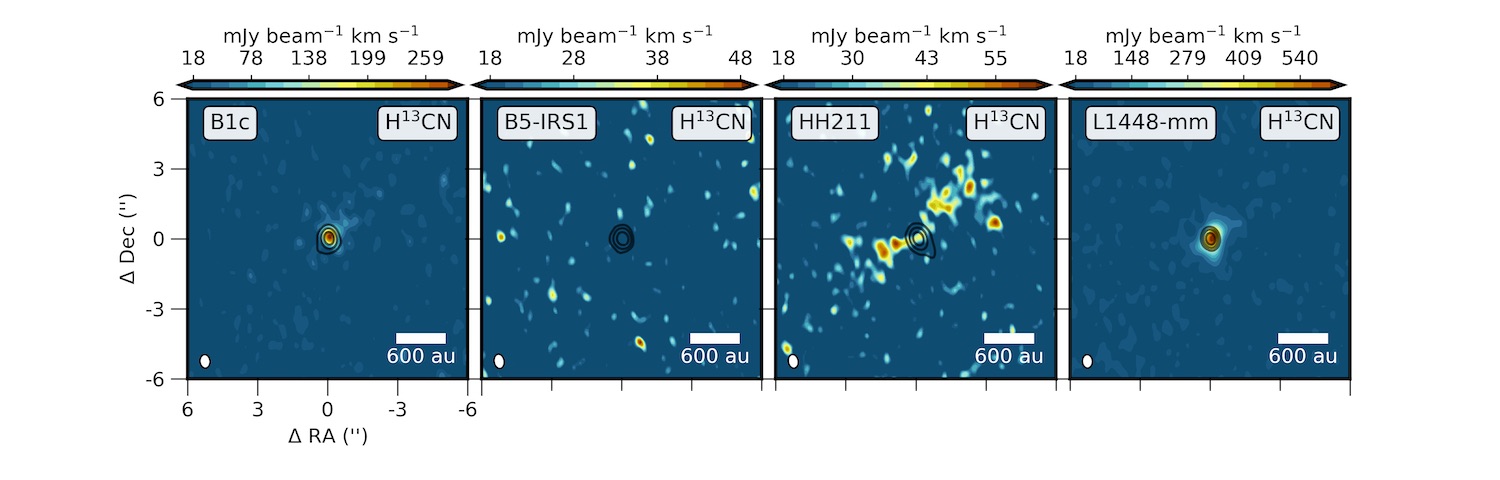}
    \caption{Moment maps of H$^{13}$CN  2--1 ($E_{\rm up}$=14 K) at 0\farcs4 resolution. Contours: continuum, color scale: moment 0 map integrated from -7 to 7 km s$^{-1}$ w.r.t  $\varv_{\rm sys}$ }

 \label{fig:h13cn}
 
 \end{figure*}

\begin{figure}
\centering
  \includegraphics[width=0.85\linewidth,trim={0cm 0cm 0cm 0cm}]
  {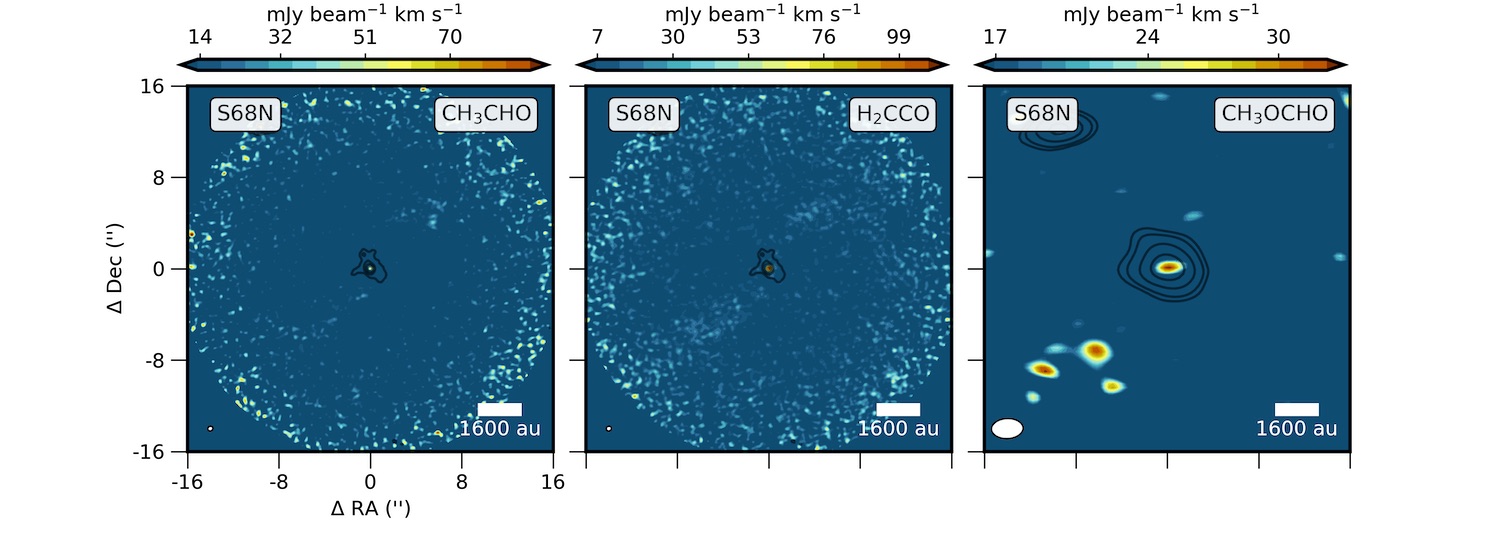}
  \caption{Additional plots of sputtering products for S68N. CH$_3$CHO 14$_{0,14}$--13$_{0,13}$($E_{\rm up}=96$ K), 
 H$_2$CCO 13$_{1,13}$--12$_{1,12}$ ($E_{\rm up}$=101 K)  which are separated by 4 km s$^{-1}$, CH$_3$OCHO 10$_{0,10}$--9$_{0,9}$ ($E_{\rm up}$=30 K). } 
 \label{fig:fig9_extra}
\end{figure}

\begin{figure}
\centering
  \includegraphics[width=0.95\linewidth,trim={0cm 0cm 0cm 0cm}]
  {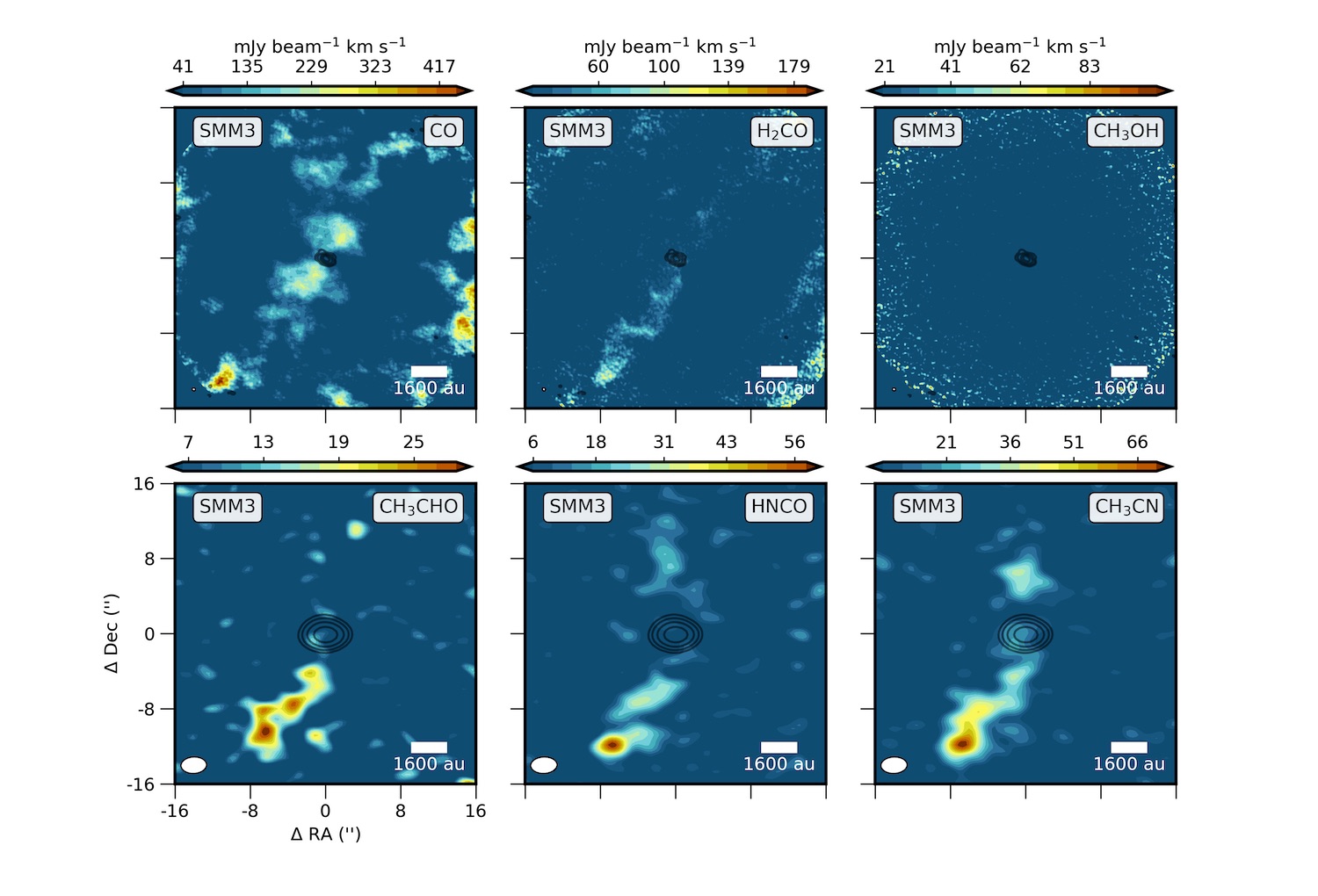}
  \caption{ Moment 0 map in colourscale and 1.3 mm continuum in contours of SMM3.  CO 2--1 ($E_{\rm up}$=17 K), H$_2$CO 3$_{2,1}$-2$_{2,0}$ $E_{\rm up}$= 68 K and CH$_3$OH  4$_{2}$--5$_{1}$ $E_{\rm up}=$60 K moment 0 maps obtained in Band 6 at 0\farcs5 resolution. 
  CH$_3$CHO 6$_{1,6,1}$--5$_{1,5,1}$ $E_{\rm up}$=21 K, HNCO 5$_{0,5}$--4$_{0,4}$ $E_{\rm up}$=16 K, CH$_3$CN 6$_0$--5$_0$ $E_{\rm up}$=19 K, and CH$_3$OCHO 9$_{5,5}$--8 $_{5,4}$ $E_{\rm up}$=43 K. Map of the sputtering products emission toward SMM3 in Band 3 at 3\arcsec\   resolution. Moment 0 map integrated from  -4 to 4 km s$^{-1}$ w.r.t 
 - $\varv_{\rm sys}$. } 
 \label{fig:161}
\end{figure}

\begin{figure}
\centering
  \includegraphics[width=0.95\linewidth,trim={0cm 0cm 0cm 0cm}]
  {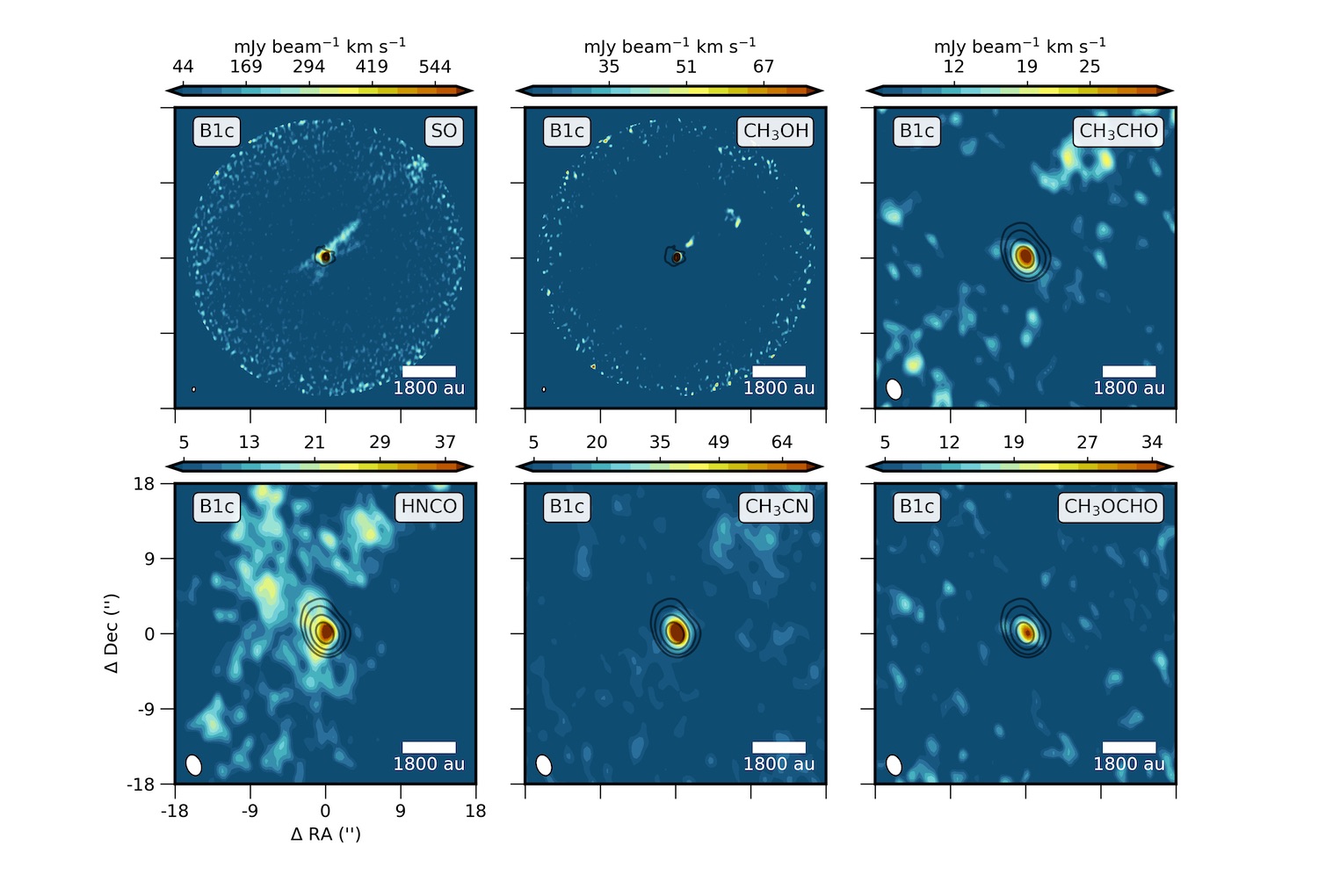}
  \caption{ Moment 0 map in colourscale and 1.3 mm continuum in contours of B1-c.  CO 2--1 $E_{\rm up}$=48 K,  CH$_3$OH  4$_{2}$--5$_{1}$ E$_{up}$=60 K moment 0 maps obtained in Band 6 at 0\farcs5 resolution. Sputtering products emission toward SMM3 in Band 3 at 3\arcsec\   resolution  CH$_3$CHO 6$_{1,6,1}$--5$_{1,5,1}$ E$_{\rm up}$=21 K, HNCO 5$_{0,5}$--4$_{0,4}$ E$_{\rm up}$=16 K, and CH$_3$CN 6$_0$-5$_0$ E$_{\rm up}$=19 K and CH$_3$OCHO 9$_{5,5}$--8$_{5,4}$  E$_{\rm up}$=43 K. Moment 0 map integrated from  -10 to 10 km s$^{-1}$ w.r.t  $\varv_{\rm sys} $ for SO and from -2 to 2 km s$^{-1}$ w.r.t  $\varv_{\rm sys}$ for the other molecules.} 
 \label{fig:151}
\end{figure}

\clearpage

\section{Cavity walls}

\begin{figure*}[h]
\centering
      \includegraphics[width=0.99\linewidth,trim={2cm 0cm 2cm 0cm}]
  {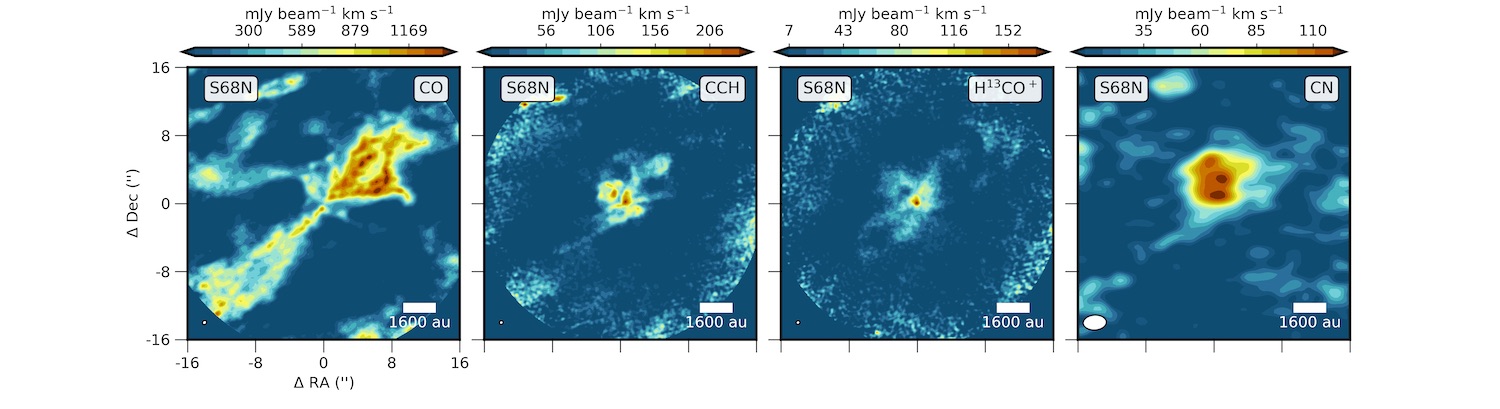}
  \includegraphics[width=0.99\linewidth,trim={2cm 0cm 2cm 0cm}]
  {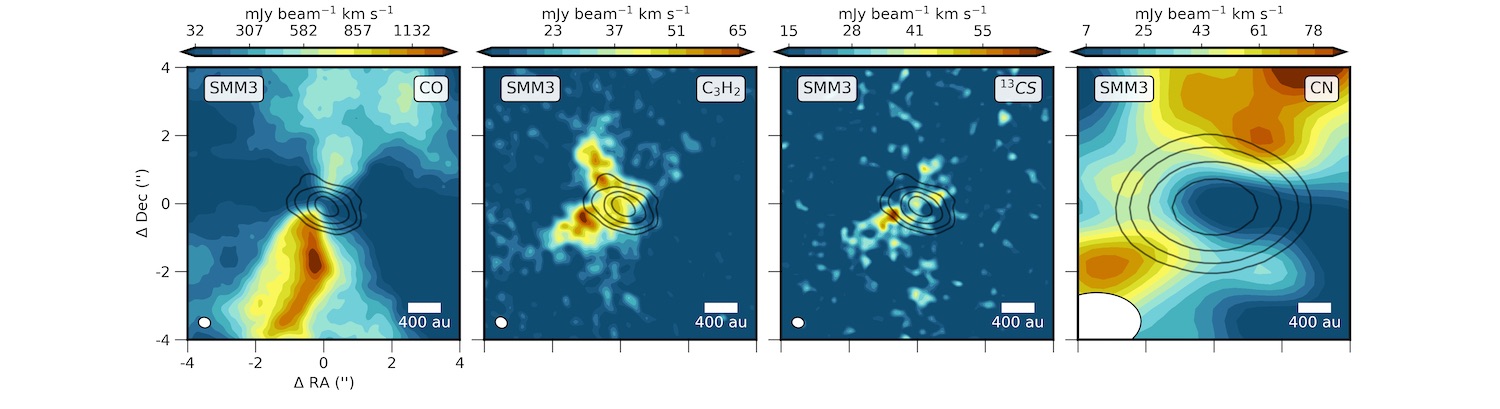}
    \includegraphics[width=0.78\linewidth,trim={2cm 0cm 2cm 0cm}]
  {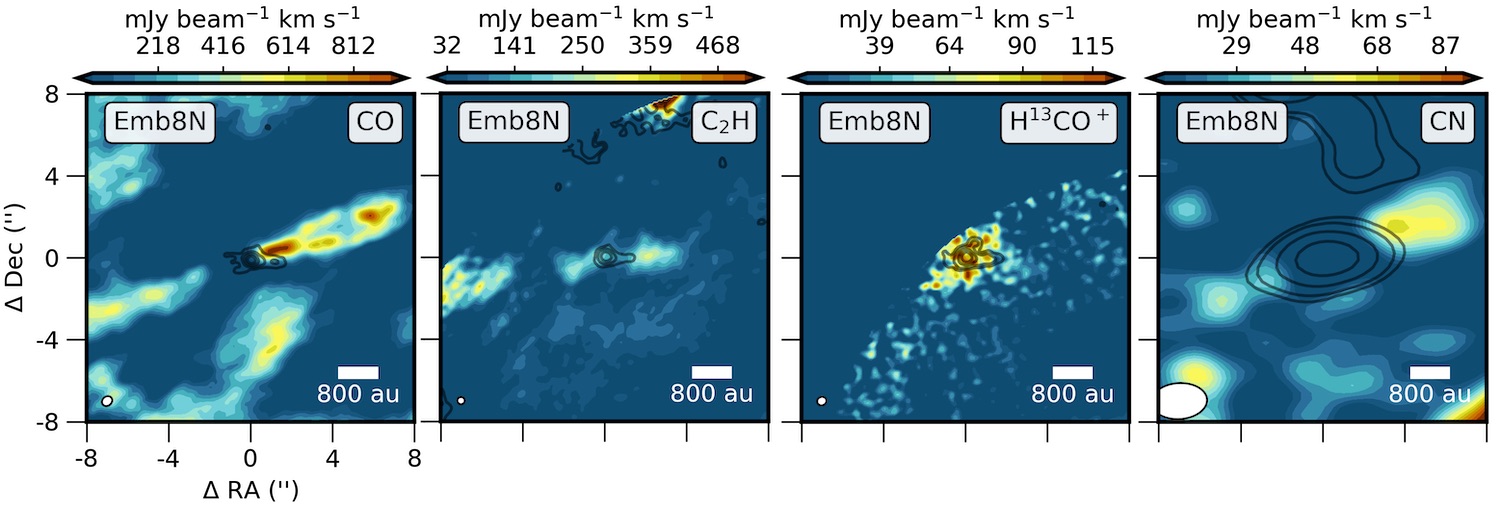}
  
\caption{Maps of the outflow cavity wall tracers toward S68N, TMC1, and Emb8N with the CO low-velocity outflow map for reference. {\it Top:} Moment 0 maps toward S68N of CO, C$_2$H, H$^{13}$CO$^+$ obtained in Band 6 at 0\farcs5 resolution and CN in Band 3 at 3\arcsec. {\it Middle:} Moment 0 maps toward TMC1 of CO, c-C$_3$H$_2$  and $^{13}$CS and CN obtained at 0\farcs5. The emission is integrated from -5 to -1 km s$^{-1}$ and from 1 to 5 km s$^{-1}$ w.r.t  $\varv_{\rm sys} $.  {\it Bottom:} Moment 0 maps toward Emb8N of CO, c-C$_2$H, H$^{13}$CO$^+$ $^{13}$CS at 0\farcs5 in Band 6. and CN in Band 3 at 3\arcsec. The emission is integrated from -5 to 5 km s$^{-1}$ w.r.t  $\varv_{\rm sys} $.}

 \label{fig:hydrocarbons_no2}
\end{figure*}

\begin{figure*}[h]
\centering
      \includegraphics[width=0.7\linewidth,trim={0cm 0cm 0cm 0cm}]
  {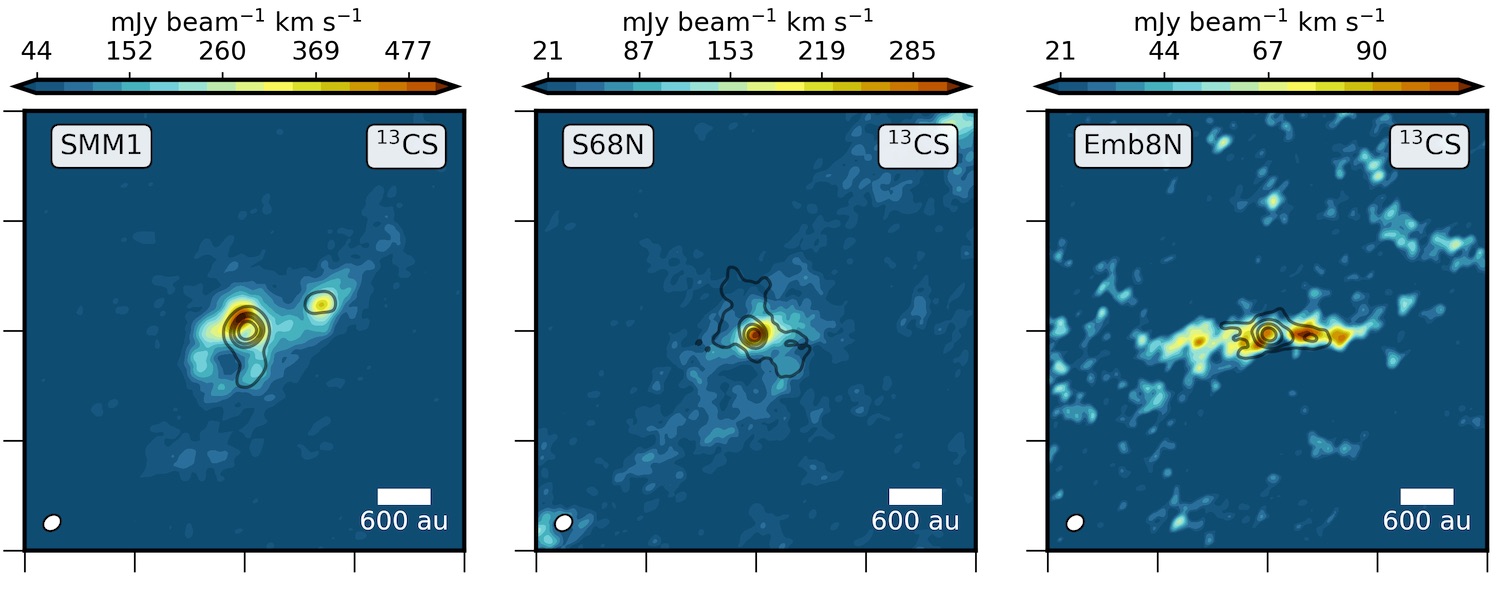}

\caption{ $^{13}$CS 5--4 ($E_{\rm up}=$33 K)  integrated from -5 to 5 km s$^{-1}$ w.r.t  $\varv_{\rm sys} $ in Band 6 at 0\farcs5.}

 \label{fig:fig_E2}
\end{figure*}

\begin{figure*}[h]
\centering
      \includegraphics[width=0.75\linewidth,trim={2cm 0cm 2cm 0cm}]
  {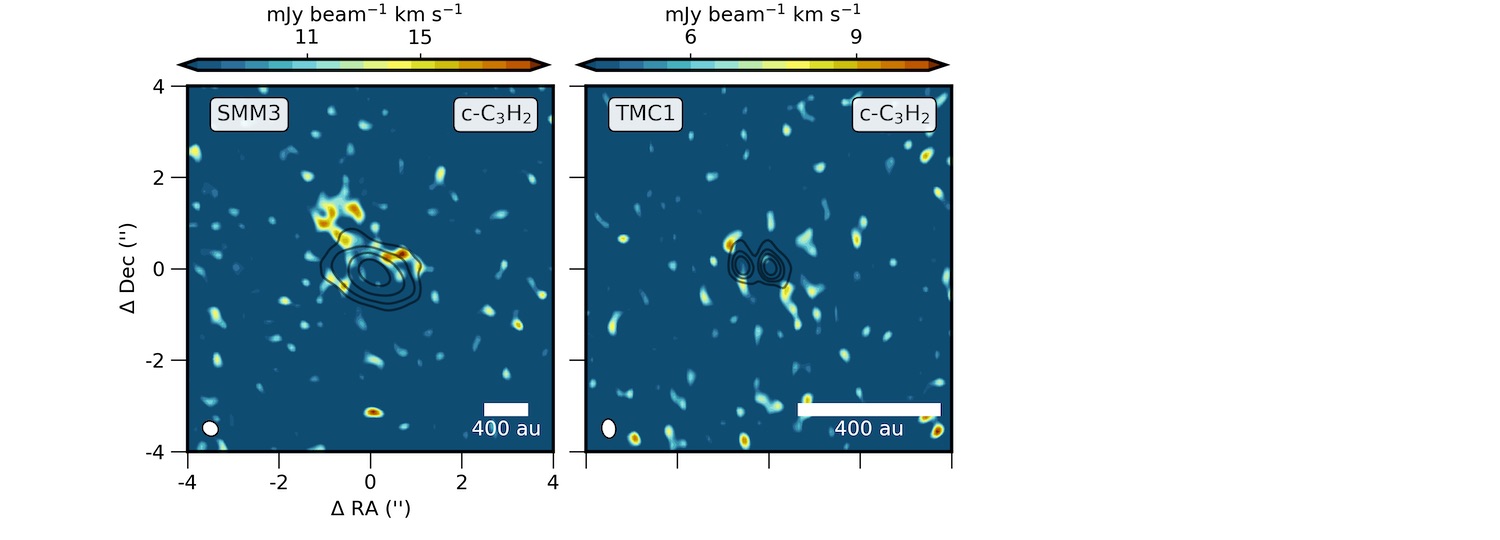}
  
\caption{ c-C$_3$H$_2$ 7$_{2,6}$--7$_{1,7}$ ($E_{\rm up}=$63 K)    integrated from -1.5 to 1.5 km s$^{-1}$ w.r.t  $\varv_{\rm sys} $.}

 \label{fig:fig132}
\end{figure*}

\clearpage

\section{Inner envelope}

\begin{figure*}[h]
\centering
      \includegraphics[width=0.8\linewidth,trim={0cm 0cm 0cm 0cm}]
  {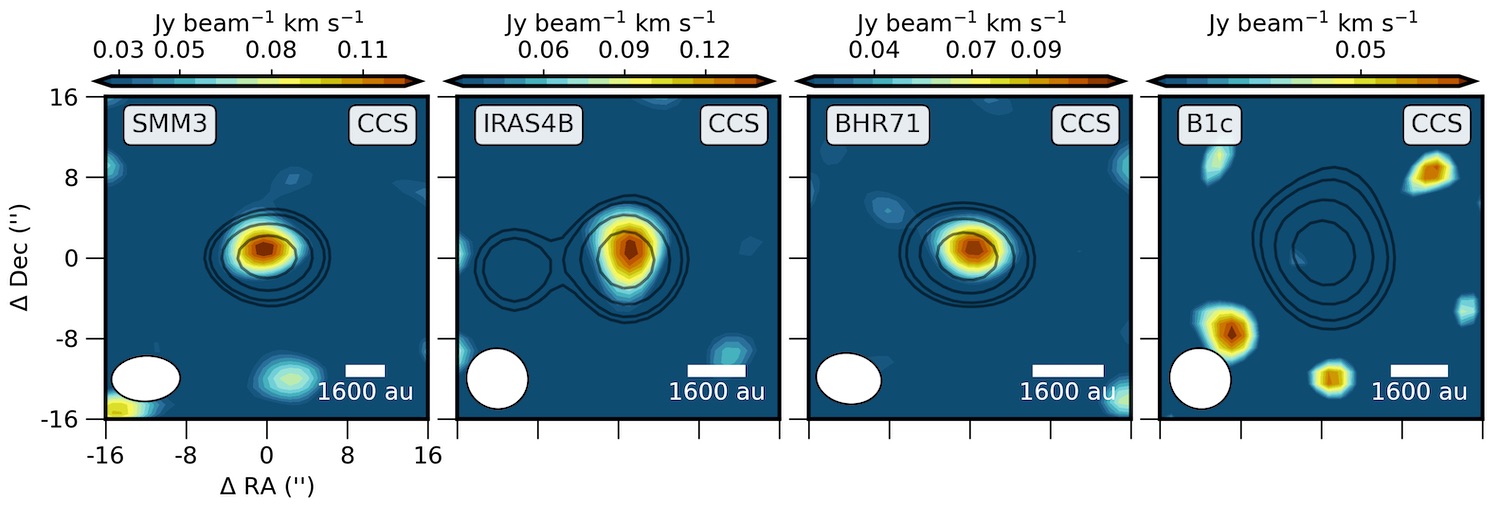}
      \includegraphics[width=0.8\linewidth,trim={0cm 0cm 0cm 0cm}]
  {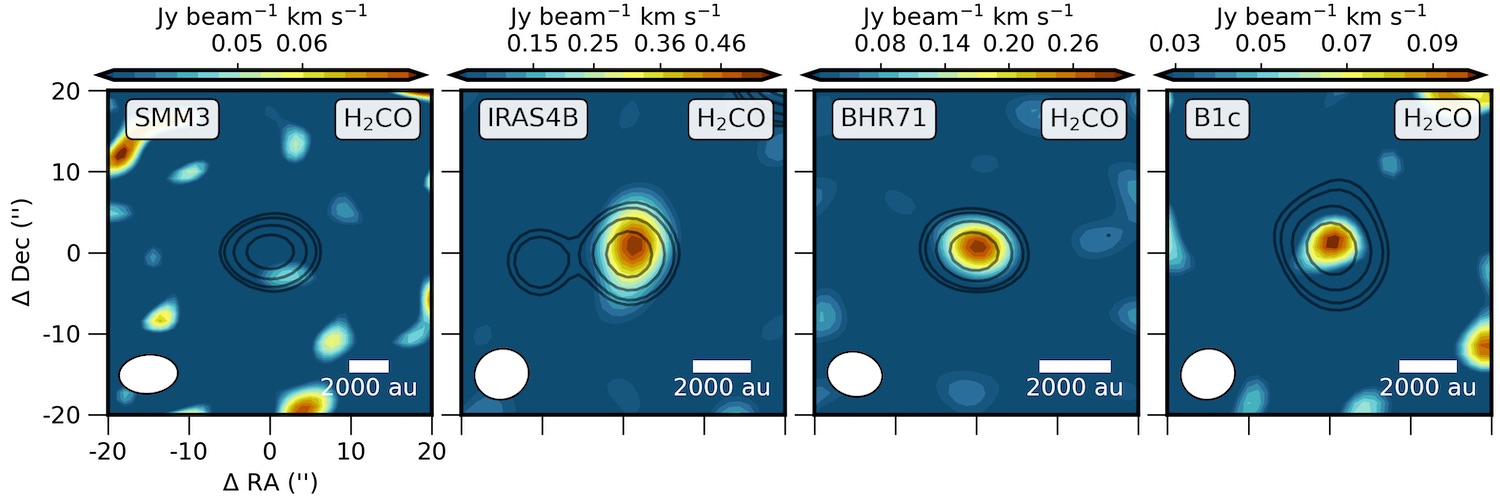}
        \includegraphics[width=0.8\linewidth,trim={0cm 0cm 0cm 0cm}]
  {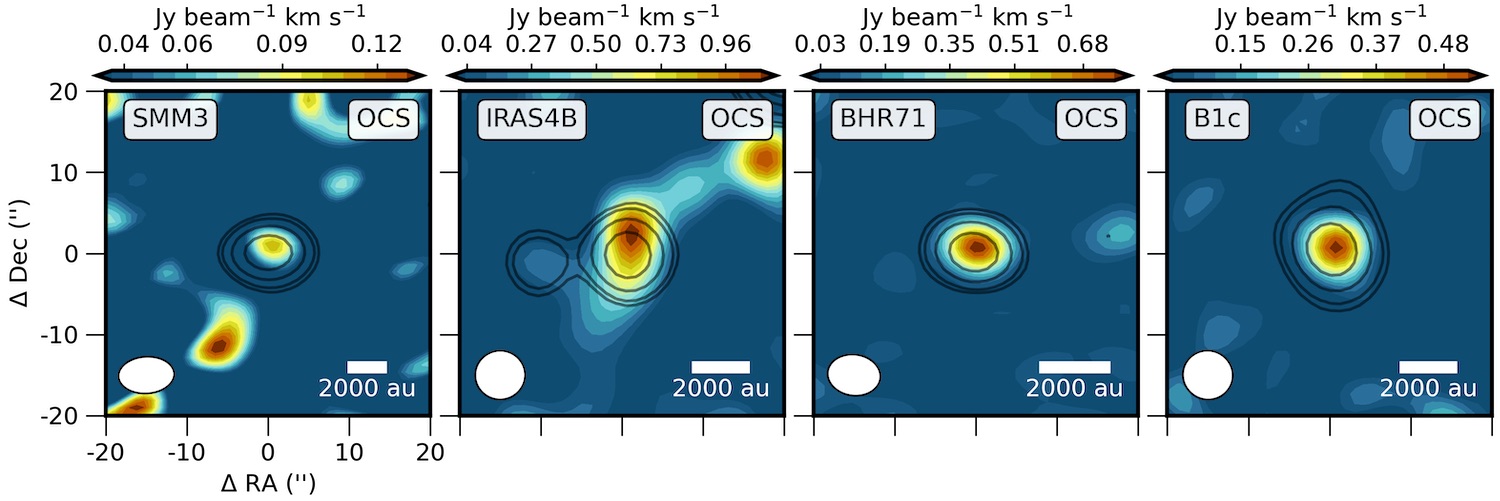}
        \includegraphics[width=0.8  \linewidth,trim={0cm 0cm 0cm 0cm}]
  {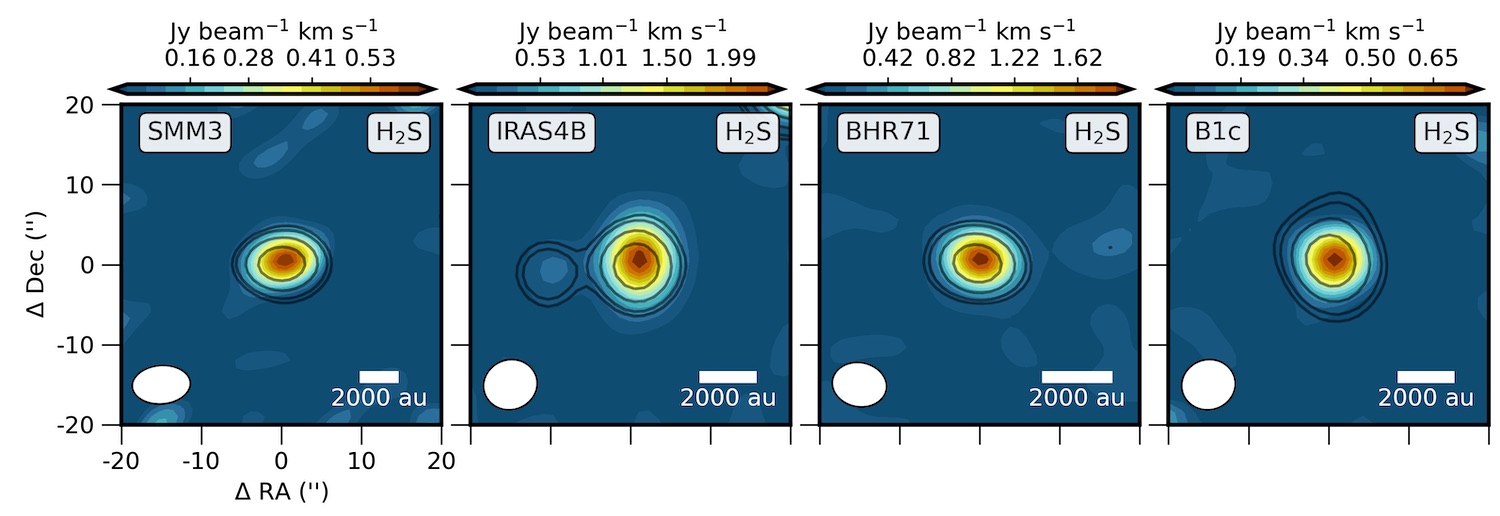}

\caption{ CCS 19$_{18}$--18$_{17}$ $E_{\rm up}$=111 K, OCS 19--18  $E_{\rm up}$=111 K, H$_2$CO 9$_{1,8}$--9$_{1,9}$  $E_{\rm up}$=174 K, H$_2$S 2$_{2}$--2$_{1}$ $E_{\rm up}$=83 K 
moment 0 maps from ACA Band 6 observations and 6\arcsec\   resolution integrated from -2 to 2 km s$^{-1}$ w.r.t  $\varv_{\rm sys} $. }

 \label{fig:102}
\end{figure*}

\begin{figure*}[h]
\centering
      \includegraphics[width=0.99\linewidth,trim={0cm 0cm 0cm 0cm}]
  {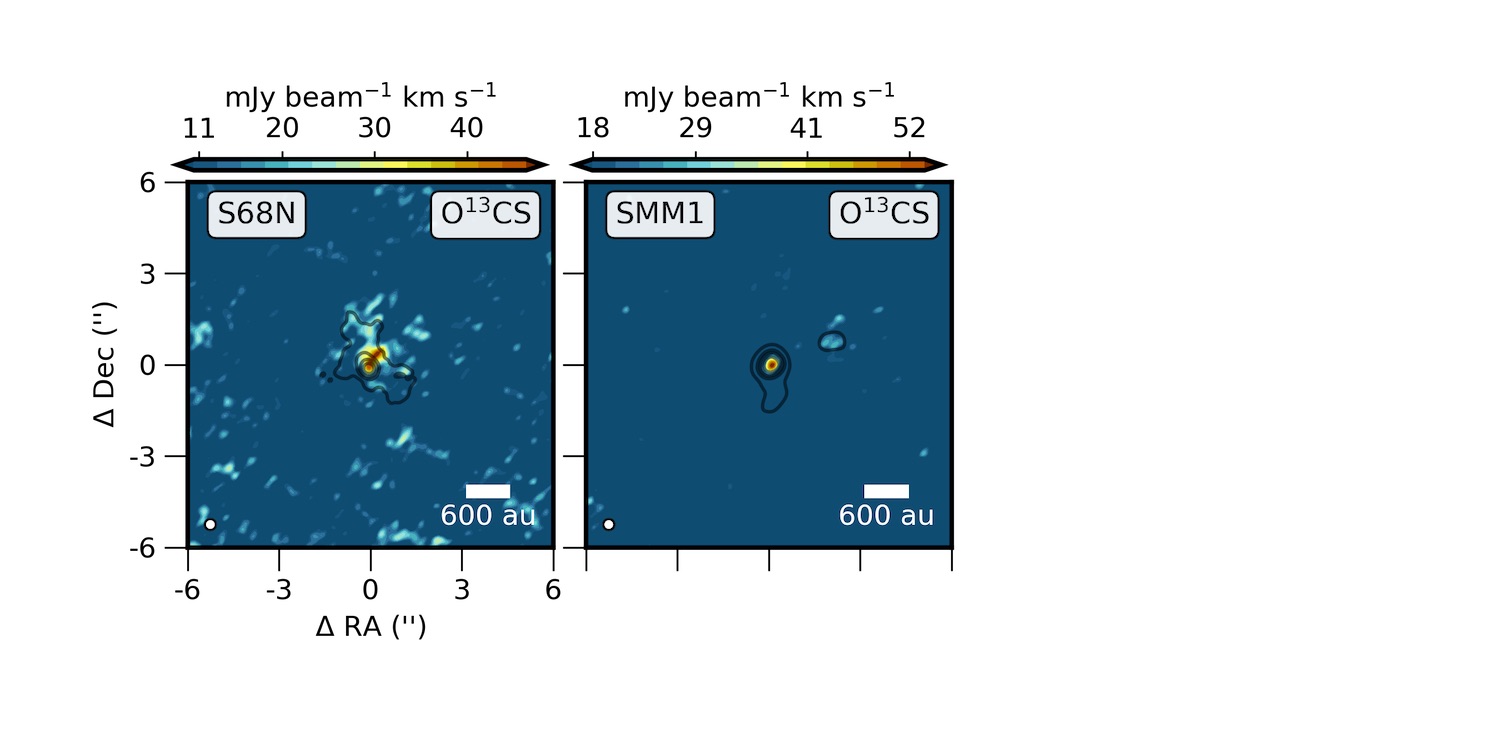}

\caption{ O$^{13}$CS 18--17 $E_{\rm up}$=99 K integrated from -3 to 3 km s$^{-1}$ w.r.t  $\varv_{\rm sys} $.}

 \label{fig:o13cs_appendix}
\end{figure*}

\section{Additional plots}

\begin{figure*}[h]
\centering
  \includegraphics[width=0.5\linewidth,trim={0cm 0cm 0cm 0cm}]
  {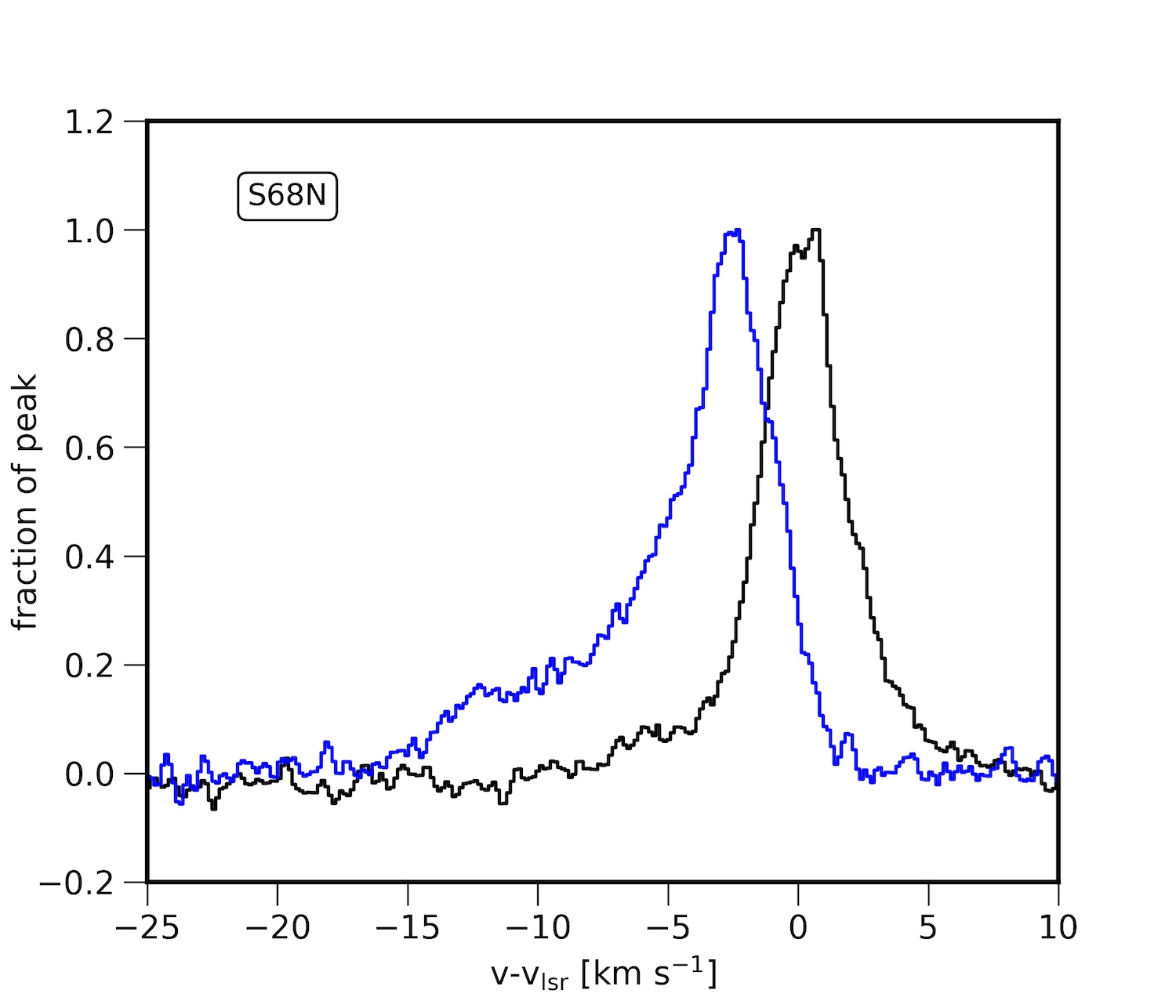}
  
  \caption{Spectra of SO 6$_7$--5$_6$ for S68N. Spectra averaged over the 0\farcs6 diameter circle on positions: central source (black), blueshifted outflow (blue). Both spectra normalized to the peak emission at the position} 
 \label{fig:so_spectra_v1}
\end{figure*}

\begin{figure}[h]
\centering
  \includegraphics[width=0.5\linewidth,trim={0cm 0cm 0cm 0cm}]
  {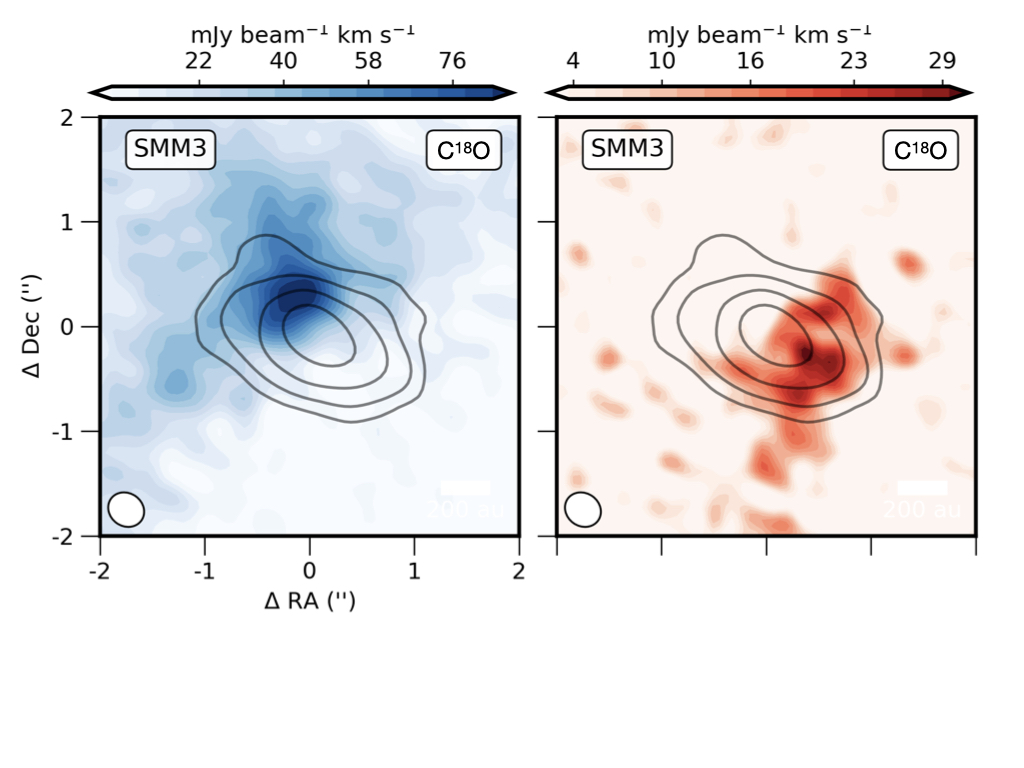}

  \caption{Map of the C$^{18}$O emission toward SMM3. Moment 0 map in colourscale and 1.3 mm continuum in contours. {\it  Left:} Moment 0 map integrated from  -3 to -0.5 km s$^{-1}$ w.r.t  $\varv_{\rm sys} $. {\it Right:} Moment 0 map integrated from 0.5 to 3 km s$^{-1}$ w.r.t  $\varv_{\rm sys} $. } 
 \label{fig:smm3_disk}
\end{figure}

\pagebreak
\newpage

\end{appendix}

f
\end{document}